# THE CHAIN OF QUANTUM MECHANICS EQUATIONS

**E.E. Perepelkin[a], B.I. Sadovnikov[a], N.G. Inozemtseva[b]**

[a] *Faculty of Physics, Lomonosov Moscow State University, Moscow, 119991 Russia*
E. Perepelkin e-mail: pevgeny@jinr.ru, B. Sadovnikov e-mail: sadovnikov@phys.msu.ru
[b] *Dubna State University, Moscow region, Moscow,141980 Russia*
N. Inozemtseva e-mail: nginozv@mail.ru

**Abstract**

The chain of quantum mechanics equations is constructed for wave functions depending on coordinate and time as well as on velocity, acceleration and acceleration of higher orders.

**Introduction**

The key concept of quantum mechanics is the wave function $\Psi(\vec{r},t)$. Knowing the wave function, one can define the probability density function $f(\vec{r},t)=|\Psi(\vec{r},t)|^2$. The function $f(\vec{r},t)$ defines the probability of a random variable $\vec{R}$ to take values $\vec{r}$.

For comprehensive consideration of systems described by random variables of the coordinate ($\vec{R}$) and velocity ($\vec{V}$), it is required to know the function $f(\vec{r},\vec{v},t)$. The consideration of the function $f(\vec{r},\vec{v},t)$ requires introduction of a new equation so that the function should satisfy it. The construction of the functions $f(\vec{r},\vec{v},t)$ from the function $f(\vec{r},t)$ was performed by E.P. Wigner and H. Weyl [1,2]. Thus, the Wigner function satisfies the Moyal equation [3].

In quantum mechanics and during construction of the Wigner function, the Fourier transform $\mathcal{F}$ is used as a «tool» for transition to the momentum representation $f(\vec{r},t) \xrightarrow{\mathcal{F}} f(\vec{p},t)$. Here we face a number of problems.

First, for purposes of the probability theory, the distribution density function $f(\vec{r},\vec{v},t)$ construction of the two random variables $\vec{R}$ and $\vec{V}$ from the distribution density function $f(\vec{r},t)$ of the random variable $\vec{R}$ requires solid argumentation. In the general case, such a construction is wrong at least because the Fourier transform is an automorphism with a period 4 [4], that is $\mathcal{F}^4[\Psi]=\Psi=\mathcal{F}^0[\Psi]$. Thus, a limitation appears at constructing the probability density function $f(\vec{r},\vec{v},\dot{\vec{v}},\ddot{\vec{v}},\dddot{\vec{v}},t)$ and higher. In addition, the distribution function $f(\vec{r},t)$ of the random variable $\vec{R}$ does not contain complete information on another random variable $\vec{V}$. In the general case, the random variables $\vec{R}$ and $\vec{V}$ may be independent.

Secondly, a question arises: «What equations should the functions $f(\vec{r},\vec{v},t)$, $f(\vec{r},\vec{v},\dot{\vec{v}},t)$, $f(\vec{r},\vec{v},\dot{\vec{v}},\ddot{\vec{v}},t)$, ... satisfy?» For the Wigner function, the Moyal equation was constructed in an «artificial» way (not from the first principles).

Thirdly, when considering the problem of a quantum harmonic oscillator the question of the positivity of the function $f(\vec{r},\vec{v},t)$ arises. Only the ground state of the oscillator provides positive probability density.

In the paper, we present a new approach to the construction of quantum mechanics and statistical physics described in kinematic variables of higher orders. The approach is based on the infinite coupled chain of Vlasov equations [5,6] for the probability density functions $f(\vec{r},t)$, $f(\vec{r},\vec{v},t)$, $f(\vec{r},\vec{v},\dot{\vec{v}},t)$, $f(\vec{r},\vec{v},\dot{\vec{v}},\ddot{\vec{v}},t)$,…. Such an approach has a number of strengths.

First, A.A. Vlasov obtained the infinite chain of equations from the first principles and it determines the infinite set of equations for the probability density functions.

Secondly, the chain of Vlasov equations agrees with the mathematical apparatus of the probability theory.

Thirdly, the chain of Vlasov equations explains the reason of negativity and positivity of a probability density function.

Fourthly, in particular cases it gives the same results as the «classical» apparatus of quantum mechanics.

Fifth, the Vlasov equation chain gives the opportunity to construct an infinite chain of equations of quantum mechanics for the wave functions $\Psi(\vec{r},t)$, $\Psi(\vec{r},\vec{v},t)$, $\Psi(\vec{r},\vec{v},\dot{\vec{v}},t)$, $\Psi(\vec{r},\vec{v},\dot{\vec{v}},\ddot{\vec{v}},t)$,…, for which $|\Psi|^2 = f$. Thus, it solves the question of positivity of the probability density functions $f = |\Psi|^2 \geq 0$.

Sixthly, the given approach allows one to extend the Lagrangian and Hamiltonian formalisms in a generalized phase space [7, 8]. The basis of this approach was laid by M.V. Ostrogradsky [9].

Seventh, a new $\Psi$-algebra is being constructed which makes it possible to write generalized equations of the electromagnetic field taking into account higher kinematic characteristics. In the particular case, generalized equations become the known Maxwell's equations.

The paper has the following structure. In §1, the $\Psi$-algebra construction is considered. The $\Psi$-algebra allows us to write in a compact form the results obtained in the following sections. The elements of the $\Psi$-algebra are $\Psi$- sets that are appropriate for describing hypercomplex numbers, RS-vectors (Riemann-Silberstein vector), spinors, bispinors and many other objects. In the $\Psi$-algebra, the operations of addition and multiplication are introduced. Differential operators $\Psi$-gradient, $\Psi$-divergence, $\Psi$-rotor are constructed, which are analogues of the gradient, divergence and rotor in the vector analysis.

In §2, based on the chain of Vlasov equations, a chain of quantum mechanics equations is constructed for the wave functions $\Psi(\vec{r},t)$, $\Psi(\vec{r},\vec{v},t)$, $\Psi(\vec{r},\vec{v},\dot{\vec{v}},t)$, $\Psi(\vec{r},\vec{v},\dot{\vec{v}},\ddot{\vec{v}},t)$,…, for which $|\Psi|^2 = f$. The construction is performed with the use of the approach we described in [10]. The equation for the wave function $\Psi(\vec{r},t)$ coincides with the Schrödinger equation. The equations for the wave functions $\Psi(\vec{r},\vec{v},t)$, $\Psi(\vec{r},\vec{v},\dot{\vec{v}},t)$, $\Psi(\vec{r},\vec{v},\dot{\vec{v}},\ddot{\vec{v}},t)$,… are new ones [16,17].

In §3, we consider the extension of the Lagrangian and Hamiltonian formalisms to the case of higher kinematic characteristics. The expression for the generalized potential energy $U(\vec{r},t)$, $U(\vec{r},\vec{v},t)$, $U(\vec{r},\vec{v},\dot{\vec{v}},t)$, $U(\vec{r},\vec{v},\dot{\vec{v}},\ddot{\vec{v}},t)$,… is obtained. The energy is contained in new wave equations for the functions $\Psi(\vec{r},t)$, $\Psi(\vec{r},\vec{v},t)$, $\Psi(\vec{r},\vec{v},\dot{\vec{v}},t)$, $\Psi(\vec{r},\vec{v},\dot{\vec{v}},\ddot{\vec{v}},t)$,… The concept of the quantum potential $Q(\vec{r},t)$ is expanded, which is used in the de Broglie-Bohm «wave-pilot» theory [11,12]. The result is the obtained relations for $Q(\vec{r},\vec{v},t)$, $Q(\vec{r},\vec{v},\dot{\vec{v}},t)$, $Q(\vec{r},\vec{v},\dot{\vec{v}},\ddot{\vec{v}},t)$,… and consideration of their relation to the classical potentials. The expression for the generalized Lagrangian and Hamiltonian is obtained as well as for the Legendre transformation linking them. The Lagrangian in the general case has a complex representation [13]. The complex form of the action $Z$ is obtained and the map $\Psi = e^Z$ is considered.

In §4 the generalized equations of motion are obtained for the change of the mean kinematic variables $\langle \vec{v} \rangle$, $\langle \dot{\vec{v}} \rangle$, $\langle \ddot{\vec{v}} \rangle$,…. In form, the generalized equations resemble the classical motion equations in the electromagnetic field. There is an analogue of «the electric and magnetic» field in the equations. In the particular case, the generalized motion equations become the classical motion equations. The higher-order kinematic variables taking into consideration modify the fields and its equations. The hydrodynamic representation of generalized motion equations is considered. The expression for the generalized quantum pressure is obtained.

In §5 the equations for analogues of «electromagnetic» fields are constructed. The obtained equations can be divided into 4 groups. In the particular case each group transforms into one of equations from the Maxwell's equations system.

In §6 the example of the quantum harmonic oscillator is considered. The probability density functions $f_n(\vec{r},t)$, $f_n(\vec{r},\vec{v},t)$, $f_n(\vec{r},\vec{v},\dot{\vec{v}},t)$, $f_n(\vec{r},\vec{v},\dot{\vec{v}},\ddot{\vec{v}},t)$,… are obtained, where $n$ - is the state number. The function $f_n(\vec{r},t)$ coincides with the known solution of the Schrödinger equation for the harmonic oscillator. The function $f_n(\vec{r},\vec{v},t)$ coincides with the Wigner function for the harmonic oscillator. The functions $f_n(\vec{r},\vec{v},\dot{\vec{v}},t)$, $f_n(\vec{r},\vec{v},\dot{\vec{v}},\ddot{\vec{v}},t)$,… are the new functions. As an example, the solution of the wave equation for the function $\Psi(\vec{r},\vec{v},t)$ is given. It is shown that the wave equation for the function $\Psi(\vec{r},\vec{v},t)$ has only one solution satisfying the ground state. The expression for the potential $U(\vec{r},\vec{v},t)$ and the motion equation are obtained. The example of calculating the mean values of velocity, accelerations and accelerations of higher orders and their standard deviations is given.

In conclusion, the main results of the work are given.

## §1 $\Psi$ –algebra

Let us construct $\Psi$ -algebra. Let $\mathbb{N}_2$ be a set of natural numbers of the form

$$\mathbb{N}_2 = \{n_k \in \mathbb{N} : n_{k+1} = 2n_k, n_1 = 1, k \in \mathbb{N}\}, \tag{1.1}$$

that is

$$\mathbb{N}_2 = \{1, 2, 4, 8, 16, 32, ...\}.$$

**Definition 1.** *Let us consider a finite ordered set $n \in \mathbb{N}_2$ of the scalar variables $\left\{ {}^{(r)}\chi, {}^{(v)}\chi, {}^{(\dot{v})}\chi, ..., {}^{\binom{(n-1)}{v}}\chi \right\}$, where ${}^{(\bullet)}\chi$ in the general case are hypercomplex numbers.*
*Let us call the set of values*

$$_n\underline{\chi} = \begin{pmatrix} {}^{(r)}\chi \\ {}^{(v)}\chi \\ ... \\ {}^{\binom{(n-1)}{v}}\chi \end{pmatrix}, \quad {}^1_n\underline{\chi} \stackrel{det}{=} {}^{(r)}\chi, \; {}^2_n\underline{\chi} \stackrel{det}{=} {}^{(v)}\chi, ..., {}^n_n\underline{\chi} \stackrel{det}{=} {}^{\binom{(n-1)}{v}}\chi, \qquad (1.2)$$

$\Psi$ - scalar of the order $n$ or $\Psi$-set (PSI-set) of the order $n$ and rank 0. Let us denote the set of elements $_n\underline{\chi}$ as $_n\Omega^{(0)}$, that is $_n\underline{\chi} \in {}_n\Omega^{(0)}$.

Let is denote the addition operation as «+» and the multiplication «∘» for the elements $_n\underline{\chi}, {}_n\underline{\eta} \in {}_n\Omega^{(0)}$.

**Definition 2.** *The element $_n\underline{\lambda} \in {}_n\Omega^{(0)}$ is called the sum $_n\underline{\chi} + {}_n\underline{\eta}$ of the elements $_n\underline{\chi}, {}_n\underline{\eta} \in {}_n\Omega^{(0)}$*

$$_n\underline{\lambda} = {}_n\underline{\chi} + {}_n\underline{\eta} = \begin{pmatrix} {}^{(r)}\chi \\ {}^{(v)}\chi \\ ... \\ {}^{\binom{(n-1)}{v}}\chi \end{pmatrix} + \begin{pmatrix} {}^{(r)}\lambda \\ {}^{(v)}\lambda \\ ... \\ {}^{\binom{(n-1)}{v}}\lambda \end{pmatrix} = \begin{pmatrix} {}^{(r)}\chi + {}^{(r)}\lambda \\ {}^{(v)}\chi + {}^{(v)}\lambda \\ ... \\ {}^{\binom{(n-1)}{v}}\chi + {}^{\binom{(n-1)}{v}}\lambda \end{pmatrix}. \qquad (1.3)$$

To define the multiplication operation «∘», let us introduce the notion of the correspondence matrix.

**Definition 3.** *Let $_n\underline{\chi}, {}_n\underline{\eta} \in {}_n\Omega^{(0)}$. The matrix $_nM(\underline{\chi\eta})$ of the size $n \times n$ of the form*

$$_n\underline{\chi} \circ {}_n\underline{\eta} \mapsto {}_nM(\underline{\chi\eta}) = \begin{pmatrix} {}^{(r)}\chi^{(r)}\eta & {}^{(r)}\chi^{(v)}\eta & ... & {}^{(r)}\chi^{\binom{(n-1)}{v}}\eta \\ {}^{(v)}\chi^{(r)}\eta & {}^{(v)}\chi^{(v)}\eta & ... & {}^{(v)}\chi^{\binom{(n-1)}{v}}\eta \\ ... & ... & ... & ... \\ {}^{\binom{(n-1)}{v}}\chi^{(r)}\eta & {}^{\binom{(n-1)}{v}}\chi^{(v)}\eta & ... & {}^{\binom{(n-1)}{v}}\chi^{\binom{(n-1)}{v}}\eta \end{pmatrix} \qquad (1.4)$$

*we call the correspondence matrix of the elements $_n\underline{\chi} \circ {}_n\underline{\eta}$.*

**Corollary 1.** *If $_nM(\underline{\chi\eta})$ is the correspondence matrix for $_n\underline{\chi} \circ {}_n\underline{\eta}$, where $_n\underline{\chi}, {}_n\underline{\eta} \in {}_n\Omega^{(0)}$, then $_nM^T(\underline{\chi\eta}) = {}_nM(\underline{\eta\chi})$ is the correspondence matrix for $_n\underline{\eta} \circ {}_n\underline{\chi}$.*

**Definition 4.**

The convolution operation of the matrix ${}_{n_k}M$

$$
{}_{n_k}M = \{\mu_{rs}\} = \begin{pmatrix}
\boxed{2\times2} & \cdots & \boxed{2\times2} & \overbrace{\boxed{2\times2} \quad \cdots \quad \boxed{2\times2}}^{n_{k-1}=2n_{k-2}} \\
\vdots & \ddots & \vdots & \vdots & \ddots & \vdots \\
\boxed{2\times2} & \cdots & \boxed{2\times2} & \boxed{2\times2} & \cdots & \boxed{2\times2} \\
\boxed{2\times2} & \cdots & \boxed{2\times2} & \boxed{2\times2} & \cdots & \boxed{2\times2} \\
\vdots & \ddots & \vdots & \vdots & \ddots & \vdots \\
\boxed{2\times2} & \cdots & \boxed{2\times2} & \boxed{2\times2} & \cdots & \boxed{2\times2}
\end{pmatrix}, \qquad (1.5)
$$
$$\underbrace{\phantom{}}_{n_k = 2n_{k-1}}$$

$\boxed{2\times2}$ *is a matrix of the size* $2\times2$,

*is the operation of transforming the matrix* ${}_{n_k}M$ *in* $k-1$, $k \in \mathbb{N}$ *steps in a column* ${}_{n_k}^{*}M$ *of* $n_k \in \mathbb{N}_2$ *elements according to the following rule*

*for* $n_1 = 1$

$$
{}_1M = \{\mu_{rs}\} = (\mu_{11}) \stackrel{\text{det}}{=} {}_1^{*}M,
$$

*for* $n_2 = 2$

$$
{}_2M = \{\mu_{rs}\} = \begin{pmatrix} \mu_{11} & \mu_{12} \\ \mu_{21} & \mu_{22} \end{pmatrix} \stackrel{\text{step 1}}{\sim} \begin{pmatrix} \mu_{11} + \mu_{22} \\ \mu_{21} + \mu_{12} \end{pmatrix} \stackrel{\text{det}}{=} {}_2^{*}M,
$$

*for* $n_3 = 4$

$$
{}_4M = \begin{pmatrix} \boxed{\begin{matrix}\mu_{11} & \mu_{12} \\ \mu_{21} & \mu_{22}\end{matrix}} & \boxed{\begin{matrix}\mu_{13} & \mu_{14} \\ \mu_{23} & \mu_{24}\end{matrix}} \\ \boxed{\begin{matrix}\mu_{31} & \mu_{32} \\ \mu_{41} & \mu_{42}\end{matrix}} & \boxed{\begin{matrix}\mu_{33} & \mu_{34} \\ \mu_{43} & \mu_{44}\end{matrix}} \end{pmatrix} \stackrel{\text{step 1}}{\sim} \begin{pmatrix} \begin{pmatrix}\mu_{11}+\mu_{22} \\ \mu_{21}+\mu_{12}\end{pmatrix} & \begin{pmatrix}\mu_{13}+\mu_{24} \\ \mu_{23}+\mu_{14}\end{pmatrix} \\ \begin{pmatrix}\mu_{31}+\mu_{32} \\ \mu_{41}+\mu_{12}\end{pmatrix} & \begin{pmatrix}\mu_{33}+\mu_{44} \\ \mu_{43}+\mu_{34}\end{pmatrix} \end{pmatrix} \stackrel{\text{step 2}}{\sim}
$$

$$
\stackrel{\text{step 2}}{\sim} \begin{pmatrix} \mu_{11}+\mu_{22}+\mu_{33}+\mu_{44} \\ \mu_{21}+\mu_{12}+\mu_{43}+\mu_{34} \\ \mu_{31}+\mu_{32}+\mu_{13}+\mu_{24} \\ \mu_{41}+\mu_{12}+\mu_{23}+\mu_{14} \end{pmatrix} \stackrel{\text{det}}{=} {}_4^{*}M,
$$

*for* $n_4 = 8$

$$_8M = \begin{pmatrix} \begin{bmatrix} \mu_{11} & \mu_{12} \\ \mu_{21} & \mu_{22} \end{bmatrix} & \begin{bmatrix} \mu_{13} & \mu_{14} \\ \mu_{23} & \mu_{24} \end{bmatrix} & \begin{bmatrix} \mu_{15} & \mu_{16} \\ \mu_{25} & \mu_{26} \end{bmatrix} & \begin{bmatrix} \mu_{17} & \mu_{18} \\ \mu_{27} & \mu_{28} \end{bmatrix} \\ \begin{bmatrix} \mu_{31} & \mu_{32} \\ \mu_{41} & \mu_{42} \end{bmatrix} & \begin{bmatrix} \mu_{33} & \mu_{34} \\ \mu_{43} & \mu_{44} \end{bmatrix} & \begin{bmatrix} \mu_{35} & \mu_{36} \\ \mu_{45} & \mu_{46} \end{bmatrix} & \begin{bmatrix} \mu_{37} & \mu_{38} \\ \mu_{47} & \mu_{48} \end{bmatrix} \\ \begin{bmatrix} \mu_{51} & \mu_{52} \\ \mu_{61} & \mu_{62} \end{bmatrix} & \begin{bmatrix} \mu_{53} & \mu_{54} \\ \mu_{63} & \mu_{64} \end{bmatrix} & \begin{bmatrix} \mu_{55} & \mu_{56} \\ \mu_{65} & \mu_{66} \end{bmatrix} & \begin{bmatrix} \mu_{57} & \mu_{58} \\ \mu_{67} & \mu_{68} \end{bmatrix} \\ \begin{bmatrix} \mu_{71} & \mu_{72} \\ \mu_{81} & \mu_{82} \end{bmatrix} & \begin{bmatrix} \mu_{73} & \mu_{74} \\ \mu_{83} & \mu_{84} \end{bmatrix} & \begin{bmatrix} \mu_{75} & \mu_{76} \\ \mu_{85} & \mu_{86} \end{bmatrix} & \begin{bmatrix} \mu_{77} & \mu_{78} \\ \mu_{87} & \mu_{88} \end{bmatrix} \end{pmatrix} \overset{\text{step 1}}{\sim}$$

$$\overset{\text{step 1}}{\sim} \begin{pmatrix} \begin{pmatrix} \mu_{11}+\mu_{22} \\ \mu_{21}+\mu_{12} \end{pmatrix} & \begin{pmatrix} \mu_{13}+\mu_{24} \\ \mu_{23}+\mu_{14} \end{pmatrix} & \begin{pmatrix} \mu_{15}+\mu_{26} \\ \mu_{25}+\mu_{16} \end{pmatrix} & \begin{pmatrix} \mu_{17}+\mu_{28} \\ \mu_{27}+\mu_{18} \end{pmatrix} \\ \begin{pmatrix} \mu_{31}+\mu_{42} \\ \mu_{41}+\mu_{32} \end{pmatrix} & \begin{pmatrix} \mu_{33}+\mu_{44} \\ \mu_{43}+\mu_{34} \end{pmatrix} & \begin{pmatrix} \mu_{35}+\mu_{46} \\ \mu_{45}+\mu_{36} \end{pmatrix} & \begin{pmatrix} \mu_{37}+\mu_{48} \\ \mu_{47}+\mu_{38} \end{pmatrix} \\ \begin{pmatrix} \mu_{51}+\mu_{62} \\ \mu_{61}+\mu_{52} \end{pmatrix} & \begin{pmatrix} \mu_{53}+\mu_{64} \\ \mu_{63}+\mu_{54} \end{pmatrix} & \begin{pmatrix} \mu_{55}+\mu_{66} \\ \mu_{65}+\mu_{56} \end{pmatrix} & \begin{pmatrix} \mu_{57}+\mu_{68} \\ \mu_{67}+\mu_{58} \end{pmatrix} \\ \begin{pmatrix} \mu_{71}+\mu_{82} \\ \mu_{81}+\mu_{72} \end{pmatrix} & \begin{pmatrix} \mu_{73}+\mu_{84} \\ \mu_{83}+\mu_{74} \end{pmatrix} & \begin{pmatrix} \mu_{75}+\mu_{86} \\ \mu_{85}+\mu_{76} \end{pmatrix} & \begin{pmatrix} \mu_{77}+\mu_{88} \\ \mu_{87}+\mu_{78} \end{pmatrix} \end{pmatrix} \overset{\text{step 2}}{\sim}$$

$$\overset{\text{step 2}}{\sim} \begin{pmatrix} \begin{pmatrix} \mu_{11}+\mu_{22}+\mu_{33}+\mu_{44} \\ \mu_{21}+\mu_{12}+\mu_{43}+\mu_{34} \\ \mu_{31}+\mu_{42}+\mu_{13}+\mu_{24} \\ \mu_{41}+\mu_{32}+\mu_{23}+\mu_{14} \end{pmatrix} & \begin{pmatrix} \mu_{15}+\mu_{26}+\mu_{37}+\mu_{48} \\ \mu_{25}+\mu_{16}+\mu_{47}+\mu_{38} \\ \mu_{35}+\mu_{46}+\mu_{17}+\mu_{28} \\ \mu_{45}+\mu_{36}+\mu_{27}+\mu_{18} \end{pmatrix} \\ \begin{pmatrix} \mu_{51}+\mu_{62}+\mu_{73}+\mu_{84} \\ \mu_{61}+\mu_{52}+\mu_{83}+\mu_{74} \\ \mu_{71}+\mu_{82}+\mu_{53}+\mu_{64} \\ \mu_{81}+\mu_{72}+\mu_{63}+\mu_{54} \end{pmatrix} & \begin{pmatrix} \mu_{55}+\mu_{66}+\mu_{77}+\mu_{88} \\ \mu_{65}+\mu_{56}+\mu_{87}+\mu_{78} \\ \mu_{75}+\mu_{86}+\mu_{57}+\mu_{68} \\ \mu_{85}+\mu_{76}+\mu_{67}+\mu_{58} \end{pmatrix} \end{pmatrix} \overset{\text{step 3}}{\sim}$$

$$\overset{\text{step 3}}{\sim} \begin{pmatrix} \mu_{11}+\mu_{22}+\mu_{33}+\mu_{44}+\mu_{55}+\mu_{66}+\mu_{77}+\mu_{88} \\ \mu_{21}+\mu_{12}+\mu_{43}+\mu_{34}+\mu_{65}+\mu_{56}+\mu_{87}+\mu_{78} \\ \mu_{31}+\mu_{42}+\mu_{13}+\mu_{24}+\mu_{75}+\mu_{86}+\mu_{57}+\mu_{68} \\ \mu_{41}+\mu_{32}+\mu_{23}+\mu_{14}+\mu_{85}+\mu_{76}+\mu_{67}+\mu_{58} \\ \mu_{51}+\mu_{62}+\mu_{73}+\mu_{84}+\mu_{15}+\mu_{26}+\mu_{37}+\mu_{48} \\ \mu_{61}+\mu_{52}+\mu_{83}+\mu_{74}+\mu_{25}+\mu_{16}+\mu_{47}+\mu_{38} \\ \mu_{71}+\mu_{82}+\mu_{53}+\mu_{64}+\mu_{35}+\mu_{46}+\mu_{17}+\mu_{28} \\ \mu_{81}+\mu_{72}+\mu_{63}+\mu_{54}+\mu_{45}+\mu_{36}+\mu_{27}+\mu_{18} \end{pmatrix} \overset{\text{det}}{=} {_8^*M}.$$

By analogy, the matrix convolution $_{n_k}^{*}M$, $n_k \in \mathbb{N}_2$ is defined. The correspondence matrix $_{n_k}M$ of the size $n_k \times n_k$ factors in $k-1$ steps onto matrices of the size $n_{k-1} \times n_{k-1}, n_{k-2} \times n_{k-2}, \ldots, 2 \times 2$.

**Remark**

Let us pay attention to the structure of the matrix convolution $_n^*M$. *The first element in a column of the convolution $_n^*M$ is the trace of the correspondence matrix $_nM$, that is $\mathrm{Sp}\,_nM$. As is known, a matrix trace is the invariant in the transition from one coordinate system to another, that is $\mathrm{Sp}\,_nM = \mathrm{inv}$.*

The second element consists of the sum of symmetric elements $\mu_{ij} + \mu_{ji}$ of the matrix $_nM$, which indices $(i,j)$ pairwise differ by 1, that is $(1,2),(3,4),(5,6),(7,8),\ldots$.

The third element also consists of the sum $\mu_{ij} + \mu_{ji}$, and the indices $(i,j)$ differ by 2, that is $(1,3),(2,4),(5,7),(6,8),\ldots$.

In the fourth line, the difference of the indices equals 3, etc. In the last line, the difference between the indices is maximal and equals $n-1$.

**Definition 5.**

Let $_n\underline{\chi}, _n\underline{\eta} \in {_n\Omega^{(0)}}$, $n \in \mathbb{N}_2$. By the product $_n\underline{\chi} \circ _n\underline{\eta}$ or $_n\underline{\chi}\,_n\underline{\eta}$ of the $\Psi$-scalar $_n\underline{\chi}$ on the $\Psi$-scalar $_n\underline{\eta}$ we mean the $\Psi$-scalar $_n\underline{\lambda} \in {_n\Omega^{(0)}}$ which is equal to the convolution $_n^*M(\underline{\chi\eta})$ of the correspondence matrix $_nM(\underline{\chi\eta})$ of the $\Psi$-scalars $_n\underline{\chi}$ and $_n\underline{\eta}$, that is

$$_n\underline{\chi} \circ _n\underline{\eta} \stackrel{\text{det}}{=} {_n\underline{\chi}}\,_n\underline{\eta} \stackrel{\text{det}}{=} {_n^*M} \stackrel{\text{det}}{=} {_n\underline{\lambda}}. \qquad (1.6)$$

**Lemma 1.** *Let $_n\underline{\chi}, _n\underline{\eta} \in {_n\Omega^{(0)}}$ and $_nM(\underline{\chi\eta})$ are the correspondence matrix of their product, then*
$$_n^*M(\underline{\chi\eta}) = {_n^*M^T}(\underline{\chi\eta}).$$

*Proof of Lemma 1*

By Definition 4, the elements of the convolution $_n^*M(\underline{\chi\eta})$ consist of components $\mu_{ij} + \mu_{ji}$, and the elements of the convolution $_n^*M^T(\underline{\chi\eta})$ consist of components $\mu_{ji} + \mu_{ij}$. As $\mu_{ij} + \mu_{ji} = \mu_{ji} + \mu_{ij}$, then $_n^*M(\underline{\chi\eta}) = {_n^*M^T}(\underline{\chi\eta})$, which was to be proved.

*Lemma 2*

*For $\Psi$-scalars the multiplication is a commutative operation, that is $_n\underline{\chi}\,_n\underline{\eta} = {_n\underline{\eta}}\,_n\underline{\chi}$.*

*Proof of Lemma 2*

Let $_nM(\underline{\chi\eta})$ and $_nM(\underline{\eta\chi})$ are the correspondence matrices for products $_n\underline{\chi} \circ _n\underline{\eta}$ and $_n\underline{\eta} \circ _n\underline{\chi}$, respectively. As $_nM(\underline{\eta\chi}) = {_nM^T}(\underline{\chi\eta})$, then from Lemma 1 it follows that

$$_n^*M(\underline{\chi\eta}) = {_n^*M^T}(\underline{\chi\eta}) = {_n^*M}(\underline{\eta\chi}),$$

which was to be proved.

**Examples**

1. Product $_2\underline{\chi} \circ {_2\underline{\eta}}$

$$_2\underline{\chi} \circ {_2\underline{\eta}} \mapsto {_2M} = \begin{pmatrix} {^{(r)}\chi}{^{(r)}\eta} & {^{(r)}\chi}{^{(v)}\eta} \\ {^{(v)}\chi}{^{(r)}\eta} & {^{(v)}\chi}{^{(v)}\eta} \end{pmatrix},$$

$$_2\underline{\chi} \circ {_2\underline{\eta}} = {_2\underline{\chi}} {_2\underline{\eta}} = {_2^*M} = {_2^*M^T} = \begin{pmatrix} {^{(r)}\chi}{^{(r)}\eta} + {^{(v)}\chi}{^{(v)}\eta} \\ {^{(v)}\chi}{^{(r)}\eta} + {^{(r)}\chi}{^{(v)}\eta} \end{pmatrix} = \begin{pmatrix} {^{(r)}\lambda} \\ {^{(v)}\lambda} \end{pmatrix} = {_2\underline{\lambda}},$$

2. Product $_4\underline{\chi} \circ {_4\underline{\eta}}$

$$_4\underline{\chi} \circ {_4\underline{\eta}} \mapsto {_4M} = \begin{pmatrix} {^{(r)}\chi}{^{(r)}\eta} & {^{(r)}\chi}{^{(v)}\eta} & {^{(r)}\chi}{^{(\dot{v})}\eta} & {^{(r)}\chi}{^{(\ddot{v})}\eta} \\ {^{(v)}\chi}{^{(r)}\eta} & {^{(v)}\chi}{^{(v)}\eta} & {^{(v)}\chi}{^{(\dot{v})}\eta} & {^{(v)}\chi}{^{(\ddot{v})}\eta} \\ {^{(\dot{v})}\chi}{^{(r)}\eta} & {^{(\dot{v})}\chi}{^{(v)}\eta} & {^{(\dot{v})}\chi}{^{(\dot{v})}\eta} & {^{(\dot{v})}\chi}{^{(\ddot{v})}\eta} \\ {^{(\ddot{v})}\chi}{^{(r)}\eta} & {^{(\ddot{v})}\chi}{^{(v)}\eta} & {^{(\ddot{v})}\chi}{^{(\dot{v})}\eta} & {^{(\ddot{v})}\chi}{^{(\ddot{v})}\eta} \end{pmatrix} = \begin{pmatrix} \lambda^{(11)} & \lambda^{(12)} \\ \lambda^{(21)} & \lambda^{(22)} \end{pmatrix},$$

where

$$\lambda^{(11)} = \begin{pmatrix} {^{(r)}\chi}{^{(r)}\eta} & {^{(r)}\chi}{^{(v)}\eta} \\ {^{(v)}\chi}{^{(r)}\eta} & {^{(v)}\chi}{^{(v)}\eta} \end{pmatrix}, \quad \lambda^{(12)} = \begin{pmatrix} {^{(r)}\chi}{^{(\dot{v})}\eta} & {^{(r)}\chi}{^{(\ddot{v})}\eta} \\ {^{(v)}\chi}{^{(\dot{v})}\eta} & {^{(v)}\chi}{^{(\ddot{v})}\eta} \end{pmatrix},$$

$$\lambda^{(21)} = \begin{pmatrix} {^{(\dot{v})}\chi}{^{(r)}\eta} & {^{(\dot{v})}\chi}{^{(v)}\eta} \\ {^{(\ddot{v})}\chi}{^{(r)}\eta} & {^{(\ddot{v})}\chi}{^{(v)}\eta} \end{pmatrix}, \quad \lambda^{(22)} = \begin{pmatrix} {^{(\dot{v})}\chi}{^{(\dot{v})}\eta} & {^{(\dot{v})}\chi}{^{(\ddot{v})}\eta} \\ {^{(\ddot{v})}\chi}{^{(\dot{v})}\eta} & {^{(\ddot{v})}\chi}{^{(\ddot{v})}\eta} \end{pmatrix}.$$

For the correspondence matrices $\lambda^{11}$, $\lambda^{12}$, $\lambda^{21}$, $\lambda^{22}$, let us calculate the elements

$$_2\underline{\lambda}^{(11)} = \begin{pmatrix} {^{(r)}\chi}{^{(r)}\eta} + {^{(v)}\chi}{^{(v)}\eta} \\ {^{(v)}\chi}{^{(r)}\eta} + {^{(r)}\chi}{^{(v)}\eta} \end{pmatrix}, \quad _2\underline{\lambda}^{(12)} = \begin{pmatrix} {^{(r)}\chi}{^{(\dot{v})}\eta} + {^{(v)}\chi}{^{(\ddot{v})}\eta} \\ {^{(v)}\chi}{^{(\dot{v})}\eta} + {^{(r)}\chi}{^{(\ddot{v})}\eta} \end{pmatrix},$$

$$_2\underline{\lambda}^{(21)} = \begin{pmatrix} {^{(\dot{v})}\chi}{^{(r)}\eta} + {^{(\ddot{v})}\chi}{^{(v)}\eta} \\ {^{(\ddot{v})}\chi}{^{(r)}\eta} + {^{(\dot{v})}\chi}{^{(v)}\eta} \end{pmatrix}, \quad _2\underline{\lambda}^{(22)} = \begin{pmatrix} {^{(\dot{v})}\chi}{^{(\dot{v})}\eta} + {^{(\ddot{v})}\chi}{^{(\ddot{v})}\eta} \\ {^{(\ddot{v})}\chi}{^{(\dot{v})}\eta} + {^{(\dot{v})}\chi}{^{(\ddot{v})}\eta} \end{pmatrix},$$

Which give a new correspondence matrix of the order $2 \times 2$

$$_4\underline{\chi} \circ {_4\underline{\eta}} \mapsto \begin{pmatrix} _2\underline{\lambda}^{(11)} & _2\underline{\lambda}^{(12)} \\ _2\underline{\lambda}^{(21)} & _2\underline{\lambda}^{(11)} \end{pmatrix},$$

as a result,

$$_4\underline{\chi} {_4\underline{\eta}} = \begin{pmatrix} _2\underline{\lambda}^{(11)} + {_2\underline{\lambda}^{(22)}} \\ _2\underline{\lambda}^{(21)} + {_2\underline{\lambda}^{(12)}} \end{pmatrix} = \begin{pmatrix} {^{(r)}\chi}{^{(r)}\eta} + {^{(v)}\chi}{^{(v)}\eta} + {^{(\dot{v})}\chi}{^{(\dot{v})}\eta} + {^{(\ddot{v})}\chi}{^{(\ddot{v})}\eta} \\ {^{(v)}\chi}{^{(r)}\eta} + {^{(r)}\chi}{^{(v)}\eta} + {^{(\ddot{v})}\chi}{^{(\dot{v})}\eta} + {^{(\dot{v})}\chi}{^{(\ddot{v})}\eta} \\ {^{(\dot{v})}\chi}{^{(r)}\eta} + {^{(\ddot{v})}\chi}{^{(v)}\eta} + {^{(r)}\chi}{^{(\dot{v})}\eta} + {^{(v)}\chi}{^{(\ddot{v})}\eta} \\ {^{(\ddot{v})}\chi}{^{(r)}\eta} + {^{(\dot{v})}\chi}{^{(v)}\eta} + {^{(v)}\chi}{^{(\dot{v})}\eta} + {^{(r)}\chi}{^{(\ddot{v})}\eta} \end{pmatrix} = \begin{pmatrix} {^{(r)}\lambda} \\ {^{(v)}\lambda} \\ {^{(\dot{v})}\lambda} \\ {^{(\ddot{v})}\lambda} \end{pmatrix} = {_4\underline{\lambda}}.$$

3. Complex $\Psi$-scalars. Let $z_1, z_2 \in \mathbb{C}$, then the following relations are true

$$_2\underline{Z}_1 = \begin{pmatrix} x_1 \\ iy_1 \end{pmatrix}, \quad _2\underline{Z}_2 = \begin{pmatrix} x_2 \\ iy_2 \end{pmatrix},$$

the sum and product are of the form

$$_2\underline{Z}_1 + {}_2\underline{Z}_2 = \begin{pmatrix} x_1 + x_2 \\ i(y_1 + y_2) \end{pmatrix},$$

$$_2\underline{Z}_1 \circ {}_2\underline{Z}_2 \mapsto {}_2M = \begin{pmatrix} x_1 x_2 & ix_1 y_2 \\ iy_1 x_2 & -y_1 y_2 \end{pmatrix}, \quad _2\underline{Z}_1 \circ {}_2\underline{Z}_2 = \begin{pmatrix} x_1 x_2 - y_1 y_2 \\ i(y_1 x_2 + x_1 y_2) \end{pmatrix},$$

$$_2\underline{Z} \circ {}_2\underline{\overline{Z}} = \begin{pmatrix} x^2 + y^2 \\ i2xy \end{pmatrix} = \begin{pmatrix} |z|^2 \\ i2xy \end{pmatrix}.$$

Note, that $\operatorname{Sp} {}_2M = |z|^2 = \operatorname{inv}$.

4. Wave function representations in quantum mechanics.
   a. Spinor representation

$$_2\underline{\Psi} = \begin{pmatrix} \psi_1 \\ \psi_2 \end{pmatrix}, \quad _2\underline{\overline{\Psi}} = \begin{pmatrix} \overline{\psi}_1 \\ \overline{\psi}_2 \end{pmatrix}, \quad \psi_1, \psi_2 \in \mathbb{C}$$

$$_2\underline{\Psi} \circ {}_2\underline{\overline{\Psi}} \mapsto {}_2M = \begin{pmatrix} \psi_1 \overline{\psi}_1 & \psi_1 \overline{\psi}_2 \\ \psi_2 \overline{\psi}_1 & \psi_2 \overline{\psi}_2 \end{pmatrix}, \quad _2\underline{\Psi} \circ {}_2\underline{\overline{\Psi}} = \begin{pmatrix} |\psi_1|^2 + |\psi_2|^2 \\ \psi_2 \overline{\psi}_1 + \psi_1 \overline{\psi}_2 \end{pmatrix}.$$

   b. Bispinor representation

$$_4\underline{\Psi} = \begin{pmatrix} \psi_1 \\ \psi_2 \\ \psi_3 \\ \psi_4 \end{pmatrix}, \quad _4\underline{\overline{\Psi}} = \begin{pmatrix} \overline{\psi}_1 \\ \overline{\psi}_2 \\ \overline{\psi}_3 \\ \overline{\psi}_4 \end{pmatrix}, \quad \psi_1, \psi_2, \psi_3, \psi_4 \in \mathbb{C},$$

$$_4\underline{\Psi} \circ {}_4\underline{\overline{\Psi}} \mapsto {}_4M = \begin{pmatrix} \psi_1 \overline{\psi}_1 & \psi_1 \overline{\psi}_2 & \psi_1 \overline{\psi}_3 & \psi_1 \overline{\psi}_4 \\ \psi_2 \overline{\psi}_1 & \psi_2 \overline{\psi}_2 & \psi_2 \overline{\psi}_3 & \psi_2 \overline{\psi}_4 \\ \psi_3 \overline{\psi}_1 & \psi_3 \overline{\psi}_2 & \psi_3 \overline{\psi}_3 & \psi_3 \overline{\psi}_4 \\ \psi_4 \overline{\psi}_1 & \psi_4 \overline{\psi}_2 & \psi_4 \overline{\psi}_3 & \psi_4 \overline{\psi}_4 \end{pmatrix}, \quad _4\underline{\Psi} \circ {}_4\underline{\overline{\Psi}} = \begin{pmatrix} |\psi_1|^2 + |\psi_2|^2 + |\psi_3|^2 + |\psi_4|^2 \\ \psi_2 \overline{\psi}_1 + \psi_1 \overline{\psi}_2 + \psi_4 \overline{\psi}_3 + \psi_3 \overline{\psi}_4 \\ \psi_3 \overline{\psi}_1 + \psi_4 \overline{\psi}_2 + \psi_1 \overline{\psi}_3 + \psi_2 \overline{\psi}_4 \\ \psi_4 \overline{\psi}_1 + \psi_3 \overline{\psi}_2 + \psi_2 \overline{\psi}_3 + \psi_1 \overline{\psi}_4 \end{pmatrix}.$$

Note, that in both cases when the spinor and bispinor are considered

$$\operatorname{Sp} {}_2M = |\psi_1|^2 + |\psi_2|^2 = \operatorname{inv},$$
$$\operatorname{Sp} {}_4M = |\psi_1|^2 + |\psi_2|^2 + |\psi_3|^2 + |\psi_4|^2 = \operatorname{inv},$$

that is the probability density remains an invariant variable.

5. Hypercomplex number representation: quaternions, octanions, sedenions, etc. For the quaternions $q_1, q_2 \in \mathbb{Q}$

$$q_1 = (a; \vec{u}) = a + iu_1 + ju_2 + ku_3,$$
$$q_2 = (b; \vec{v}) = b + iv_1 + jv_2 + kv_3,$$

the following representations are true

$$_4\underline{q_1} = \begin{pmatrix} a \\ iu_1 \\ ju_2 \\ ku_3 \end{pmatrix}, \quad _4\underline{q_2} = \begin{pmatrix} b \\ iv_1 \\ jv_2 \\ kv_3 \end{pmatrix}, \quad _4\underline{q_1} \circ {}_4\underline{q_2} \mapsto {}_4M = \begin{pmatrix} ab & iav_1 & jav_2 & kav_3 \\ iu_1b & -u_1v_1 & ku_1v_2 & -ju_1v_3 \\ ju_2b & -ku_2v_1 & -u_2v_2 & iu_2v_3 \\ ku_3b & ju_3v_1 & -iu_3v_2 & -u_3v_3 \end{pmatrix},$$

$$_4\underline{q_1} \circ {}_4\underline{q_2} = {}_4\underline{q_1} \, {}_4\underline{q_2} = \begin{pmatrix} ab - u_1v_1 - u_2v_2 - u_3v_3 \\ i(av_1 + bu_1 - u_3v_2 + u_2v_3) \\ j(av_2 + bu_2 + u_3v_1 - u_1v_3) \\ k(av_3 + bu_3 - u_2v_1 + u_1v_2) \end{pmatrix},$$

which corresponds to a quaternion

$$q_1q_2 = \left(ab - (\vec{u}, \vec{v}); a\vec{v} + b\vec{u} + [\vec{u}, \vec{v}]\right).$$

Thus, in the quaternions product only scalar part of the quaternion remains invariant

$$\text{Sp}\,{}_4M = ab - (\vec{u}, \vec{v}) = \text{inv}.$$

From this, it is natural to consider the invariant of the interval in the field theory.

6. Field theory. In the field theory, the interval has a representation

$$_4\underline{S} = \begin{pmatrix} c\Delta t \\ i\Delta x \\ j\Delta y \\ k\Delta z \end{pmatrix}, \quad _4\underline{S}\,{}_4\underline{S} = \begin{pmatrix} c^2\Delta t^2 - \Delta x^2 - \Delta y^2 - \Delta z^2 \\ 2ic\Delta t\Delta x \\ 2jc\Delta t\Delta y \\ 2kc\Delta t\Delta z \end{pmatrix},$$

it follows from here that

$$\text{Sp}\,{}_4M = c^2\Delta t^2 - \Delta x^2 - \Delta y^2 - \Delta z^2 = \text{inv}.$$

Similar relations can be obtained for the octanions ${}_8\underline{\chi}$, sedenions ${}_{16}\underline{\chi}$, etc.

**Remark**

Note, that the $\Psi$ - set space dimension coincides with the space dimension Cayley's algebra, according to the Cayley-Dickson procedure.

In the case of real and complex numbers, the operation of addition and multiplication are commutative and associative.

The $\Psi$ - set can consist not only of scalar variables (1.2), but also of vectors.

**Definition 6.** *The $\Psi$-set (1.7) will be called the $\Psi$-vector of the order $n \in \mathbb{N}_2$ or $\Psi$-set (PSI-set) of the order $n$ and rank 1*

$$_n\underline{\vec{u}} = \begin{pmatrix} {}^{(r)}\vec{u} \\ {}^{(v)}\vec{u} \\ ... \\ {}^{\binom{(n-1)}{v}}\vec{u} \end{pmatrix}, \quad {}^1_n\underline{\vec{u}} \stackrel{\text{det}}{=} {}^{(r)}\vec{u}, \quad {}^2_n\underline{\vec{u}} \stackrel{\text{det}}{=} {}^{(v)}\vec{u},..., {}^n_n\underline{\vec{u}} \stackrel{\text{det}}{=} {}^{\binom{(n-1)}{v}}\vec{u}. \quad (1.7)$$

*The set of $\Psi$-vectors (1.7) we denote as ${}_n\Omega^{(1)}$, that is ${}_n\underline{\vec{u}} \in {}_n\Omega^{(1)}$.*

For $\Psi$ - vectors, the addition «+» and multiplication «∘» operations can be defined.

**Definition 7.** *We will call the $\Psi$-vector ${}_n\underline{\vec{d}} \in {}_n\Omega^{(1)}$ the sum ${}_n\underline{\vec{u}} + {}_n\underline{\vec{s}}$ of the $\Psi$- vectors ${}_n\underline{\vec{u}}, {}_n\underline{\vec{s}} \in {}_n\Omega^{(1)}$*

$$_n\underline{\vec{d}} = {}_n\underline{\vec{u}} + {}_n\underline{\vec{s}} = \begin{pmatrix} {}^{(r)}\vec{u} \\ {}^{(v)}\vec{u} \\ ... \\ {}^{\binom{(n-1)}{v}}\vec{u} \end{pmatrix} + \begin{pmatrix} {}^{(r)}\vec{s} \\ {}^{(v)}\vec{s} \\ ... \\ {}^{\binom{(n-1)}{v}}\vec{s} \end{pmatrix} = \begin{pmatrix} {}^{(r)}\vec{u} + {}^{(r)}\vec{s} \\ {}^{(v)}\vec{u} + {}^{(v)}\vec{s} \\ ... \\ {}^{\binom{(n-1)}{v}}\vec{u} + {}^{\binom{(n-1)}{v}}\vec{s} \end{pmatrix}. \quad (1.8)$$

The multiplication operation «∘» allows two variants: scalar multiplication and vector one.

**Definition 8.** Let $_n\vec{\underline{u}}, {}_n\vec{\underline{s}} \in \Omega_n^{(1)}$. The matrix of the size $n \times n$ ($n \in \mathbb{N}_2$) $_nM(\vec{\underline{u}} \cdot \vec{\underline{s}})$

$$(_n\vec{\underline{u}}, {}_n\vec{\underline{s}}) \mapsto {}_nM(\vec{\underline{u}} \cdot \vec{\underline{s}}) = \begin{pmatrix} \left(^{(r)}\vec{u}, {}^{(r)}\vec{s}\right) & \left(^{(r)}\vec{u}, {}^{(v)}\vec{s}\right) & \ldots & \left(^{\binom{n-1}{v}}\vec{u}, {}^{\binom{n-1}{v}}\vec{s}\right) \\ \left(^{(v)}\vec{u}, {}^{(r)}\vec{s}\right) & \left(^{(v)}\vec{u}, {}^{(v)}\vec{s}\right) & \ldots & \left(^{(v)}\vec{u}, {}^{\binom{n-1}{v}}\vec{s}\right) \\ \ldots & \ldots & \ldots & \ldots \\ \left(^{\binom{n-1}{v}}\vec{u}, {}^{(r)}\vec{s}\right) & \left(^{\binom{n-1}{v}}\vec{u}, {}^{(v)}\vec{s}\right) & \ldots & \left(^{\binom{n-1}{v}}\vec{u}, {}^{\binom{n-1}{v}}\vec{s}\right) \end{pmatrix}. \quad (1.9)$$

is called the correspondence matrix of the scalar product of the $\Psi$ - vectors $_n\vec{\underline{u}}$ and $_n\vec{\underline{s}}$. The matrix elements (1.9) are the scalar products of the vectors $(\bullet, \bullet)$.

The correspondence matrix $_nM(\vec{\underline{u}} \times \vec{\underline{s}})$ of the vector product of the $\Psi$ - vector $_n\vec{\underline{u}}$ by the $\Psi$ - vector $_n\vec{\underline{s}}$ is of the form

$$[_n\vec{\underline{u}}, {}_n\vec{\underline{s}}] \mapsto {}_nM(\vec{\underline{u}} \times \vec{\underline{s}}) = \begin{pmatrix} \left[^{(r)}\vec{u}, {}^{(r)}\vec{s}\right] & \left[^{(r)}\vec{u}, {}^{(v)}\vec{s}\right] & \ldots & \left[^{\binom{n-1}{v}}\vec{u}, {}^{\binom{n-1}{v}}\vec{s}\right] \\ \left[^{(v)}\vec{u}, {}^{(r)}\vec{s}\right] & \left[^{(v)}\vec{u}, {}^{(v)}\vec{s}\right] & \ldots & \left[^{(v)}\vec{u}, {}^{\binom{n-1}{v}}\vec{s}\right] \\ \ldots & \ldots & \ldots & \ldots \\ \left[^{\binom{n-1}{v}}\vec{u}, {}^{(r)}\vec{s}\right] & \left[^{\binom{n-1}{v}}\vec{u}, {}^{(v)}\vec{s}\right] & \ldots & \left[^{\binom{n-1}{v}}\vec{u}, {}^{\binom{n-1}{v}}\vec{s}\right] \end{pmatrix}. \quad (1.10)$$

The elements of the matrix (1.10) are the vector products of the vectors $[\bullet, \bullet]$.

**Definition 8.** We will call the convolution of the correspondence matrix $_n^*M(\vec{\underline{u}} \cdot \vec{\underline{s}})$ the scalar product of the $\Psi$ - vectors $(_n\vec{\underline{u}}, {}_n\vec{\underline{s}})$. We will call the convolution of the correspondence matrix $_n^*M(\vec{\underline{u}} \times \vec{\underline{s}})$ the vector product of the $\Psi$ - vectors $[_n\vec{\underline{u}}, {}_n\vec{\underline{s}}]$.

**Remark**

The scalar product of complex ($\mathbb{C}$) $\Psi$-vectors is commutative. For hypercomplex ($\mathbb{Q}, \mathbb{O}, \mathbb{S},\ldots$) $\Psi$-vectors (quaternions, octanions, sedenions, etc.), commutativity is violated in the general case. The vector product is not commutative in the general case.

For the complex ($\mathbb{C}$) $\Psi$-vectors, the vector product is anti-commutative, that is $[_n\vec{\underline{u}}, {}_n\vec{\underline{s}}] = -[_n\vec{\underline{s}}, {}_n\vec{\underline{u}}]$ and $_nM(\vec{\underline{u}} \times \vec{\underline{s}}) = -{}_nM^T(\vec{\underline{u}} \times \vec{\underline{s}}) = -{}_nM(\vec{\underline{s}} \times \vec{\underline{u}})$.

**Examples**

1. The Riemann-Silberstein vector (RS-vector). The vector product of the $\Psi$-vectors $_2\vec{F}$ and $_2\vec{F}^*$ is of the form

$$_2\vec{F} = \begin{pmatrix} \vec{E} \\ i\vec{B} \end{pmatrix}, \quad _2\vec{F}^* = \begin{pmatrix} \vec{E} \\ -i\vec{B} \end{pmatrix}, \quad \begin{aligned} \vec{F}_2 &= \vec{E} + i\vec{B}, \\ \vec{F}_2^* &= \vec{E} - i\vec{B}, \end{aligned}$$

$$\left[_2\vec{F}, _2\vec{F}^*\right] \mapsto {_2M} = \begin{pmatrix} [\vec{E},\vec{E}] & -i[\vec{E},\vec{B}] \\ i[\vec{B},\vec{E}] & [\vec{B},\vec{B}] \end{pmatrix} = \begin{pmatrix} 0 & -i[\vec{E},\vec{B}] \\ -i[\vec{E},\vec{B}] & 0 \end{pmatrix},$$

$$\left[_2\vec{F}, _2\vec{F}^*\right] = \begin{pmatrix} 0 \\ -2i\vec{S} \end{pmatrix},$$

where $\vec{S}$ is the Poynting vector. Thus, the Poynting vector $\vec{S}$ is not an invariant. Indeed, as is known from the field theory, for a definite choice of the coordinate system it is possible to achieve the absence of electric or magnetic field.

The scalar product of the $\Psi$-vectors $_2\vec{F}$ and $_2\vec{F}^*$ is of the form

$$\left(_2\vec{F}, \ \vec{F}^*\right) \mapsto {_2M} = \begin{pmatrix} (\vec{E},\vec{E}) & -i(\vec{E},\vec{B}) \\ i(\vec{B},\vec{E}) & (\vec{B},\vec{B}) \end{pmatrix} = \begin{pmatrix} |\vec{E}|^2 & -i(\vec{E},\vec{B}) \\ i(\vec{E},\vec{B}) & |\vec{B}|^2 \end{pmatrix},$$

$$\left(_2\vec{F}, _2\vec{F}^*\right) = \begin{pmatrix} w \\ 0 \end{pmatrix},$$

$$\text{Sp}\,_2M = w = \text{inv},$$

where $w = |\vec{E}|^2 + |\vec{B}|^2$ is the density of electromagnetic field energy. Consequently, the energy density $w$ is an invariant.

2. The product of the $\Psi$-scalar $_n\underline{\chi}$ by the $\Psi$-vector $_n\vec{\underline{u}}$. The product result is a $\Psi$-vector. For example, at $n=2$, we obtain

$$_2\underline{\chi}\,_2\vec{\underline{u}} \mapsto \begin{pmatrix} {}^{(r)}\chi\,{}^{(r)}\vec{u} & {}^{(r)}\chi\,{}^{(v)}\vec{u} \\ {}^{(v)}\chi\,{}^{(r)}\vec{u} & {}^{(v)}\chi\,{}^{(v)}\vec{u} \end{pmatrix}, \quad _2\underline{\chi}\,_2\vec{\underline{u}} = \begin{pmatrix} {}^{(r)}\chi\,{}^{(r)}\vec{u} + {}^{(v)}\chi\,{}^{(v)}\vec{u} \\ {}^{(v)}\chi\,{}^{(r)}\vec{u} + {}^{(r)}\chi\,{}^{(v)}\vec{u} \end{pmatrix},$$

$$_2\vec{\underline{u}}\,_2\underline{\chi} \mapsto \begin{pmatrix} {}^{(r)}\vec{u}\,{}^{(r)}\chi & {}^{(r)}\vec{u}\,{}^{(v)}\chi \\ {}^{(v)}\vec{u}\,{}^{(r)}\chi & {}^{(v)}\vec{u}\,{}^{(v)}\chi \end{pmatrix}, \quad _2\vec{\underline{u}}\,_2\underline{\chi} = \begin{pmatrix} {}^{(r)}\vec{u}\,{}^{(r)}\chi + {}^{(v)}\vec{u}\,{}^{(v)}\chi \\ {}^{(r)}\chi\,{}^{(v)}\vec{u} + {}^{(v)}\chi\,{}^{(r)}\vec{u} \end{pmatrix},$$

from here

$$_2\underline{\chi}\,_2\vec{\underline{u}} = {_2\vec{\underline{u}}\,_2\underline{\chi}}.$$

**Lemma 3.** *For $_n\underline{\chi} \in {_n}\Omega^{(0)}$ and $_n\vec{\underline{u}} \in {_n}\Omega^{(1)}$ at $n \in \mathbb{N}_2$ it is true that*

$$_n\underline{\chi}\,_n\vec{\underline{u}} = {_n}\vec{\underline{u}}\,_n\underline{\chi}.$$

*Proof of Lemma 3.*
The proof of Lemma 3 is analog to the proof of Lemma 2.

The elements of the $\Psi$-set are matrices or tensors, in the general case.

**Definition 8.** *We call the $\Psi$-set (1.11) the $\Psi$-tensor of the order $n \in \mathbb{N}_2$ and the rank $(k+l)$*

$$_nT_k^l = \begin{pmatrix} {^{(r)}}\tau_{r_1\ldots r_k}^{q_1\ldots q_l} \\ {^{(v)}}\tau_{r_1\ldots r_k}^{q_1\ldots q_l} \\ \ldots \\ {\left({^{(n-1)}_v}\right)}\tau_{r_1\ldots r_k}^{q_1\ldots q_l} \end{pmatrix}, \quad {_n^1}T_k^l \stackrel{det}{=} {^{(r)}}\tau_{r_1\ldots r_k}^{q_1\ldots q_l},\ {_n^2}T_k^l \stackrel{det}{=} {^{(v)}}\tau_{r_1\ldots r_k}^{q_1\ldots q_l},\ \ldots,\ {_n^n}T_k^l \stackrel{det}{=} {^{\left({^{(n-1)}_v}\right)}}\tau_{r_1\ldots r_k}^{q_1\ldots q_l}. \quad (1.11)$$

*The elements of the $\Psi$-set (1.11) are k-times covariant and l-times contrvariant tensors. We denote the set of the $\Psi$-tensors (1.11) as $_n\Omega^{(k+l)}$, that is $_nT_k^l \in {_n}\Omega^{(k+l)}$.*

**Remark**
The $\Psi$-tensor (1.11) of the zero rank corresponds to the $\Psi$-scalar (1.2). The $\Psi$-tensor (1.11) of the first rank corresponds to the $\Psi$-vector (1.7), etc. The addition and multiplication operations for the $\Psi$-tensors of the second rank are performed with account of the rules of matrix addition and multiplication. Operations with the $\Psi$-tensors of the third rank and higher are performed on the base of the tensor calculus rules.

Let us consider an example of the $\Psi$-matrix. The Dirac equation for the bispinor $\psi$ is of the form

$$i\hbar\frac{\partial}{\partial t}\psi = \left[mc^2\alpha_0 + c\sum_{j=1}^{3}\alpha_j p_j\right]\psi, \quad (1.12)$$

where

$$p_j = -i\hbar\frac{\partial}{\partial x_j}, \quad \psi = \begin{pmatrix} \psi_1 \\ \psi_2 \\ \psi_3 \\ \psi_4 \end{pmatrix}, \quad j=1,2,3,$$

$$\alpha_0 = \begin{pmatrix} \sigma_0 & 0 \\ 0 & -\sigma_0 \end{pmatrix}, \quad \alpha_j = \begin{pmatrix} 0 & \sigma_j \\ \sigma_j & 0 \end{pmatrix}, \quad j=1,2,3$$

$$\sigma_0 = \begin{pmatrix} 1 & 0 \\ 0 & 1 \end{pmatrix}, \quad \sigma_1 = \begin{pmatrix} 0 & 1 \\ 1 & 0 \end{pmatrix}, \quad \sigma_2 = \begin{pmatrix} 0 & -i \\ i & 0 \end{pmatrix}, \quad \sigma_3 = \begin{pmatrix} 1 & 0 \\ 0 & -1 \end{pmatrix}.$$

Let us define the $\Psi$-matrix $_4\underline{A}$ and the $\Psi$-scalar $_4\underline{p}$ and consider their product $_4\underline{A}\,_4\underline{p}$

$$_4\underline{A} \stackrel{det}{=} \begin{pmatrix} c\alpha_0 \\ c\alpha_1 \\ c\alpha_2 \\ c\alpha_3 \end{pmatrix}, \quad _4\underline{p} \stackrel{det}{=} \begin{pmatrix} mc \\ p_1 \\ p_2 \\ p_3 \end{pmatrix}, \tag{1.13}$$

$$_4\underline{A} \,_4\underline{p} \mapsto \begin{pmatrix} mc^2\alpha_0 & c\alpha_0 p_1 & c\alpha_0 p_2 & c\alpha_0 p_3 \\ mc^2\alpha_1 & c\alpha_1 p_1 & c\alpha_1 p_2 & c\alpha_1 p_3 \\ mc^2\alpha_2 & c\alpha_2 p_1 & c\alpha_2 p_2 & c\alpha_2 p_3 \\ mc^2\alpha_3 & c\alpha_3 p_1 & c\alpha_3 p_2 & c\alpha_3 p_3 \end{pmatrix},$$

$$_4\underline{A} \,_4\underline{p} = \begin{pmatrix} mc^2\alpha_0 + c\alpha_1 p_1 + c\alpha_2 p_2 + c\alpha_3 p_3 \\ c\alpha_0 p_1 + mc^2\alpha_1 + c\alpha_3 p_2 + c\alpha_2 p_3 \\ c\alpha_3 p_1 + c\alpha_0 p_2 + mc^2\alpha_2 + c\alpha_1 p_3 \\ c\alpha_2 p_1 + c\alpha_1 p_2 + c\alpha_0 p_3 + mc^2\alpha_3 \end{pmatrix} \stackrel{det}{=} \begin{pmatrix} _1\mu \\ _2\mu \\ _3\mu \\ _4\mu \end{pmatrix},$$

where $_s\mu$ ($s = 1...4$) is a matrix of the size $4 \times 4$. The variable $_4\underline{A}\,_4\underline{p}$ is the $\Psi$-matrix. Let us define the $\Psi$-vector $_4\vec{\underline{\psi}}$ of the form

$$_4\vec{\underline{\psi}} = \begin{pmatrix} ^{(1)}\vec{\psi} \\ ^{(2)}\vec{\psi} \\ ^{(3)}\vec{\psi} \\ ^{(4)}\vec{\psi} \end{pmatrix}, \quad \begin{array}{l} ^{(1)}\vec{\psi} = \{\psi_1, \psi_2, \psi_3, \psi_4\}, \\ ^{(j)}\vec{\psi} = \{0,0,0,0\}, j = 2,3,4. \end{array} \tag{1.14}$$

We multiply the $\Psi$-matrix $_4\underline{A}\,_4\underline{p}$ by the $\Psi$-vector $_4\vec{\underline{\psi}}$ and obtain the $\Psi$-vector $_4\underline{A}\,_4\underline{p}\,_4\vec{\underline{\psi}}$

$$_4\underline{A}\,_4\underline{p}\,_4\vec{\underline{\psi}} = \begin{pmatrix} _1\mu\,^{(1)}\vec{\psi} + _2\mu\,^{(2)}\vec{\psi} + _3\mu\,^{(3)}\vec{\psi} + _4\mu\,^{(4)}\vec{\psi} \\ _2\mu\,^{(1)}\vec{\psi} + _1\mu\,^{(2)}\vec{\psi} + _4\mu\,^{(3)}\vec{\psi} + _3\mu\,^{(4)}\vec{\psi} \\ _1\mu\,^{(3)}\vec{\psi} + _3\mu\,^{(1)}\vec{\psi} + _4\mu\,^{(2)}\vec{\psi} + _2\mu\,^{(4)}\vec{\psi} \\ _1\mu\,^{(4)}\vec{\psi} + _2\mu\,^{(3)}\vec{\psi} + _3\mu\,^{(2)}\vec{\psi} + _4\mu\,^{(1)}\vec{\psi} \end{pmatrix} = \begin{pmatrix} _1\mu\,^{(1)}\vec{\psi} \\ _2\mu\,^{(1)}\vec{\psi} \\ _3\mu\,^{(1)}\vec{\psi} \\ _4\mu\,^{(1)}\vec{\psi} \end{pmatrix} =$$

$$= \begin{pmatrix} \left[ mc^2\alpha_0 + c\alpha_1 p_1 + c\alpha_2 p_2 + c\alpha_3 p_3 \right]\,^{(1)}\vec{\psi} \\ \left[ c\alpha_0 p_1 + mc^2\alpha_1 + c\alpha_2 p_3 + c\alpha_3 p_2 \right]\,^{(1)}\vec{\psi} \\ \left[ c\alpha_0 p_2 + c\alpha_1 p_3 + mc^2\alpha_2 + c\alpha_3 p_1 \right]\,^{(1)}\vec{\psi} \\ \left[ c\alpha_0 p_3 + c\alpha_1 p_2 + c\alpha_2 p_1 + mc^2\alpha_3 \right]\,^{(1)}\vec{\psi} \end{pmatrix} \tag{1.15}$$

The first component of the $\Psi$-vector $_4\underline{A}\,_4\underline{p}\,_4\vec{\underline{\psi}}$ corresponds to the Dirac equation (1.12) for the bispinor $_4^1\vec{\underline{\psi}} = {}^{(1)}\vec{\psi}$

$$i\hbar \frac{\partial}{\partial t}\,_4^1\vec{\underline{\psi}} = {}^1\left( _4\underline{A}\,_4\underline{p}\,_4\vec{\underline{\psi}} \right). \tag{1.16}$$

Consider the question of the values of the rest components of the $\Psi$-vector (1.15). Let there be initial boundary conditions, for which the Dirac equation (1.12), (1.16) has a unique solution $^{(1)}\vec{\psi}_1$, then the $\Psi$-vector $_4\underline{\vec{\psi}}_1$ (1.14) is uniquely found. The rest three lines $^j\left(_4\underline{A}\,_4\underline{p}\,_4\underline{\vec{\psi}}_1\right)$, $j=2,3,4$ in the expression (1.15) give some values. If we take another initial boundary conditions, for which the Dirac equation also has a unique solution, for example, $^{(1)}\vec{\psi}_2$, then the $\Psi$-vector $_4\underline{\vec{\psi}}_2$ (1.14) will be defined uniquely as well. The rest three lines $^j\left(_4\underline{A}\,_4\underline{p}\,_4\underline{\vec{\psi}}_2\right)$, $j=2,3,4$ in the expression (1.15) give different values.

**Thus, only the first component (1.16) of the $\Psi$-vector (1.15) is always unchanged for any initial boundary conditions. All other components $^j\left(_4\underline{A}\,_4\underline{p}\,_4\underline{\vec{\psi}}\right)$, $j=2,3,4$ vary depending on the initial conditions. Such an asymmetry is caused by the fact that the first component is the trace of the correspondence matrix $\mathrm{Sp}\,_n M$, and the trace of the matrix is an invariant, that is $\mathrm{Sp}\,_n M = \mathrm{inv}$. Therefore, it makes sense to impose conditions only on the first component of the $\Psi$-sets as it is an invariant.**

**Remark**

The Dirac equation (1.16) can be «expanded». We remove condition (1.14), and then we obtain four equations for four bispinors

$$i\hbar\frac{\partial}{\partial t}{}^{(1)}\vec{\psi} = \left[mc^2\alpha_0 + c\alpha_1 p_1 + c\alpha_2 p_2 + c\alpha_3 p_3\right]{}^{(1)}\vec{\psi},$$

$$i\hbar\frac{\partial}{\partial t}{}^{(2)}\vec{\psi} = \left[c\alpha_0 p_1 + mc^2\alpha_1 + c\alpha_2 p_3 + c\alpha_3 p_2\right]{}^{(2)}\vec{\psi},$$

$$i\hbar\frac{\partial}{\partial t}{}^{(3)}\vec{\psi} = \left[c\alpha_0 p_2 + c\alpha_1 p_3 + mc^2\alpha_2 + c\alpha_3 p_1\right]{}^{(3)}\vec{\psi},$$

$$i\hbar\frac{\partial}{\partial t}{}^{(4)}\vec{\psi} = \left[c\alpha_0 p_3 + c\alpha_1 p_2 + c\alpha_2 p_1 + mc^2\alpha_3\right]{}^{(4)}\vec{\psi}.$$

The first of the four equations is the Dirac equation for the bispinor $^{(1)}\vec{\psi}$.

*Differential $\Psi$-sets.*

Differential operators can be used as $\Psi$-vectors. Without limiting the generality, let us consider the case $n=4$

$$_4\underline{\nabla} = \begin{pmatrix} \nabla_r \\ \nabla_v \\ \nabla_{\dot{v}} \\ \nabla_{\ddot{v}} \end{pmatrix}. \tag{1.17}$$

1. $\Psi$-gradient. The $\Psi$-gradient is a $\Psi$-vector.

$$_4\underline{\nabla} = \begin{pmatrix} \nabla_r \\ \nabla_v \\ \nabla_{\dot{v}} \\ \nabla_{\ddot{v}} \end{pmatrix}, \quad _4\underline{\chi} = \begin{pmatrix} {}^{(r)}\chi \\ {}^{(v)}\chi \\ {}^{(\dot{v})}\chi \\ {}^{(\ddot{v})}\chi \end{pmatrix}, \quad (_4\underline{\nabla}\,_4\underline{\chi}) \mapsto {}_4M = \begin{pmatrix} \nabla_r{}^{(r)}\chi & \nabla_r{}^{(v)}\chi & \nabla_r{}^{(\dot{v})}\chi & \nabla_r{}^{(\ddot{v})}\chi \\ \nabla_v{}^{(r)}\chi & \nabla_v{}^{(v)}\chi & \nabla_v{}^{(\dot{v})}\chi & \nabla_v{}^{(\ddot{v})}\chi \\ \nabla_{\dot{v}}{}^{(r)}\chi & \nabla_{\dot{v}}{}^{(v)}\chi & \nabla_{\dot{v}}{}^{(\dot{v})}\chi & \nabla_{\dot{v}}{}^{(\ddot{v})}\chi \\ \nabla_{\ddot{v}}{}^{(r)}\chi & \nabla_{\ddot{v}}{}^{(v)}\chi & \nabla_{\ddot{v}}{}^{(\dot{v})}\chi & \nabla_{\ddot{v}}{}^{(\ddot{v})}\chi \end{pmatrix},$$

$$(_4\underline{\nabla}\,_4\underline{\chi}) = \begin{pmatrix} \nabla_r{}^{(r)}\chi + \nabla_v{}^{(v)}\chi + \nabla_{\dot{v}}{}^{(\dot{v})}\chi + \nabla_{\ddot{v}}{}^{(\ddot{v})}\chi \\ \nabla_r{}^{(v)}\chi + \nabla_v{}^{(r)}\chi + \nabla_{\dot{v}}{}^{(\ddot{v})}\chi + \nabla_{\ddot{v}}{}^{(\dot{v})}\chi \\ \nabla_r{}^{(\dot{v})}\chi + \nabla_v{}^{(\ddot{v})}\chi + \nabla_{\dot{v}}{}^{(r)}\chi + \nabla_{\ddot{v}}{}^{(v)}\chi \\ \nabla_r{}^{(\ddot{v})}\chi + \nabla_v{}^{(\dot{v})}\chi + \nabla_{\dot{v}}{}^{(v)}\chi + \nabla_{\ddot{v}}{}^{(r)}\chi \end{pmatrix},$$

$$\mathrm{Sp}\,_4M = \nabla_r{}^{(r)}\chi + \nabla_v{}^{(v)}\chi + \nabla_{\dot{v}}{}^{(\dot{v})}\chi + \nabla_{\ddot{v}}{}^{(\ddot{v})}\chi. \tag{1.18}$$

2. $\Psi$-divergence. The $\Psi$-divergence is a $\Psi$-scalar.

$$_4\underline{\vec{u}} = \begin{pmatrix} {}^{(r)}\vec{u} \\ {}^{(v)}\vec{u} \\ {}^{(\dot{v})}\vec{u} \\ {}^{(\ddot{v})}\vec{u} \end{pmatrix}, \quad (_4\underline{\nabla},\,_4\underline{\vec{u}}) \mapsto {}_4M = \begin{pmatrix} (\nabla_r,{}^{(r)}\vec{u}) & (\nabla_r,{}^{(v)}\vec{u}) & (\nabla_r,{}^{(\dot{v})}\vec{u}) & (\nabla_r,{}^{(\ddot{v})}\vec{u}) \\ (\nabla_v,{}^{(r)}\vec{u}) & (\nabla_v,{}^{(v)}\vec{u}) & (\nabla_v,{}^{(\dot{v})}\vec{u}) & (\nabla_v,{}^{(\ddot{v})}\vec{u}) \\ (\nabla_{\dot{v}},{}^{(r)}\vec{u}) & (\nabla_{\dot{v}},{}^{(v)}\vec{u}) & (\nabla_{\dot{v}},{}^{(\dot{v})}\vec{u}) & (\nabla_{\dot{v}},{}^{(\ddot{v})}\vec{u}) \\ (\nabla_{\ddot{v}},{}^{(r)}\vec{u}) & (\nabla_{\ddot{v}},{}^{(v)}\vec{u}) & (\nabla_{\ddot{v}},{}^{(\dot{v})}\vec{u}) & (\nabla_{\ddot{v}},{}^{(\ddot{v})}\vec{u}) \end{pmatrix},$$

$$(_4\underline{\nabla},\,_4\underline{\vec{u}}) = \begin{pmatrix} \mathrm{div}_r{}^{(r)}\vec{u} + \mathrm{div}_v{}^{(v)}\vec{u} + \mathrm{div}_{\dot{v}}{}^{(\dot{v})}\vec{u} + \mathrm{div}_{\ddot{v}}{}^{(\ddot{v})}\vec{u} \\ \mathrm{div}_r{}^{(v)}\vec{u} + \mathrm{div}_v{}^{(r)}\vec{u} + \mathrm{div}_{\dot{v}}{}^{(\ddot{v})}\vec{u} + \mathrm{div}_{\ddot{v}}{}^{(\dot{v})}\vec{u} \\ \mathrm{div}_r{}^{(\dot{v})}\vec{u} + \mathrm{div}_v{}^{(\ddot{v})}\vec{u} + \mathrm{div}_{\dot{v}}{}^{(r)}\vec{u} + \mathrm{div}_{\ddot{v}}{}^{(v)}\vec{u} \\ \mathrm{div}_r{}^{(\ddot{v})}\vec{u} + \mathrm{div}_v{}^{(\dot{v})}\vec{u} + \mathrm{div}_{\dot{v}}{}^{(v)}\vec{u} + \mathrm{div}_{\ddot{v}}{}^{(r)}\vec{u} \end{pmatrix}.$$

$$\mathrm{Sp}\,_4M = \mathrm{div}_r{}^{(r)}\vec{u} + \mathrm{div}_v{}^{(v)}\vec{u} + \mathrm{div}_{\dot{v}}{}^{(\dot{v})}\vec{u} + \mathrm{div}_{\ddot{v}}{}^{(\ddot{v})}\vec{u}. \tag{1.19}$$

3. $\Psi$-rotor. The $\Psi$-rotor is a $\Psi$-vector.

$$[_4\underline{\nabla},\,_4\underline{\vec{u}}] \mapsto {}_4M = \begin{pmatrix} [\nabla_r,{}^{(r)}\vec{u}] & [\nabla_r,{}^{(v)}\vec{u}] & [\nabla_r,{}^{(\dot{v})}\vec{u}] & [\nabla_r,{}^{(\ddot{v})}\vec{u}] \\ [\nabla_v,{}^{(r)}\vec{u}] & [\nabla_v,{}^{(v)}\vec{u}] & [\nabla_v,{}^{(\dot{v})}\vec{u}] & [\nabla_v,{}^{(\ddot{v})}\vec{u}] \\ [\nabla_{\dot{v}},{}^{(r)}\vec{u}] & [\nabla_{\dot{v}},{}^{(v)}\vec{u}] & [\nabla_{\dot{v}},{}^{(\dot{v})}\vec{u}] & [\nabla_{\dot{v}},{}^{(\ddot{v})}\vec{u}] \\ [\nabla_{\ddot{v}},{}^{(r)}\vec{u}] & [\nabla_{\ddot{v}},{}^{(v)}\vec{u}] & [\nabla_{\ddot{v}},{}^{(\dot{v})}\vec{u}] & [\nabla_{\ddot{v}},{}^{(\ddot{v})}\vec{u}] \end{pmatrix},$$

$$[_4\underline{\nabla},\,_4\underline{\vec{u}}] = \begin{pmatrix} \mathrm{rot}_r{}^{(r)}\vec{u} + \mathrm{rot}_v{}^{(v)}\vec{u} + \mathrm{rot}_{\dot{v}}{}^{(\dot{v})}\vec{u} + \mathrm{rot}_{\ddot{v}}{}^{(\ddot{v})}\vec{u} \\ \mathrm{rot}_r{}^{(v)}\vec{u} + \mathrm{rot}_v{}^{(r)}\vec{u} + \mathrm{rot}_{\dot{v}}{}^{(\ddot{v})}\vec{u} + \mathrm{rot}_{\ddot{v}}{}^{(\dot{v})}\vec{u} \\ \mathrm{rot}_r{}^{(\dot{v})}\vec{u} + \mathrm{rot}_v{}^{(\ddot{v})}\vec{u} + \mathrm{rot}_{\dot{v}}{}^{(r)}\vec{u} + \mathrm{rot}_{\ddot{v}}{}^{(v)}\vec{u} \\ \mathrm{rot}_r{}^{(\ddot{v})}\vec{u} + \mathrm{rot}_v{}^{(\dot{v})}\vec{u} + \mathrm{rot}_{\dot{v}}{}^{(v)}\vec{u} + \mathrm{rot}_{\ddot{v}}{}^{(r)}\vec{u} \end{pmatrix}.$$

$$\mathrm{Sp}_4 M = \mathrm{rot}_r{}^{(r)}\vec{u} + \mathrm{rot}_v{}^{(v)}\vec{u} + \mathrm{rot}_{\dot{v}}{}^{(\dot{v})}\vec{u} + \mathrm{rot}_{\ddot{v}}{}^{(\ddot{v})}\vec{u}. \qquad (1.20)$$

**Remark**

Previously in monograph [7, 8] we obtained the expressions (1.18-20) for the generalized phase space for consideration of the Vlasov equation chain. It is the expressions (1.18-20) that, due to their «invariance», makes sense when constructing kinetic equations such as the Vlasov equation chain.

In [7, 8] the representations (1.18-20) had the notations $\nabla_\xi$, $\mathrm{div}_\xi$, $\mathrm{rot}_\xi$ respectively. In the given paper, for brevity, we also use the mentioned notations.

4. The properties of the $\Psi$-rotor, divergence and gradient. The following expressions are true
    a. The $\Psi$-rotor from the $\Psi$-gradient equals to the zero $\Psi$-vector, that is

$$\left[{}_n\nabla, {}_n\nabla {}_n\underline{\chi}\right] = \begin{pmatrix} 0 \\ \ldots \\ 0 \end{pmatrix}. \qquad (1.18)$$

    b. the $\Psi$-divergence from the $\Psi$-rotor equals to the zero $\Psi$-scalar, that is

$$\left({}_n\nabla, \left[{}_n\nabla, {}_n\underline{\vec{u}}\right]\right) = \begin{pmatrix} 0 \\ \ldots \\ 0 \end{pmatrix}. \qquad (1.19)$$

Let us verify the given identical equations for the case $n=2$. For the expression (1.19), we obtain

$${}_2\underline{\vec{u}} = \begin{pmatrix} {}^{(r)}\vec{u} \\ {}^{(v)}\vec{u} \end{pmatrix}, \quad \left[{}_2\nabla, {}_2\underline{\vec{u}}\right] \mapsto \begin{pmatrix} \left[\nabla_r, {}^{(r)}\vec{u}\right] & \left[\nabla_r, {}^{(v)}\vec{u}\right] \\ \left[\nabla_v, {}^{(r)}\vec{u}\right] & \left[\nabla_v, {}^{(v)}\vec{u}\right] \end{pmatrix},$$

$$\left[{}_2\nabla, {}_2\underline{\vec{u}}\right] = \begin{pmatrix} \mathrm{rot}_r{}^{(r)}\vec{u} + \mathrm{rot}_v{}^{(v)}\vec{u} \\ \mathrm{rot}_v{}^{(r)}\vec{u} + \mathrm{rot}_r{}^{(v)}\vec{u} \end{pmatrix},$$

$$\left({}_2\nabla, \left[{}_2\nabla, {}_2\underline{\vec{u}}\right]\right) \mapsto \begin{pmatrix} \mathrm{div}_r\,\mathrm{rot}_v{}^{(v)}\vec{u} & \mathrm{div}_r\,\mathrm{rot}_v{}^{(r)}\vec{u} \\ \mathrm{div}_v\,\mathrm{rot}_r{}^{(r)}\vec{u} & \mathrm{div}_v\,\mathrm{rot}_r{}^{(v)}\vec{u} \end{pmatrix}.$$

Taking into consideration the properties of the scalar triple product

$$\mathrm{div}_r\,\mathrm{rot}_v\,\vec{a} = \left(\nabla_r, [\nabla_v, \vec{a}]\right) = \left([\nabla_r, \nabla_v], \vec{a}\right) = -\left([\nabla_v, \nabla_r], \vec{a}\right) =$$
$$= -\left(\nabla_v, [\nabla_r, \vec{a}]\right) = -\mathrm{div}_v\,\mathrm{rot}_r\,\vec{a}, \qquad (1.20)$$

we obtain

$$\left( {}_2\underline{\nabla}, [{}_2\underline{\nabla}, {}_2\vec{\underline{u}}] \right) \mapsto \begin{pmatrix} \operatorname{div}_r \operatorname{rot}_v \vec{u}^{(v)} & -\operatorname{div}_v \operatorname{rot}_r \vec{u}^{(r)} \\ \operatorname{div}_v \operatorname{rot}_r \vec{u}^{(r)} & -\operatorname{div}_r \operatorname{rot}_v \vec{u}^{(v)} \end{pmatrix}.$$

As a result,

$$\left( {}_2\underline{\nabla}, [{}_2\underline{\nabla}, {}_2\vec{\underline{u}}] \right) = \begin{pmatrix} 0 \\ 0 \end{pmatrix}.$$

By analogy, for the expression (1.18), we obtain

$$\left( {}_2\underline{\nabla}\, {}_2\underline{\chi} \right) = \begin{pmatrix} \nabla_r{}^{(r)}\chi + \nabla_v{}^{(v)}\chi \\ \nabla_v{}^{(r)}\chi + \nabla_r{}^{(v)}\chi \end{pmatrix},$$

$$\left[ {}_2\underline{\nabla}, {}_2\underline{\nabla}\, {}_2\underline{\chi} \right] = \begin{pmatrix} \operatorname{rot}_r\left(\nabla_r{}^{(r)}\chi + \nabla_v{}^{(v)}\chi\right) + \operatorname{rot}_v\left(\nabla_v{}^{(r)}\chi + \nabla_r{}^{(v)}\chi\right) \\ \operatorname{rot}_v\left(\nabla_r{}^{(r)}\chi + \nabla_v{}^{(v)}\chi\right) + \operatorname{rot}_r\left(\nabla_v{}^{(r)}\chi + \nabla_r{}^{(v)}\chi\right) \end{pmatrix} =$$

$$= \begin{pmatrix} \operatorname{rot}_r \nabla_v{}^{(v)}\chi + \operatorname{rot}_v \nabla_r{}^{(v)}\chi \\ \operatorname{rot}_v \nabla_r{}^{(r)}\chi + \operatorname{rot}_r \nabla_v{}^{(r)}\chi \end{pmatrix} = \begin{pmatrix} 0 \\ 0 \end{pmatrix},$$

where it is taken into account that

$$\operatorname{rot}_r \nabla_v \lambda = [\nabla_r, \nabla_v \lambda] = [\nabla_r, \nabla_v]\lambda = -[\nabla_v, \nabla_r]\lambda = -[\nabla_v, \nabla_r \lambda] = -\operatorname{rot}_v \nabla_r \lambda. \qquad (1.21)$$

Thus, we can define the $\Psi$-algebra $\left\langle {}_n\Omega^{(k+l)}, +, \circ \right\rangle$. The ${}_n\Omega^{(k+l)}$ is the set of the $\Psi$-algebra. The operations «+» and «∘» are the binary operation of the $\Psi$-algebra. For each set of $n, k, l$ one can consider the existence of an identity element, an inverse element, a commutativity and associativity properties.

## §2 Chain of quantum mechanics equations

Let us consider an infinite coupled chain of Vlasov equations [5,6,14] for the probability density function $f_n(\vec{r}, \vec{v}, \dot{\vec{v}}..., t)$

$$\Pi_n S_n = -Q_n, \quad n \in \mathbb{N}, \qquad (2.1)$$

where

$$\Pi_n \stackrel{\mathrm{det}}{=} \frac{d_n}{dt} \stackrel{\mathrm{det}}{=} \frac{\partial}{\partial t} + \left(\dot{\vec{r}}, \nabla_r\right) + \ldots + \left(\overset{(n-1)}{\vec{r}}, \nabla_{\overset{(n-2)}{r}}\right) + \left(\left\langle \overset{(n)}{\vec{r}} \right\rangle, \nabla_{\overset{(n-1)}{r}}\right) = \frac{\partial}{\partial t} + \left({}_n\underline{\vec{u}}, {}_n\underline{\nabla}\right)_\xi,$$

$$_n\vec{u}(\vec{\xi}_n,t) = \begin{pmatrix} \dot{\vec{r}} \\ \ddot{\vec{r}} \\ \dots \\ \left\langle \overset{(n)}{\vec{r}} \right\rangle(\vec{\xi}_n,t) \end{pmatrix} = \begin{pmatrix} \vec{v} \\ \dot{\vec{v}} \\ \dots \\ \left\langle \overset{(n-1)}{\vec{v}} \right\rangle(\vec{\xi}_n,t) \end{pmatrix}, \quad \vec{\xi}_n = \begin{pmatrix} \vec{r} \\ \vec{v} \\ \dots \\ \overset{(n-2)}{\vec{v}} \end{pmatrix}, \quad _n\underline{\nabla} = \begin{pmatrix} \nabla_r \\ \nabla_{\dot{r}} \\ \dots \\ \nabla_{\overset{(n-1)}{r}} \end{pmatrix} = \begin{pmatrix} \nabla \\ \nabla_v \\ \dots \\ \nabla_{\overset{(n-2)}{v}} \end{pmatrix}, \quad (2.2)$$

$$S_n(\vec{\xi}_n,t) \overset{\text{det}}{=} \ln f_n(\vec{\xi}_n,t), \quad Q_n(\vec{\xi}_n,t) \overset{\text{det}}{=} \text{div}_{\overset{(n-1)}{r}} \left\langle \overset{(n-1)}{\vec{v}} \right\rangle(\vec{\xi}_n,t),$$

Probability density functions satisfy the relations

$$f_0(t) \overset{\text{det}}{=} N(t) = \int_{(\infty)} f_1(\vec{r},t) d^3r = \int_{(\infty)}\int_{(\infty)} f_2(\vec{r},\vec{v},t) d^3r d^3v = \dots =$$
$$= \dots = \int_{(\infty)}\int_{(\infty)}\int_{(\infty)} \dots f_\infty(\vec{r},\vec{v},\dot{\vec{v}},\dots,t) d^3r d^3v d^3\dot{v} \dots , \quad (2.3)$$

where $N(t)$ is the normalization factor or a number of particles, which can be non-integer in the general case [5,6].

The mean values are determined according to the relations [5,6]

$$f_1(\vec{r},t)\langle\vec{v}\rangle(\vec{r},t) = \int_{(\infty)} f_2(\vec{r},\vec{v},t)\vec{v} d^3v, \quad (2.4)$$

$$f_2(\vec{r},\vec{v},t)\langle\dot{\vec{v}}\rangle(\vec{r},\vec{v},t) = \int_{(\infty)} f_3(\vec{r},\vec{v},\dot{\vec{v}},t)\dot{\vec{v}} d^3\dot{v},$$

...

The chain of equations (2.1) was written initially by A.A. Vlasov in the form

$$\frac{\partial f_1}{\partial t} + \text{div}_r\left[f_1\langle\vec{v}\rangle\right] = 0,$$

$$\frac{\partial f_2}{\partial t} + (\vec{v},\nabla_r f_2) + \text{div}_v\left[f_2\langle\dot{\vec{v}}\rangle\right] = 0, \quad (2.5)$$

...

$$\frac{\partial f_n}{\partial t} + (\vec{v},\nabla_r f_n) + (\dot{\vec{v}},\nabla_v f_n) + \dots + \text{div}_{\overset{(n-2)}{v}} f_n\left\langle\overset{(n-1)}{\vec{v}}\right\rangle = 0,$$

...

In [10], from the first equation in the chain (2.5), (2.1) we obtained the Schrödinger equation for the wave function $\Psi_1(\vec{r},t)$, $f_1 = |\Psi_1|^2$. Let us construct the following quantum mechanics equations for the wave functions $\Psi_2(\vec{r},\vec{v},t)$, $\Psi_3(\vec{r},\vec{v},\dot{\vec{v}},t)$, …, corresponding to the distribution functions $f_2(\vec{r},\vec{v},t)$, $f_3(\vec{r},\vec{v},\dot{\vec{v}},t)$,… We use the procedure described in [10] for constructing.

Since the probability density function $f_n(\vec{\xi}_n, t)$ is a positive function by definition, it can be represented as a squared absolute value of some complex function $\Psi_n(\vec{\xi}_n, t)$, that is:

$$f_n(\vec{\xi}_n, t) = |\Psi_n(\vec{\xi}_n, t)|^2 = \Psi_n \bar{\Psi}_n \geq 0. \tag{2.6}$$

By the Helmholtz decomposition, let us represent the vector field $\left\langle \overset{(n-1)}{\vec{v}} \right\rangle$ in the form of superposition of the fields:

$$\left\langle \overset{(n-1)}{\vec{v}} \right\rangle (\vec{\xi}_n, t) = -\alpha_n \nabla_{\underset{\vec{v}}{(n-2)}} \Phi_n(\vec{\xi}_n, t) + \gamma_n \vec{A}_n(\vec{\xi}_n, t), \quad \text{div}_{\underset{\vec{v}}{(n-2)}} \vec{A}_n = 0. \tag{2.7}$$

where $\alpha_n, \gamma_n$ are some real constants, $\Phi_n(\vec{\xi}_n, t)$ is the scalar potential of the field $\left\langle \overset{(n-1)}{\vec{v}} \right\rangle$ and $\vec{A}_n(\vec{\xi}_n, t)$ is a vortex component of the field $\left\langle \overset{(n-1)}{\vec{v}} \right\rangle$. Let us rewrite the expression (2.7)

$$\left\langle \overset{(n-1)}{\vec{v}} \right\rangle = -\alpha_n \nabla_{\underset{v}{(n-2)}} \Phi_n + \gamma_n \vec{A}_n = i^2 \alpha_n \nabla_{\underset{v}{(n-2)}} \Phi_n + \gamma_n \vec{A}_n = $$
$$= i\alpha_n \nabla_{\underset{v}{(n-2)}} (0 + i\Phi_n) + \gamma_n \vec{A}_n = i\alpha_n \nabla_{\underset{v}{(n-2)}} \left( \ln \left| \frac{\Psi_n}{\bar{\Psi}_n} \right| + i\Phi_n \right) + \gamma_n \vec{A}_n. \tag{2.8}$$

Since the function $\Psi_n(\vec{\xi}_n, t)$ is complex, it satisfies the exponential form of the representation of the form:

$$\Psi_n(\vec{\xi}_n, t) = |\Psi_n(\vec{\xi}_n, t)| e^{i\varphi_n(\vec{\xi}_n, t)}, \tag{2.9}$$

where $\varphi_n(\vec{\xi}_n, t)$ is a phase. Taking into account the representation (2.9), the function $\frac{\Psi_n}{\bar{\Psi}_n}$ is of the form:

$$\frac{\Psi_n}{\bar{\Psi}_n} = e^{i2\varphi_n(\vec{\xi}_n, t)} \Rightarrow \text{Arg}\left[ \frac{\Psi_n}{\bar{\Psi}_n} \right] = 2\varphi_n + 2\pi k \overset{\text{det}}{=} \Phi_n. \tag{2.10}$$

Thus, the potential $\Phi_n$. of the field $\left\langle \overset{(n-1)}{\vec{v}} \right\rangle$ in (2.10) is defined as an argument of the complex function $\frac{\Psi_n}{\bar{\Psi}_n}$. Considering the notation (2.10), we rewrite the expression (2.8) and obtain:

$$\left\langle \vec{v} \right\rangle^{(n-1)}\left(\vec{\xi}_n,t\right) = i\alpha_n \nabla_{\substack{(n-2)\\v}}\left(\ln\left|\frac{\Psi_n}{\bar{\Psi}_n}\right| + i\operatorname{Arg}\left[\frac{\Psi_n}{\bar{\Psi}_n}\right]\right) + \gamma_n \vec{A}_n = i\alpha_n \nabla_{\substack{(n-2)\\v}}\operatorname{Ln}\left[\frac{\Psi_n}{\bar{\Psi}_n}\right] + \gamma_n \vec{A}_n =$$

$$= i\alpha_n \nabla_{\substack{(n-2)\\v}}\left[\operatorname{Ln}(\Psi_n) - \operatorname{Ln}(\bar{\Psi}_n)\right] + \gamma_n \vec{A}_n = i\alpha_n\left[\frac{\nabla_{\substack{(n-2)\\v}}\Psi_n}{\Psi_n} - \frac{\nabla_{\substack{(n-2)\\v}}\bar{\Psi}_n}{\bar{\Psi}_n}\right] + \gamma_n \vec{A}_n. \qquad (2.11)$$

Considering (2.6) and (2.10), we rewrite the equation (2.1)/(2.5)

$$\bar{\Psi}_n \frac{\partial \Psi_n}{\partial t} + \Psi_n \frac{\partial \bar{\Psi}_n}{\partial t} + \bar{\Psi}_n\left(\vec{v}, \nabla_r \Psi_n\right) + \Psi_n\left(\vec{v}, \nabla_r \bar{\Psi}_n\right) + \bar{\Psi}_n\left(\dot{\vec{v}}, \nabla_v \Psi_n\right) + \Psi_n\left(\dot{\vec{v}}, \nabla_v \bar{\Psi}_n\right) + \ldots +$$

$$+ i\alpha_n \operatorname{div}_{\substack{(n-2)\\v}}\left[\Psi_n \bar{\Psi}_n\left(\frac{\nabla_{\substack{(n-2)\\v}}\Psi_n}{\Psi_n} - \frac{\nabla_{\substack{(n-2)\\v}}\bar{\Psi}_n}{\bar{\Psi}_n} - i\frac{\gamma_n}{\alpha_n}\vec{A}_n\right)\right] = 0,$$

or

$$\bar{\Psi}_n \frac{\partial \Psi_n}{\partial t} + \Psi_n \frac{\partial \bar{\Psi}_n}{\partial t} + \bar{\Psi}_n\left(\vec{v}, \nabla_r \Psi_n\right) + \Psi_n\left(\vec{v}, \nabla_r \bar{\Psi}_n\right) + \bar{\Psi}_n\left(\dot{\vec{v}}, \nabla_v \Psi_n\right) + \Psi_n\left(\dot{\vec{v}}, \nabla_v \bar{\Psi}_n\right) + \ldots +$$

$$+ i\alpha_n \operatorname{div}_{\substack{(n-2)\\v}}\left[\bar{\Psi}_n \nabla_{\substack{(n-2)\\v}}\Psi_n - \Psi_n \nabla_{\substack{(n-2)\\v}}\bar{\Psi}_n - i\frac{\gamma_n}{\alpha_n}\Psi_n \bar{\Psi}_n \vec{A}_n\right] = 0, \qquad (2.12)$$

Let us write out the expressions $\operatorname{div}_{\substack{(n-2)\\v}}\left[\bar{\Psi}_n \nabla_{\substack{(n-2)\\v}}\Psi_n - \Psi_n \nabla_{\substack{(n-2)\\v}}\bar{\Psi}_n\right]$ and $\operatorname{div}_{\substack{(n-2)\\v}}\left[\Psi_n \bar{\Psi}_n \vec{A}_n\right]$, we obtain:

$$\operatorname{div}_{\substack{(n-2)\\v}}\left[\bar{\Psi}_n \nabla_{\substack{(n-2)\\v}}\Psi_n - \Psi_n \nabla_{\substack{(n-2)\\v}}\bar{\Psi}_n\right] = \nabla_{\substack{(n-2)\\v}}\bar{\Psi}_n \nabla_{\substack{(n-2)\\v}}\Psi_n + \bar{\Psi}_n \Delta_{\substack{(n-2)\\v}}\Psi_n - \nabla_{\substack{(n-2)\\v}}\Psi_n \nabla_{\substack{(n-2)\\v}}\bar{\Psi}_n -$$
$$-\Psi_n \Delta_{\substack{(n-2)\\v}}\bar{\Psi}_n = \bar{\Psi}_n \Delta_{\substack{(n-2)\\v}}\Psi_n - \Psi_n \Delta_{\substack{(n-2)\\v}}\bar{\Psi}_n, \qquad (2.13)$$
$$\operatorname{div}_{\substack{(n-2)\\v}}\left[\Psi_n \bar{\Psi}_n \vec{A}_n\right] = \left(\nabla_{\substack{(n-2)\\v}}\Psi_n \bar{\Psi}_n, \vec{A}_n\right) + \Psi_n \bar{\Psi}_n\left(\nabla_{\substack{(n-2)\\v}}, \vec{A}_n\right) =$$
$$= \bar{\Psi}_n\left(\vec{A}_n, \nabla_{\substack{(n-2)\\v}}\Psi_n\right) + \Psi_n\left(\vec{A}_n, \nabla_{\substack{(n-2)\\v}}\bar{\Psi}_n\right).$$

In the latter expression (2.13) there is considered (2.7) that $\operatorname{div}_{\substack{(n-2)\\v}}\vec{A}_n = 0$. Let us substitute (2.13) into (2.12) and obtain

$$\bar{\Psi}_n \frac{\partial \Psi_n}{\partial t} + \Psi_n \frac{\partial \bar{\Psi}_n}{\partial t} + \bar{\Psi}_n\left(\vec{v}, \nabla_r \Psi_n\right) + \Psi_n\left(\vec{v}, \nabla_r \bar{\Psi}_n\right) + \bar{\Psi}_n\left(\dot{\vec{v}}, \nabla_v \Psi_n\right) +$$
$$+ \Psi_n\left(\dot{\vec{v}}, \nabla_v \bar{\Psi}_n\right) + \ldots + i\alpha_n\left(\bar{\Psi}_n \Delta_{\substack{(n-2)\\v}}\Psi_n - \Psi_n \Delta_{\substack{(n-2)\\v}}\bar{\Psi}_n\right) + \qquad (2.14)$$
$$+ \bar{\Psi}_n\left(\gamma_n \vec{A}_n, \nabla_{\substack{(n-2)\\v}}\Psi_n\right) + \Psi_n\left(\gamma \vec{A}_n, \nabla_{\substack{(n-2)\\v}}\bar{\Psi}_n\right) = 0.$$

We group the terms of the expression (2.14) and obtain

$$\overline{\Psi}_n \left[ \frac{\partial \Psi_n}{\partial t} + (\vec{v}, \nabla_r \Psi_n) + (\dot{\vec{v}}, \nabla_v \Psi_n) + ... + \left( \overset{(n-2)}{\vec{v}}, \nabla_{\overset{(n-3)}{v}} \Psi_n \right) + i\alpha_n \Delta_{\overset{(n-2)}{v}} \Psi_n + \gamma_n \left( \vec{A}_n, \nabla_{\overset{(n-2)}{v}} \Psi_n \right) \right] +$$

$$+ \Psi_n \left[ \frac{\partial \overline{\Psi}_n}{\partial t} + (\vec{v}, \nabla_r \overline{\Psi}_n) + (\dot{\vec{v}}, \nabla_v \overline{\Psi}_n) + ... + \left( \overset{(n-2)}{\vec{v}}, \nabla_{\overset{(n-3)}{v}} \overline{\Psi}_n \right) - i\alpha_n \Delta_{\overset{(n-2)}{v}} \overline{\Psi}_n + \gamma_n \left( \vec{A}_n, \nabla_{\overset{(n-2)}{v}} \overline{\Psi}_n \right) \right] = 0.$$

(2.15)

Let us introduce the notation of the differential operator:

$$\hat{p}_1 \overset{\text{det}}{=} -\frac{i}{\beta_1} \nabla_r, \quad \overline{\hat{p}}_1 = \frac{i}{\beta_1} \nabla_r, \quad \hat{p}_1^2 = \overline{\hat{p}}_1^2 = -\frac{1}{\beta_1^2} \Delta_r,$$

$$\nabla_r = i\beta_1 \hat{p}_1, \quad \nabla_r = -i\beta_1 \overline{\hat{p}}_1, \quad \Delta_r = -\beta_1^2 \hat{p}_1^2 = -\beta_1^2 \overline{\hat{p}}_1^2,$$

(2.16)

for $n > 1$

$$\hat{p}_n \overset{\text{det}}{=} -\frac{i}{\beta_n} \nabla_{\overset{(n-2)}{v}}, \quad \overline{\hat{p}}_n = \frac{i}{\beta_n} \nabla_{\overset{(n-2)}{v}}, \quad \hat{p}_n^2 = \overline{\hat{p}}_n^2 = -\frac{1}{\beta_n^2} \Delta_{\overset{(n-2)}{v}},$$

$$\nabla_{\overset{(n-2)}{v}} = i\beta_n \hat{p}_n, \quad \nabla_{\overset{(n-2)}{v}} = -i\beta_n \overline{\hat{p}}_n, \quad \Delta_{\overset{(n-2)}{v}} = -\beta_n^2 \hat{p}_n^2 = -\beta_n^2 \overline{\hat{p}}_n^2,$$

where $\beta_n \in \mathbb{R}$, $\beta_n \neq 0$. Then the expression (2.15) is of the form:

$$\overline{\Psi}_n \left[ \frac{1}{\beta_n} \frac{\partial}{\partial t} + i\frac{\beta_1}{\beta_n}(\vec{v}, \hat{p}_1) + i\frac{\beta_2}{\beta_n}(\dot{\vec{v}}, \hat{p}_2) + ... + i\frac{\beta_{n-1}}{\beta_n}\left(\overset{(n-2)}{\vec{v}}, \hat{p}_{n-1}\right) - i\alpha_n \beta_n \left( \hat{p}_n^2 - \frac{\gamma_n}{\alpha_n \beta_n}(\vec{A}_n, \hat{p}_n) \right) \right] \Psi_n +$$

$$+ \Psi_n \left[ \frac{1}{\beta_n} \frac{\partial}{\partial t} - i\frac{\beta_1}{\beta_n}(\vec{v}, \overline{\hat{p}}_1) - i\frac{\beta_2}{\beta_n}(\dot{\vec{v}}, \overline{\hat{p}}_2) - ... - i\frac{\beta_{n-1}}{\beta_n}\left(\overset{(n-2)}{\vec{v}}, \overline{\hat{p}}_{n-1}\right) + i\alpha \beta_n \left( \overline{\hat{p}}_n^2 - \frac{\gamma_n}{\alpha_n \beta_n}(\vec{A}_n, \overline{\hat{p}}_n) \right) \right] \overline{\Psi}_n = 0.$$

(2.17)

The expression (2.17) can be written in another form, taking into consideration that

$$\hat{p}_n^2 - \frac{\gamma_n}{\alpha_n \beta_n}(\vec{A}_n, \hat{p}_n) = \left( \hat{p}_n - \frac{\gamma_n}{2\alpha_n \beta_n} \vec{A}_n \right)^2 - \frac{\gamma_n^2}{4\alpha_n^2 \beta_n^2} |\vec{A}_n|^2,$$

$$\overline{\hat{p}}_n^2 - \frac{\gamma_n}{\alpha_n \beta_n}(\vec{A}_n, \overline{\hat{p}}_n) = \left( \overline{\hat{p}}_n - \frac{\gamma_n}{2\alpha_n \beta_n} \vec{A}_n \right)^2 - \frac{\gamma_n^2}{4\alpha_n^2 \beta_n^2} |\vec{A}_n|^2.$$

(2.18)

since

$$\left( \hat{p}_n - \frac{\gamma_n}{2\alpha_n \beta_n} \vec{A}_n \right)^2 \Psi = \left( \hat{p}_n - \frac{\gamma_n}{2\alpha_n \beta_n} \vec{A}_n \right) \left( \hat{p}_n - \frac{\gamma_n}{2\alpha_n \beta_n} \vec{A}_n \right) \Psi =$$

$$= \left( \hat{p}_n - \frac{\gamma_n}{2\alpha_n \beta_n} \vec{A}_n \right) \left( \hat{p}_n \Psi - \frac{\gamma_n}{2\alpha_n \beta_n} \vec{A}_n \Psi \right) = \hat{p}_n^2 \Psi - \frac{\gamma_n}{2\alpha_n \beta_n}(\hat{p}_n, \vec{A}_n \Psi) - \frac{\gamma_n}{2\alpha_n \beta_n}(\vec{A}_n, \hat{p}_n \Psi) +$$

$$+ \frac{\gamma_n^2}{4\alpha_n^2 \beta_n^2} |\vec{A}_n|^2 \Psi = \hat{p}_n^2 \Psi - \frac{\gamma_n}{2\alpha_n \beta_n} \Psi (\hat{p}_n, \vec{A}_n) - \frac{\gamma_n}{2\alpha_n \beta_n} (\vec{A}_n, \hat{p}_n \Psi) - \frac{\gamma_n}{2\alpha_n \beta_n} (\vec{A}_n, \hat{p}_n \Psi) +$$

$$+\frac{\gamma_n^2}{4\alpha_n^2\beta_n^2}\left|\vec{A}_n\right|^2 = \hat{p}_n^2\Psi - \frac{\gamma_n}{\alpha_n\beta_n}\left(\vec{A}_n,\hat{p}_n\right)\Psi + \frac{\gamma_n^2}{4\alpha_n^2\beta_n^2}\left|\vec{A}_n\right|^2\Psi.$$

Considering (2.18), the expression (2.17) is of the form:

$$\bar{\Psi}_n\left[\frac{1}{\beta_n}\frac{\partial}{\partial t}+i\frac{\beta_1}{\beta_n}\left(\vec{v},\hat{p}_1\right)+i\frac{\beta_2}{\beta_n}\left(\dot{\vec{v}},\hat{p}_2\right)+\ldots+i\frac{\beta_{n-1}}{\beta_n^2}\left(\overset{(n-2)}{\vec{v}},\hat{p}_{n-1}\right)-\right.$$
$$\left.-i\alpha_n\beta_n\left(\hat{p}_n-\frac{\gamma_n}{2\alpha_n\beta_n}\vec{A}_n\right)^2+i\frac{\gamma_n^2}{4\alpha_n\beta_n}\left|\vec{A}_n\right|^2\right]\Psi_n+\Psi_n\left[i\alpha_n\beta_n\left(\bar{\hat{p}}_n-\frac{\gamma_n}{2\alpha_n\beta_n}\vec{A}_n\right)^2-i\frac{\gamma_n^2}{4\alpha_n\beta_n}\left|\vec{A}_n\right|^2+\right.$$
$$\left.\frac{1}{\beta_n}\frac{\partial}{\partial t}-i\frac{\beta_1}{\beta_n}\left(\vec{v},\bar{\hat{p}}_1\right)-i\frac{\beta_2}{\beta_n}\left(\dot{\vec{v}},\bar{\hat{p}}_2\right)-\ldots-i\frac{\beta_{n-1}}{\beta_n^2}\left(\overset{(n-2)}{\vec{v}},\bar{\hat{p}}_{n-1}\right)\right]\bar{\Psi}_n=0.$$

or

$$\bar{\Psi}_n\left[\frac{1}{\beta_n}\frac{\partial}{\partial t}+i\frac{\beta_1}{\beta_n}\left(\vec{v},\hat{p}_1\right)+i\frac{\beta_2}{\beta_n}\left(\dot{\vec{v}},\hat{p}_2\right)+\ldots+i\frac{\beta_{n-1}}{\beta_n}\left(\overset{(n-2)}{\vec{v}},\hat{p}_{n-1}\right)-i\alpha_n\beta_n\left(\hat{p}_n-\frac{\gamma_n}{2\alpha_n\beta_n}\vec{A}_n\right)^2\right]\Psi_n+$$
$$+\Psi_n\left[\frac{1}{\beta_n}\frac{\partial}{\partial t}-i\frac{\beta_1}{\beta_n}\left(\vec{v},\bar{\hat{p}}_1\right)-i\frac{\beta_2}{\beta_n}\left(\dot{\vec{v}},\bar{\hat{p}}_2\right)-\ldots-i\frac{\beta_{n-1}}{\beta_n}\left(\overset{(n-2)}{\vec{v}},\bar{\hat{p}}_{n-1}\right)+i\alpha_n\beta_n\left(\bar{\hat{p}}_n-\frac{\gamma_n}{2\alpha_n\beta_n}\vec{A}_n\right)^2\right]\bar{\Psi}_n=0.$$
(2.19)

Thus, we can work both with the expression (2.17) and (2.19). Let us consider the expression (2.17). We introduce the notation of the linear operator $L$ as:

$$L_n=\frac{1}{\beta_n}\frac{\partial}{\partial t}+i\frac{\beta_1}{\beta_n}\left(\vec{v},\hat{p}_1\right)+i\frac{\beta_2}{\beta_n}\left(\dot{\vec{v}},\hat{p}_2\right)+\ldots+i\frac{\beta_{n-1}}{\beta_n}\left(\overset{(n-2)}{\vec{v}},\hat{p}_{n-1}\right)-i\alpha_n\beta_n\left(\hat{p}_n^2-\frac{\gamma_n}{\alpha_n\beta_n}\left(\vec{A}_n,\hat{p}_n\right)\right),$$
$$\bar{L}_n=\frac{1}{\beta_n}\frac{\partial}{\partial t}-i\frac{\beta_1}{\beta_n}\left(\vec{v},\bar{\hat{p}}_1\right)-i\frac{\beta_2}{\beta_n}\left(\dot{\vec{v}},\bar{\hat{p}}_2\right)-\ldots-i\frac{\beta_{n-1}}{\beta_n}\left(\overset{(n-2)}{\vec{v}},\bar{\hat{p}}_{n-1}\right)+i\alpha_n\beta_n\left(\bar{\hat{p}}_n^2-\frac{\gamma_n}{\alpha_n\beta_n}\left(\vec{A}_n,\bar{\hat{p}}_n\right)\right),$$
(2.20)

then expression (2.17) taking into account the introduced notation (2.20) is of the form:

$$\bar{\Psi}_n L_n \Psi_n + \Psi_n \bar{L}_n \bar{\Psi}_n = 0, \qquad (2.21)$$

or

$$K_n + \bar{K}_n = 0,$$

where $K_n = \bar{\Psi}_n L_n \Psi_n$. The expression (2.21) means that $\mathrm{Re}\, K_n = 0$. If the real part equals to zero, then $K_n$ is a pure imaginary value, that is:

$$K_n = iv_n,\ v_n \in \mathbb{R},$$

or

$$\bar{\Psi}_n L_n \Psi_n = iv_n,\ L_n\Psi_n = i\frac{v_n}{\bar{\Psi}_n} = i\frac{v_n}{\bar{\Psi}_n}\frac{\Psi_n}{\Psi_n} = i\frac{v_n}{|\Psi_n|^2}\Psi_n = i\mu_n\Psi_n, \mu_n \in \mathbb{R},$$
$$L_n\Psi_n = i\mu_n\Psi_n. \qquad (2.22)$$

Note that the operator $L_n$ is anti-Hermitian. We consider the scalar product in the Hilbert space above the field of complex numbers with the use of (2.21):

$$\langle \Psi_n, L_n \Psi_n \rangle = \int \Psi_n \overline{L_n \Psi_n} d^n\omega = -\int \overline{\Psi}_n L_n \Psi_n d^n\omega =$$
$$= -\int (L_n \Psi_n) \overline{\Psi}_n d^n\omega = -\langle L_n \Psi_n, \Psi_n \rangle,$$

that is

$$\langle \Psi_n, L_n \Psi_n \rangle = \langle \overline{L}_n \Psi_n, \Psi_n \rangle \Rightarrow \langle \overline{L}_n \Psi_n, \Psi_n \rangle = -\langle L_n \Psi_n, \Psi_n \rangle. \qquad (2.23)$$

As is known, the anti-Hermitian operator has pure imaginary eigenvalue, which corresponds to the expression (2.22).

Let us return to the expression (2.17) and divide it by the value $i$, we obtain:

$$\overline{\Psi}_n \left[ -\frac{i}{\beta_n} \frac{\partial}{\partial t} + \frac{\beta_1}{\beta_n}(\vec{v}, \hat{p}_1) + \frac{\beta_2}{\beta_n}(\dot{\vec{v}}, \hat{p}_2) + \ldots + \frac{\beta_{n-1}}{\beta_n}\left(\overset{(n-2)}{\vec{v}}, \hat{p}_{n-1}\right) - \alpha_n \beta_n \left( \hat{p}_n^2 - \frac{\gamma_n}{\alpha_n \beta_n}(\vec{A}_n, \hat{p}_n) \right) \right] \Psi_n -$$
$$-\Psi_n \left[ \frac{i}{\beta_n} \frac{\partial}{\partial t} + \frac{\beta_1}{\beta_n}(\vec{v}, \overline{\hat{p}}_1) + \frac{\beta_2}{\beta_n}(\dot{\vec{v}}, \overline{\hat{p}}_2) + \ldots + \frac{\beta_{n-1}}{\beta_n}\left(\overset{(n-2)}{\vec{v}}, \overline{\hat{p}}_{n-1}\right) - \alpha_n \beta_n \left( \overline{\hat{p}}_n^2 - \frac{\gamma_n}{\alpha_n \beta_n}(\vec{A}_n, \overline{\hat{p}}_n) \right) \right] \overline{\Psi}_n = 0.$$
$$(2.24)$$

Let us introduce the notation of the linear operator $\mathcal{L}_n$ as follows:

$$\mathcal{L}_n = -\frac{i}{\beta_n}\frac{\partial}{\partial t} + \frac{\beta_1}{\beta_n}(\vec{v}, \hat{p}_1) + \frac{\beta_2}{\beta_n}(\dot{\vec{v}}, \hat{p}_2) + \ldots + \frac{\beta_{n-1}}{\beta_n}\left(\overset{(n-2)}{\vec{v}}, \hat{p}_{n-1}\right) - \alpha_n \beta_n \left( \hat{p}_n^2 - \frac{\gamma_n}{\alpha_n \beta_n}(\vec{A}_n, \hat{p}_n) \right),$$
$$\overline{\mathcal{L}}_n = \frac{i}{\beta_n}\frac{\partial}{\partial t} + \frac{\beta_1}{\beta_n}(\vec{v}, \overline{\hat{p}}_1) + \frac{\beta_2}{\beta_n}(\dot{\vec{v}}, \overline{\hat{p}}_2) + \ldots + \frac{\beta_{n-1}}{\beta_n}\left(\overset{(n-2)}{\vec{v}}, \overline{\hat{p}}_{n-1}\right) - \alpha_n \beta_n \left( \overline{\hat{p}}_n^2 - \frac{\gamma_n}{\alpha_n \beta_n}(\vec{A}_n, \overline{\hat{p}}_n) \right).$$
$$(2.25)$$

Using the notation (2.25), the equation (2.24) is of the form:

$$\overline{\Psi}_n \mathcal{L}_n \Psi_n - \Psi_n \overline{\mathcal{L}}_n \overline{\Psi}_n = 0,$$

or

$$M_n - \overline{M}_n = 0, \qquad (2.26)$$

where $M_n = \overline{\Psi}_n \mathcal{L}_n \Psi_n$. The expression (2.26) means that $\text{Im} M_n = 0$. If the imaginary part equals zero, then M is a real value, that is:

$$M_n = m_n, \ m_n \in \mathbb{R},$$

or

$$\overline{\Psi}_n \mathcal{L}_n \Psi_n = m_n, \ \mathcal{L}_n \Psi_n = \frac{m_n}{\overline{\Psi}_n} = \frac{m_n}{\overline{\Psi}_n} \frac{\Psi_n}{\Psi_n} = \frac{m_n}{|\Psi_n|^2} \Psi_n = -U_n \Psi_n, \ U_n(\vec{\xi}_n, t) \in \mathbb{R},$$
$$\mathcal{L}_n \Psi_n = -U_n \Psi_n. \qquad (2.27)$$

Note, that the operator $\mathcal{L}_n$ is Hermitian. Indeed, let us consider the scalar product in the Hilbert space above the field of complex numbers with the use of (2.26):

$$\langle \Psi_n, \mathcal{L}_n \Psi_n \rangle = \int \Psi_n \overline{\mathcal{L}_n \Psi_n} d^n\omega = \int \overline{\Psi}_n \mathcal{L}_n \Psi_n d^n\omega = \int (\mathcal{L}_n \Psi_n) \overline{\Psi}_n d^n\omega = \langle \mathcal{L}_n \Psi_n, \Psi_n \rangle,$$

that is

$$\langle \Psi_n, \mathcal{L}_n \Psi_n \rangle = \langle \overline{\mathcal{L}}_n \Psi_n, \Psi_n \rangle \Rightarrow \langle \overline{\mathcal{L}}_n \Psi_n, \Psi_n \rangle = \langle \mathcal{L}_n \Psi_n, \Psi_n \rangle. \tag{2.28}$$

From comparison of the operators $L_n$ and $\mathcal{L}_n$ (2.20) and (2.25) it follows that

$$L_n \Psi_n = i \mathcal{L}_n \Psi_n. \tag{2.29}$$

Considering the expressions (2.22) and (2.27), the equation (2.29) is of the form:

$$i\mu_n \Psi_n = -iU_n \Psi_n,$$

from here it follows that

$$\mu_n = -U_n. \tag{2.30}$$

As a result of the equation (2.22) and (2.27) taking into account (2.30) is of the form:

$$L_n \Psi_n + iU_n \Psi_n = 0, \qquad \mathcal{L}_n \Psi_n + U_n \Psi_n = 0.$$

or

$$\frac{i}{\beta_n} \frac{\partial \Psi_n}{\partial t} = \frac{\beta_1}{\beta_n}(\vec{v}, \hat{p}_1) \Psi_n + \frac{\beta_2}{\beta_n}(\dot{\vec{v}}, \hat{p}_2) \Psi_n + \ldots + \frac{\beta_{n-1}}{\beta_n}\left(\overset{(n-2)}{\vec{v}}, \hat{p}_{n-1}\right) \Psi_n -$$
$$- \alpha_n \beta_n \left( \hat{p}_n^2 - \frac{\gamma_n}{\alpha_n \beta_n}(\vec{A}_n, \hat{p}_n) \right) \Psi_n + U_n \Psi_n. \tag{2.31}$$

Considering (2.18) the equation (2.31) is of the form

$$\frac{i}{\beta_n} \frac{\partial \Psi_n}{\partial t} = \frac{\beta_1}{\beta_n}(\vec{v}, \hat{p}_1) \Psi_n + \frac{\beta_2}{\beta_n}(\dot{\vec{v}}, \hat{p}_2) \Psi_n + \ldots + \frac{\beta_{n-1}}{\beta_n}\left(\overset{(n-2)}{\vec{v}}, \hat{p}_{n-1}\right) \Psi_n -$$
$$- \alpha_n \beta_n \left( \hat{p}_n - \frac{\gamma_n}{2\alpha_n \beta_n} \vec{A}_n \right)^2 \Psi_n + \frac{1}{2\alpha_n \beta_n} \frac{\left|\gamma_n \vec{A}_n\right|^2}{2} \Psi_n + U_n \Psi_n. \tag{2.32}$$

The equation (2.31)/(2.32) is the required equation for the wave function $\Psi_n(\vec{r}, \vec{v}, \dot{\vec{v}}, \ldots, t)$. We define the constant values $\alpha_n, \beta_n, \gamma_n$ as follows

$$\alpha_n \overset{\text{det}}{=} -\frac{\hbar_n}{2m}, \qquad \beta_n \overset{\text{det}}{=} \frac{1}{\hbar_n}, \qquad \gamma_n = -\frac{e}{m}, \tag{2.33}$$

where $\hbar_1 \stackrel{\text{det}}{=} \hbar$ is the reduced Planck constant. In this case, the Heisenberg uncertainty principle for the phase space is fulfilled $(\vec{r}, \vec{v})$

$$\sigma_r \sigma_v \geq |\alpha_1| = \frac{\hbar_1}{2m}. \tag{2.34}$$

The values of the rest constants $\hbar_n$, $n>1$ we also associate with the generalized uncertainty principle [7,15] for the rest projections of the generalized phase space [7,8]

$$\sigma_v \sigma_{\dot{v}} \geq |\alpha_2| = \frac{\hbar_2}{2m}, \tag{2.35}$$

$$\sigma_{\dot{v}} \sigma_{\ddot{v}} \geq |\alpha_3| = \frac{\hbar_3}{2m},$$

...

In the particular case, at $n=1$, the equation (2.31)/(2.32) becomes the known Schrödinger equation for the wave function $\Psi_1(\vec{r},t)$ [10]

$$\frac{i}{\beta_1} \frac{\partial \Psi_1}{\partial t} = -\alpha_1 \beta_1 \left( \hat{p}_1 - \frac{\gamma_1}{2\alpha_1 \beta_1} \vec{A}_1 \right)^2 \Psi_1 + \frac{1}{2\alpha_1 \beta_1} \frac{|\gamma_1 \vec{A}_1|^2}{2} \Psi_1 + U_1 \Psi_1, \tag{2.36}$$

In the case of $n=2$ we obtain a new equation for the wave function $\Psi_2(\vec{r}, \vec{v}, t)$

$$\frac{i}{\beta_2} \frac{\partial \Psi_2}{\partial t} = -\alpha_2 \beta_2 \left( \hat{p}_2 - \frac{\gamma_2}{2\alpha_2 \beta_2} \vec{A}_2 \right)^2 \Psi_2 + \frac{1}{2\alpha_2 \beta_2} \frac{|\gamma_2 \vec{A}_2|^2}{2} \Psi_2 + \frac{\beta_1}{\beta_2} (\vec{v}, \hat{p}_1) \Psi_2 + U_2 \Psi_2, \tag{2.37}$$

According to (2.27), the expression for the potential $U_n$ is of the form

$$U_n(\vec{\xi}_n, t) = -\frac{\bar{\Psi}_n \mathcal{L}_n \Psi_n}{f_n(\vec{\xi}_n, t)}. \tag{2.38}$$

The expression (2.38) determines the generalized Hamilton-Jacobi equation, which is considered in the next section.

### §3 Lagrangian and Hamiltonian formalism

Proceeding form (2.38) and (2.25), we obtain the expression for the potential $U_n$. Taking into consideration (2.16), the operator $\mathcal{L}_n$ can be written as follows

$$\mathcal{L}_n = -\frac{i}{\beta_n} \frac{\partial}{\partial t} - \frac{i}{\beta_n}(\vec{v}, \nabla_r) - \frac{i}{\beta_n}(\dot{\vec{v}}, \nabla_v) - \ldots - \frac{i}{\beta_n} \left( \overset{(n-2)}{\vec{v}}, \nabla_{\overset{(n-3)}{v}} \right) - \frac{\alpha_n}{\beta_n} \left( -\Delta_{\overset{(n-2)}{v}} + \frac{i\gamma_n}{\alpha_n} \left( \vec{A}_n, \nabla_{\overset{(n-2)}{v}} \right) \right). \tag{3.1}$$

Let us perform intermediate transformations

$$\Delta_{(n-2) \atop v}\Psi_n = \left(\nabla_{(n-2) \atop v}, e^{i\varphi_n}\nabla_{(n-2) \atop v}|\Psi_n|\right) + i\left(\nabla_{(n-2) \atop v}, e^{i\varphi}|\Psi_n|\nabla_{(n-2) \atop v}\varphi_n\right) = ie^{i\varphi_n}\left(\nabla_{(n-2) \atop v}\varphi_n, \nabla_{(n-2) \atop v}|\Psi_n|\right) +$$

$$+ e^{i\varphi_n}\Delta_{(n-2) \atop v}|\Psi_n| - e^{i\varphi}|\Psi_n|\left(\nabla_{(n-2) \atop v}\varphi_n, \nabla_{(n-2) \atop v}\varphi_n\right) + ie^{i\varphi_n}\left(\nabla_{(n-2) \atop v}|\Psi_n|, \nabla_{(n-2) \atop v}\varphi_n\right) + ie^{i\varphi_n}|\Psi_n|\Delta_{(n-2) \atop v}\varphi_n,$$

$$\Delta_{(n-2) \atop v}\Psi_n = e^{i\varphi_n}\left\{\Delta_{(n-2) \atop v}|\Psi_n| - |\Psi_n|\left|\nabla_{(n-2) \atop v}\varphi_n\right|^2 + i\left[2\left(\nabla_{(n-2) \atop v}\varphi_n, \nabla_{(n-2) \atop v}|\Psi_n|\right) + |\Psi_n|\Delta_{(n-2) \atop v}\varphi_n\right]\right\}. \qquad (3.2)$$

Substituting (3.2) into (3.1), we obtain

$$\bar{\Psi}_n \mathcal{L}_n \Psi_n = -\frac{i}{\beta_n}\bar{\Psi}_n\frac{\partial\Psi_n}{\partial t} - \frac{i}{\beta_n}\bar{\Psi}_n\left(\vec{v},\nabla_r\Psi_n\right) - \frac{i}{\beta_n}\bar{\Psi}_n\left(\dot{\vec{v}},\nabla_v\Psi_n\right) - \ldots -$$

$$-\frac{i}{\beta_n}\bar{\Psi}_n\left(\overset{(n-2)}{\vec{v}},\nabla_{(n-3) \atop v}\Psi_n\right) + \frac{\alpha_n}{\beta_n}\bar{\Psi}_n\Delta_{(n-2) \atop v}\Psi_n - \frac{i\gamma_n}{\beta_n}\bar{\Psi}_n\left(\vec{A}_n,\nabla_{(n-2) \atop v}\Psi_n\right) =$$

$$= -\frac{i}{\beta_n}|\Psi_n|\frac{\partial|\Psi_n|}{\partial t} + \frac{1}{\beta_n}|\Psi_n|^2\frac{\partial\varphi_n}{\partial t} - \frac{i}{\beta_n}|\Psi_n|\left(\vec{v},\nabla_r|\Psi_n|\right) + \frac{1}{\beta_n}|\Psi_n|^2\left(\vec{v},\nabla_r\varphi_n\right)$$

$$-\frac{i}{\beta_n}|\Psi_n|\left(\dot{\vec{v}},\nabla_v|\Psi_n|\right) + \frac{1}{\beta_n}|\Psi_n|^2\left(\dot{\vec{v}},\nabla_v\varphi_n\right) - \ldots - \frac{i}{\beta_n}|\Psi_n|\left(\overset{(n-2)}{\vec{v}},\nabla_{(n-3) \atop v}|\Psi_n|\right) + \frac{1}{\beta_n}|\Psi_n|^2\left(\overset{(n-2)}{\vec{v}},\nabla_{(n-3) \atop v}\varphi_n\right) +$$

$$+\frac{\alpha_n}{\beta_n}|\Psi_n|\left\{\Delta_{(n-2) \atop v}|\Psi_n| - |\Psi_n|\left|\nabla_{(n-2) \atop v}\varphi_n\right|^2 + i\left[2\left(\nabla_{(n-2) \atop v}\varphi_n,\nabla_{(n-2) \atop v}|\Psi_n|\right) + |\Psi_n|\Delta_{(n-2) \atop v}\varphi_n\right]\right\} -$$

$$-i\frac{\gamma_n}{\beta_n}|\Psi_n|\left(\vec{A}_n,\nabla_{(n-2) \atop v}|\Psi_n|\right) + \frac{\gamma_n}{\beta_n}|\Psi_n|^2\left(\vec{A}_n,\nabla_{(n-2) \atop v}\varphi_n\right),$$

$$\bar{\Psi}_n\mathcal{L}_n\Psi_n = \frac{1}{\beta_n}|\Psi_n|^2\frac{\partial\varphi_n}{\partial t} + \frac{1}{\beta_n}|\Psi_n|^2\left(\vec{v},\nabla_r\varphi_n\right) + \frac{1}{\beta_n}|\Psi_n|^2\left(\dot{\vec{v}},\nabla_v\varphi_n\right) + \ldots +$$

$$+\frac{1}{\beta_n}|\Psi_n|^2\left(\overset{(n-2)}{\vec{v}},\nabla_{(n-3) \atop v}\varphi_n\right) + \frac{\gamma_n}{\beta_n}|\Psi_n|^2\left(\vec{A}_n,\nabla_{(n-2) \atop v}\varphi_n\right) + \frac{\alpha_n}{\beta_n}|\Psi_n|\left(\Delta_{(n-2) \atop v}|\Psi_n| - |\Psi_n|\left|\nabla_{(n-2) \atop v}\varphi_n\right|^2\right) - \qquad (3.3)$$

$$-\frac{i}{\beta_n}|\Psi_n|\frac{\partial|\Psi_n|}{\partial t} - \frac{i}{\beta_n}|\Psi_n|\left(\vec{v},\nabla_r|\Psi_n|\right) - \frac{i}{\beta_n}|\Psi_n|\left(\dot{\vec{v}},\nabla_v|\Psi_n|\right) - \ldots - \frac{i}{\beta_n}|\Psi_n|\left(\overset{(n-2)}{\vec{v}},\nabla_{(n-3) \atop v}|\Psi_n|\right) +$$

$$+i\frac{\alpha_n}{\beta_n}|\Psi_n|\left[2\left(\nabla_{(n-2) \atop v}\varphi_n,\nabla_{(n-2) \atop v}|\Psi_n|\right) + |\Psi_n|\Delta_{(n-2) \atop v}\varphi_n\right] - i\frac{\gamma_n}{\beta_n}|\Psi_n|\left(\vec{A}_n,\nabla_{(n-2) \atop v}|\Psi_n|\right)$$

Substituting (3.3) into (2.38), we obtain

$$U_n = -\frac{\overline{\Psi}_n \mathcal{L}_n \Psi_n}{|\Psi_n|^2} = -\frac{1}{\beta_n}\frac{\partial \varphi_n}{\partial t} - \frac{1}{\beta_n}(\vec{v},\nabla_r \varphi_n) - \frac{1}{\beta_n}(\dot{\vec{v}},\nabla_v \varphi_n) - \ldots - \frac{1}{\beta_n}\left(\overset{(n-2)}{\vec{v}},\nabla_{(n-3)\atop v}\varphi_n\right) -$$
$$-\frac{\gamma_n}{\beta_n}\left(\vec{A}_n,\nabla_{(n-2)\atop v}\varphi_n\right) - \frac{\alpha_n}{\beta_n}\frac{\Delta_{(n-2)\atop v}|\Psi_n|}{|\Psi_n|} + \frac{\alpha_n}{\beta_n}\left|\nabla_{(n-2)\atop v}\varphi_n\right|^2 + i\frac{1}{\beta_n}\frac{1}{|\Psi_n|}\frac{\partial |\Psi_n|}{\partial t} +$$
$$+i\frac{\gamma_n}{\beta_n}\frac{1}{|\Psi_n|}\left(\vec{A}_n,\nabla_{(n-2)\atop v}|\Psi_n|\right) - i\frac{2\alpha_n}{\beta_n}\frac{1}{|\Psi_n|}\left(\nabla_{(n-2)\atop v}\varphi_n,\nabla_{(n-2)\atop v}|\Psi_n|\right) - i\frac{\alpha_n}{\beta_n}\Delta_{(n-2)\atop v}\varphi_n +$$
$$+\frac{i}{\beta_n}\frac{1}{|\Psi_n|}(\vec{v},\nabla_r|\Psi_n|) + \frac{i}{\beta_n}\frac{1}{|\Psi_n|}(\dot{\vec{v}},\nabla_v|\Psi_n|) + \ldots + \frac{i}{\beta_n}\frac{1}{|\Psi_n|}\left(\overset{(n-2)}{\vec{v}},\nabla_{(n-3)\atop v}|\Psi_n|\right).$$

(3.4)

According to the definition (2.27), the function $U_n$ is a real function, consequently, the imaginary side of the expression (3.4) must become zero, let us verify this condition. Indeed, considering the expressions (2.6), (2.7), (2.10), we obtain:

$$\mathrm{Im}\,U_n = \frac{1}{\beta_n}\frac{1}{|\Psi_n|^2}\left[|\Psi_n|\frac{\partial |\Psi_n|}{\partial t} + \left(\gamma_n \vec{A}_n, |\Psi_n|\nabla_{(n-2)\atop v}|\Psi_n|\right)+\right.$$
$$+\left(-2\alpha_n \nabla_{(n-2)\atop v}\varphi_n, |\Psi_n|\nabla_{(n-2)\atop v}|\Psi_n|\right) - \alpha_n|\Psi_n|^2 \Delta_{(n-2)\atop v}\varphi_n\bigg] +$$
$$+\frac{1}{\beta_n}\frac{1}{|\Psi_n|^2}\left[(\vec{v},|\Psi_n|\nabla_r|\Psi_n|) + (\dot{\vec{v}},|\Psi_n|\nabla_v|\Psi_n|) + \ldots + \left(\overset{(n-2)}{\vec{v}},|\Psi_n|\nabla_{(n-3)\atop v}|\Psi_n|\right)\right] =$$
$$= \frac{1}{2\beta_n}\frac{1}{|\Psi_n|^2}\left[\frac{\partial |\Psi_n|^2}{\partial t} + \left(\gamma_n \vec{A}_n,\nabla_{(n-2)\atop v}|\Psi_n|^2\right) + \left(-2\alpha_n \nabla_{(n-2)\atop v}\varphi_n,\nabla_{(n-2)\atop v}|\Psi_n|^2\right) - 2\alpha_n|\Psi_n|^2 \Delta_{(n-2)\atop v}\varphi_n\right] +$$
$$+\frac{1}{2\beta_n}\frac{1}{|\Psi_n|^2}\left[(\vec{v},\nabla_r|\Psi_n|^2) + (\dot{\vec{v}},\nabla_v|\Psi_n|^2) + \ldots + \left(\overset{(n-2)}{\vec{v}},\nabla_{(n-3)\atop v}|\Psi_n|^2\right)\right],$$

$$2\beta_n|\Psi_n|^2\,\mathrm{Im}\,U_n = \frac{\partial f_n}{\partial t} + \left(\gamma_n \vec{A}_n,\nabla_{(n-2)\atop v}f_n\right) + \left(-\alpha_n \nabla_{(n-2)\atop v}\Phi_n,\nabla_{(n-2)\atop v}f_n\right) - \alpha_n f_n \Delta_{(n-2)\atop v}\Phi_n +$$
$$+(\vec{v},\nabla_r f_n) + (\dot{\vec{v}},\nabla_v f_n) + \ldots + \left(\overset{(n-2)}{\vec{v}},\nabla_{(n-3)\atop v}f_n\right) =$$
$$= \frac{\partial f_n}{\partial t} + (\vec{v},\nabla_r f_n) + (\dot{\vec{v}},\nabla_v f_n) + \ldots + \left(\overset{(n-2)}{\vec{v}},\nabla_{(n-3)\atop v}f_n\right) +$$
$$+\left(-\alpha_n \nabla_{(n-2)\atop v}\Phi_n + \gamma_n \vec{A}_n, \nabla_{(n-2)\atop v}f_n\right) + f_n\left(\nabla_{(n-2)\atop v}, -\alpha_n \nabla_{(n-2)\atop v}\Phi_n + \gamma_n \vec{A}_n\right) =$$
$$= \frac{\partial f_n}{\partial t} + (\vec{v},\nabla_r f_n) + (\dot{\vec{v}},\nabla_v f_n) + \ldots + \left(\overset{(n-2)}{\vec{v}},\nabla_{(n-3)\atop v}f_n\right) + \left(\left\langle\overset{(n-1)}{\vec{v}}\right\rangle,\nabla_{(n-2)\atop v}f_n\right) + f_n\left(\nabla_{(n-2)\atop v},\left\langle\overset{(n-1)}{\vec{v}}\right\rangle\right) =$$
$$= \frac{\partial f_n}{\partial t} + (\vec{v},\nabla_r f_n) + (\dot{\vec{v}},\nabla_v f_n) + \ldots + \left(\overset{(n-2)}{\vec{v}},\nabla_{(n-3)\atop v}f_n\right) + \mathrm{div}_{(n-2)\atop v}\left(\left\langle\overset{(n-1)}{\vec{v}}\right\rangle f_n\right) = 0.$$

As a result, we obtain for the function $U_n$ the representation:

$$U_n\left(\vec{\xi}_n,t\right)=-\frac{1}{\beta_n}\frac{\partial\varphi_n}{\partial t}-Q_n+\frac{\alpha_n}{\beta_n}\left|\nabla_{v}^{(n-2)}\varphi_n\right|^2-\frac{1}{\beta_n}\left(\gamma_n\vec{A}_n,\nabla_{v}^{(n-2)}\varphi_n\right)-$$
$$-\frac{1}{\beta_n}\left[\left(\vec{v},\nabla_r\varphi_n\right)+\left(\dot{\vec{v}},\nabla_v\varphi_n\right)+\ldots+\left(\overset{(n-2)}{\vec{v}},\nabla_{v}^{(n-3)}\varphi_n\right)\right],\quad (3.5)$$

where

$$Q_n\overset{\text{det}}{=}\frac{\alpha_n}{\beta_n}\frac{\Delta_{v}^{(n-2)}|\Psi_n|}{|\Psi_n|}=\frac{\alpha_n}{2\beta_n}\left(\Delta_{v}^{(n-2)}S_n+\frac{1}{2}\left|\nabla_{v}^{(n-2)}S_n\right|^2\right),\quad (3.6)$$

where $Q_n$ is a generalized quantum potential. In the particular case at $n=1$, the generalized quantum potential $Q_n$ equals the «classical quantum potential» [11,12]

$$Q_1=Q=\frac{\alpha_1}{\beta_1}\frac{\Delta_r|\Psi_1|}{|\Psi_1|}=-\frac{\hbar^2}{2m}\frac{\Delta_r|\Psi_1|}{|\Psi_1|}.\quad (3.7)$$

In the case of $n=1$, the expression (3.5) is of the known form [10-12]

$$U_1\left(\vec{r},t\right)=-\frac{1}{\beta_1}\left\{\frac{\partial\varphi_1}{\partial t}+\alpha_1\left[\frac{\Delta_r\sqrt{f_1}}{\sqrt{f_1}}-\left|\nabla_r\varphi_1\right|^2\right]+\gamma_1\left(\vec{A}_1,\nabla_r\varphi_1\right)\right\}.\quad (3.8)$$

At $n=2$, the potential is of the form

$$U_2\left(\vec{r},\vec{v},t\right)=-\frac{1}{\beta_2}\left\{\frac{\partial\varphi_2}{\partial t}+\alpha_2\left[\frac{\Delta_v\sqrt{f_2}}{\sqrt{f_2}}-\left|\nabla_v\varphi_2\right|^2\right]+\gamma_2\left(\vec{A}_2,\nabla_v\varphi_2\right)+\left(\vec{v},\nabla_r\varphi_2\right)\right\}.\quad (3.9)$$

The expression (3.5) can be rewritten in terms of the in terms of the operator $\Pi_n\varphi_n=\dfrac{d_n\varphi_n}{dt}$, indeed

$$-\beta_n U_n\left(\vec{\xi}_n,t\right)=\frac{\partial\varphi_n}{\partial t}+\left(\vec{v},\nabla_r\varphi_n\right)+\left(\dot{\vec{v}},\nabla_v\varphi_n\right)+\ldots+\left(\overset{(n-2)}{\vec{v}},\nabla_{v}^{(n-3)}\varphi_n\right)+$$
$$+\left(-2\alpha_n\nabla_{v}^{(n-2)}\varphi_n,\nabla_{v}^{(n-2)}\varphi_n\right)+\alpha_n\left(\nabla_{v}^{(n-2)}\varphi_n,\nabla_{v}^{(n-2)}\varphi_n\right)+\left(\gamma_n\vec{A}_n,\nabla_{v}^{(n-2)}\varphi_n\right)+\alpha_n\frac{\Delta_{v}^{(n-2)}|\Psi_n|}{|\Psi_n|}=$$
$$=\frac{\partial\varphi_n}{\partial t}+\left(\vec{v},\nabla_r\varphi_n\right)+\left(\dot{\vec{v}},\nabla_v\varphi_n\right)+\ldots+\left(\overset{(n-2)}{\vec{v}},\nabla_{v}^{(n-3)}\varphi_n\right)+\left(-\alpha_n\nabla_{v}^{(n-2)}\Phi_n+\gamma_n\vec{A}_n,\nabla_{v}^{(n-2)}\varphi_n\right)+$$
$$+\alpha_n\left(\nabla_{v}^{(n-2)}\varphi_n,\nabla_{v}^{(n-2)}\varphi_n\right)+\alpha_n\frac{\Delta_{v}^{(n-2)}|\Psi_n|}{|\Psi_n|}=\alpha_n\left(\nabla_{v}^{(n-2)}\varphi_n,\nabla_{v}^{(n-2)}\varphi_n\right)+\alpha_n\frac{\Delta_{v}^{(n-2)}|\Psi_n|}{|\Psi_n|}+$$
$$+\frac{\partial\varphi_n}{\partial t}+\left(\vec{v},\nabla_r\varphi_n\right)+\left(\dot{\vec{v}},\nabla_v\varphi_n\right)+\ldots+\left(\overset{(n-2)}{\vec{v}},\nabla_{v}^{(n-3)}\varphi_n\right)+\left(\left\langle\overset{(n-1)}{\vec{v}}\right\rangle,\nabla_{v}^{(n-2)}\varphi_n\right)=$$

$$= \alpha_n \left| \nabla_{(n-2)} \varphi_n \right|^2 + \alpha_n \frac{\Delta_{(n-2)} |\Psi_n|}{|\Psi_n|} + \frac{d_n \varphi_n}{dt},$$

$$U_n(\vec{\xi}_n, t) = -\frac{1}{\beta_n} \left\{ \frac{d_n \varphi_n}{dt} + \alpha_n \left[ \frac{\Delta_{(n-2)} |\Psi_n|}{|\Psi_n|} + \left| \nabla_{(n-2)} \varphi_n \right|^2 \right] \right\}. \tag{3.10}$$

From the expression (3.10) we can obtain the representation for the generalized Lagrangian $L_n(\vec{\xi}_n, t)$, indeed

$$-\frac{1}{\beta_n} \frac{d_n \varphi_n}{dt} = \frac{1}{4\alpha_n \beta_n} \left| 2\alpha_n \nabla_{(n-2)} \varphi_n \right|^2 + U_n + Q_n = \frac{1}{4\alpha_n \beta_n} \left| \alpha_n \nabla_{(n-2)} \Phi_n \right|^2 + U_n + Q_n,$$

$$-\frac{1}{\beta_n} \frac{d_n \varphi_n}{dt} = \frac{1}{2\alpha_n \beta_n} \frac{1}{2} \left| \left\langle \vec{v}_p^{(n-1)} \right\rangle \right|^2 + U_n + Q_n \stackrel{\text{det}}{=} -L_n. \tag{3.11}$$

In this case the phase $\varphi_n$ has the meaning of a generalized action. Note that the Lagrangian $L_n$ does not contain the information on the vortex field $\vec{A}_n$. A similar situation occurs also in the classical case.

We obtain the generalized Hamilton-Jacobi equation. Let us rewrite the expression for the potential (3.5).

$$-\frac{1}{\beta_n} \frac{\partial \varphi_n}{\partial t} = \frac{1}{2\alpha_n \beta_n} \frac{1}{2} \left| \left\langle \vec{v}_p^{(n-1)} \right\rangle \right|^2 + U_n + Q_n +$$

$$+ \frac{1}{2\alpha_n \beta_n} \left[ (\vec{v}, \alpha_n \nabla_r \Phi_n) + (\dot{\vec{v}}, \alpha_n \nabla_v \Phi_n) + \ldots + \left( \vec{v}^{(n-2)}, \alpha_n \nabla_{(n-3)} \Phi_n \right) - \left( \left\langle \vec{v}^{(n-1)} \right\rangle, \left\langle \vec{v}_p^{(n-1)} \right\rangle \right) \right],$$

$$-\frac{1}{\beta_n} \frac{\partial \varphi_n}{\partial t} = -\frac{1}{2\alpha_n \beta_n} \left( \left\langle \vec{v}^{(n-1)} \right\rangle - \frac{1}{2} \left\langle \vec{v}_p^{(n-1)} \right\rangle, \left\langle \vec{v}_p^{(n-1)} \right\rangle \right) + U_n + Q_n +$$

$$+ \frac{1}{2\alpha_n \beta_n} \left[ (\vec{v}, \alpha_n \nabla_r \Phi_n) + (\dot{\vec{v}}, \alpha_n \nabla_v \Phi_n) + \ldots + \left( \vec{v}^{(n-2)}, \alpha_n \nabla_{(n-3)} \Phi_n \right) \right],$$

As a result,

$$-\frac{1}{\beta_n} \frac{\partial \varphi_n}{\partial t} = -\frac{1}{2\alpha_n \beta_n} \frac{1}{2} \left| \left\langle \vec{v}_p^{(n-1)} \right\rangle \right|^2 + U_n + Q_n - \frac{1}{2\alpha_n \beta_n} \left( \gamma_n \vec{A}_n, \left\langle \vec{v}_p^{(n-1)} \right\rangle \right) +$$

$$+ \frac{1}{2\alpha_n \beta_n} \left[ (\vec{v}, \alpha_n \nabla_r \Phi_n) + (\dot{\vec{v}}, \alpha_n \nabla_v \Phi_n) + \ldots + \left( \vec{v}^{(n-2)}, \alpha_n \nabla_{(n-3)} \Phi_n \right) \right], \tag{3.12}$$

The equation (3.12) can also be rewritten in terms of $\left\langle \vec{v}^{(n-1)} \right\rangle$. According to (2.7), $\left\langle \vec{v}^{(n-1)} \right\rangle$ can be represented in the form $\left\langle \vec{v}^{(n-1)} \right\rangle = \left\langle \vec{v}_p^{(n-1)} \right\rangle + \left\langle \vec{v}_s^{(n-1)} \right\rangle$, where $\left\langle \vec{v}_p^{(n-1)} \right\rangle = -\alpha_n \nabla_{\vec{v}^{(n-2)}} \Phi_n$, $\left\langle \vec{v}_s^{(n-1)} \right\rangle = \gamma_n \vec{A}_n$, then the following expression is true

$$\frac{1}{2}\left|\left\langle \vec{v}_p^{(n-1)} \right\rangle\right|^2 + \left(\gamma_n \vec{A}_n, \left\langle \vec{v}_p^{(n-1)} \right\rangle\right) = \frac{1}{2}\left|\left\langle \vec{v}^{(n-1)} \right\rangle\right|^2 - \frac{\left|\gamma_n \vec{A}_n\right|^2}{2}. \tag{3.13}$$

Substituting (3.13) into (3.12), we obtain

$$-\frac{1}{\beta_n}\frac{\partial \varphi_n}{\partial t} = -\frac{1}{2\alpha_n \beta_n}\frac{1}{2}\left|\left\langle \vec{v}^{(n-1)} \right\rangle\right|^2 + U_n + Q_n + \frac{1}{2\alpha_n \beta_n}\frac{\left|\gamma_n \vec{A}_n\right|^2}{2} +$$
$$+\frac{1}{2\alpha_n \beta_n}\left[(\vec{v},\alpha_n \nabla_r \Phi_n) + (\dot{\vec{v}},\alpha_n \nabla_v \Phi_n) + ... + \left(\vec{v}^{(n-2)},\alpha_n \nabla_{\vec{v}^{(n-3)}} \Phi_n\right)\right] \stackrel{\text{det}}{=} H_n, \tag{3.14}$$

where we call the function $H_n\left(\vec{\xi}_n, t\right)$ the generalized Hamiltonian function. Note that the Hamiltonian $H_n$, unlike the Lagrangian $L_n$, contains the information on the vortex field $\vec{A}_n$. A similar situation occurs also in the classical case.

We call the obtained equation (3.14) the generalized Hamilton-Jacobi equation. In the particular case at $n=1$, the equation (3.14) becomes a classical Hamilton-Jacobi equation [10-13]

$$-\frac{1}{\beta_1}\frac{\partial \varphi_1}{\partial t} = -\frac{1}{2\alpha_1 \beta_1}\frac{1}{2}\left|\left\langle \vec{v} \right\rangle\right|^2 + U_1 + Q_1 + \frac{1}{2\alpha_1 \beta_1}\frac{\left|\gamma_1 \vec{A}_1\right|^2}{2} = H_1,$$
$$-\hbar\frac{\partial \varphi_1}{\partial t} = \frac{m}{2}\left|\left\langle \vec{v} \right\rangle\right|^2 + e\chi_1 = H_1, \tag{3.15}$$

where $e\chi_1$ is a classical potential

$$e\chi_1 \stackrel{\text{det}}{=} U_1 + Q_1 + \frac{e^2}{2m}\left|\vec{A}_1\right|^2. \tag{3.16}$$

At $n=2$ we obtain «new Hamilton-Jacobi equation»

$$-\frac{1}{\beta_2}\frac{\partial \varphi_2}{\partial t} = -\frac{1}{2\alpha_2 \beta_2}\frac{1}{2}\left|\left\langle \dot{\vec{v}} \right\rangle\right|^2 + U_2 + Q_2 + \frac{1}{2\alpha_2 \beta_2}\frac{\left|\gamma_2 \vec{A}_2\right|^2}{2} + \frac{1}{2\alpha_2 \beta_2}(\vec{v},\alpha_2 \nabla_r \Phi_2) = H_2,$$
$$-\hbar_2 \frac{\partial \varphi_2}{\partial t} = \frac{m}{2}\left|\left\langle \dot{\vec{v}} \right\rangle\right|^2 + e\chi_2 = H_2, \tag{3.17}$$

where

$$e\chi_2 \stackrel{\text{det}}{=} U_2 + Q_2 + \frac{m}{2}\left|\gamma_2 \vec{A}_2\right|^2 + \hbar_2\left(\vec{v}, \nabla_r \varphi_2\right). \tag{3.18}$$

Using (3.11) and (3.14), we obtain the generalized Legendre transformation in the form

$$\beta_n L_n = \frac{d_n \varphi_n}{dt} = \frac{\partial \varphi_n}{\partial t} + \left({}_n\vec{u}, {}_n\nabla \varphi_n\right)_\xi = -\beta_n H_n + \left({}_n\vec{u}, {}_n\nabla \varphi_n\right)_\xi,$$

$$L_n + H_n + \frac{1}{\beta_n}\left({}_n\vec{u}, {}_n\nabla \varphi_n\right)_\xi. \tag{3.19}$$

In the particular case at $n=1$, the transformation (3.19) becomes a classical Legendre transformation for the Lagrangian and the Hamiltonian

$$L_1 + H_1 + m\left(\langle\vec{v}\rangle, \langle\vec{v}_p\rangle\right). \tag{3.20}$$

At $n=2$, from (3.19) we obtain the transformation which relates to $L_2$ and $H_2$

$$L_2 + H_2 = \hbar_2\left(\vec{v}, \nabla_r \varphi_2\right) + m\left(\langle\dot{\vec{v}}\rangle, \langle\dot{\vec{v}}_p\rangle\right). \tag{3.21}$$

By analogy with [13], let us construct a complex action taking into account kinematic characteristics of the higher value. We use the notation introduced in §2

$$i\Phi_n = \text{Ln}\left(\frac{\Psi_n}{\bar{\Psi}_n}\right) = \text{Ln}\,\Psi_n - \text{Ln}\,\bar{\Psi}_n, \tag{3.22}$$

$$S_n = \text{Ln}\,f_n = \text{Ln}\left(\Psi_n \bar{\Psi}_n\right) = \text{Ln}\,\Psi_n + \text{Ln}\,\bar{\Psi}_n. \tag{3.23}$$

We define the complex function $M_n$ as

$$M_n\left(\vec{\xi}_n, t\right) \stackrel{\text{det}}{=} S_n\left(\vec{\xi}_n, t\right) + i\Phi_n\left(\vec{\xi}_n, t\right). \tag{3.24}$$

The functions $S_n$ and $\Phi_n$ are real functions of the real variables $\left(\vec{\xi}_n, t\right) \left(\vec{r}, t\right)$, therefore, they respectively equal to the real and imaginary parts of the complex functions $M_n$. According to the definitions (3.22) and (3.23) to the functions (3.24), the following is correct:

$$Z_n \stackrel{\text{det}}{=} \frac{M_n}{2} = \text{Ln}\,\Psi_n, \quad \Psi_n = e^{Z_n}. \tag{3.25}$$

Taking into account the relation (2.1) and (3.11), we calculate $\Pi_n Z_n$

$$\Pi_n Z_n = \frac{1}{2}\Pi_n M_n = \frac{d_n Z_n}{dt} = \frac{1}{2}\frac{d_n S_n}{dt} + \frac{i}{2}\frac{d_n \Phi_n}{dt} = -\frac{1}{2}Q_n + \frac{i}{\hbar_n}L_n \stackrel{\text{det}}{=} \Lambda_n. \tag{3.26}$$

By analogy with [13], we call the function $\Lambda_n$ the complex Lagrangian. The imaginary side of $\operatorname{Im}\Lambda_n$ corresponds to the generalized Lagrangian $L_n$, which is related to the generalized action $\varphi_n$. The real part $\operatorname{Re}\Lambda_n$ corresponds to the sources of the flow of probability (2.2) $Q_n = -\alpha_n \Delta_{(n-2)\atop \vec{v}} \Phi_n$. Thus, the variable $Z_n$ corresponds to the complex action.

Let's obtain an expression for the generalized complex Hamiltonian $\mathcal{H}_n$ and the Legendre transformation connecting the functions $\Lambda_n$ and $\mathcal{H}_n$.

$$\nabla_{(n-2)\atop \vec{v}} Z_n = \frac{1}{2}\nabla_{(n-2)\atop \vec{v}} S_n - \frac{i}{2\alpha_n}\left\langle \overset{(n-1)}{\vec{v}_p}\right\rangle, \qquad \Delta_{(n-2)\atop \vec{v}} Z_n = \frac{1}{2}\Delta_{(n-2)\atop \vec{v}} S_n - \frac{i}{2\alpha_n}Q_n,$$

$$\left|\nabla_{(n-2)\atop \vec{v}} Z_n\right|^2 = \frac{1}{4}\left|\nabla_{(n-2)\atop \vec{v}} S_n\right|^2 + \frac{1}{4\alpha_n^2}\left|\left\langle \overset{(n-1)}{\vec{v}_p}\right\rangle\right|^2,$$

$$\Delta_{(n-2)\atop \vec{v}} Z_n + \left|\nabla_{(n-2)\atop \vec{v}} Z_n\right|^2 = \frac{1}{2}\left[\Delta_{(n-2)\atop \vec{v}} S_n + \frac{1}{2}\left|\nabla_{(n-2)\atop \vec{v}} S_n\right|^2\right] + \frac{1}{4\alpha_n^2}\left|\left\langle \overset{(n-1)}{\vec{v}_p}\right\rangle\right|^2 - \frac{i}{2\alpha_n}Q_n =$$

$$= \frac{1}{4\alpha_n^2}\left|\left\langle \overset{(n-1)}{\vec{v}_p}\right\rangle\right|^2 + \frac{\beta_n}{\alpha_n}Q_n - \frac{i}{2\alpha_n}Q_n = \frac{\beta_n}{\alpha_n}\left(\frac{1}{4\alpha_n\beta_n}\left|\left\langle \overset{(n-1)}{\vec{v}_p}\right\rangle\right|^2 + Q_n - \frac{i}{2\beta_n}Q_n\right),$$

$$-\frac{1}{4\alpha_n\beta_n}\left|\left\langle \overset{(n-1)}{\vec{v}_p}\right\rangle\right|^2 + Q_n^{(M)} = \mathcal{Q}_n, \qquad (3.27)$$

where

$$\mathcal{Q}_n \overset{\text{det}}{=} Q_n - \frac{i}{2\beta_n}Q_n,$$

$$Q_n^{(M)} \overset{\text{det}}{=} \frac{\alpha_n}{2\beta_n}\left(\Delta_{(n-2)\atop \vec{v}} M_n + \frac{1}{2}\left|\nabla_{(n-2)\atop \vec{v}} M_n\right|^2\right).$$

Taking into account (3.26), (3.11), and (3.27), we obtain

$$-\frac{i}{\beta_n}\Lambda_n = -\frac{1}{4\alpha_n\beta_n}\left|\left\langle \overset{(n-1)}{\vec{v}_p}\right\rangle\right|^2 - e\mathcal{X}_n, \qquad (3.28)$$

$$e\mathcal{X}_n \overset{\text{det}}{=} U_n + \mathcal{Q}_n,$$

or

$$\frac{i}{2\beta_n}\frac{d_n M_n}{dt} = \frac{i}{\beta_n}\Lambda_n = \frac{\alpha_n}{2\beta_n}\frac{1}{2}\left|\nabla_{(n-2)\atop \vec{v}} M_n\right|^2 + \frac{\alpha_n}{2\beta_n}\Delta_{(n-2)\atop \vec{v}} M_n + U_n,$$

or

$$\frac{i}{\beta_n}\Lambda_n = \frac{\alpha_n}{\beta_n}\left(\left|\nabla_{(n-2)\atop \vec{v}} Z_n\right|^2 + \Delta_{(n-2)\atop \vec{v}} Z_n\right) + U_n = Q_n^{(M)} + U_n \overset{\text{det}}{=} e\mathrm{X}_n^{(M)}.$$

Let's obtain an expression for the generalized complex Hamiltonian

$$\Pi_n Z_n = \frac{\partial Z_n}{\partial t} + \left(_n\underline{\vec{u}},\, _n\underline{\nabla} Z_n\right)_\xi = -\mathcal{H}_n + \left(_n\underline{\vec{u}},\, _n\underline{\nabla} Z_n\right)_\xi = \Lambda_n,$$

$$\Lambda_n + \mathcal{H}_n = \left({}_n\vec{\underline{u}}, {}_n\underline{\nabla} Z_n\right)_\xi, \qquad (3.29)$$

where $\mathcal{H}_n \stackrel{\text{det}}{=} -\dfrac{\partial Z_n}{\partial t}$ is the generalized complex Hamilton-Jacobi equation. Expression (3.29) defines the Legendre transformation for the functions $\Lambda_n$ and $\mathcal{H}_n$. Substituting (3.28) into (3.29) we obtain

$$\mathcal{H}_n = -\frac{1}{2}\left(\frac{\partial S_n}{\partial t} + i\frac{\partial \Phi_n}{\partial t}\right) = \frac{1}{2}Q_n + \frac{1}{2}\left({}_n\vec{\underline{u}}, {}_n\underline{\nabla} S_n\right)_\xi + i\beta_n \mathrm{H}_n,$$

or

$$\mathcal{H}_n = \left({}_n\vec{\underline{u}}, {}_n\underline{\nabla} Z_n\right)_\xi + i\beta_n\left[\frac{1}{4\alpha_n\beta_n}\left|\left\langle \vec{v}_p^{(n-1)}\right\rangle\right|^2 + e\mathcal{X}_n\right] = \left({}_n\vec{\underline{u}}, {}_n\underline{\nabla} Z_n\right)_\xi + i\beta_n e \mathrm{X}_n^{(M)}.$$

Thus, the wave function $\Psi_n$ is fully determined by the introduction of the function $Z_n$. From (3.25) we can see that there is a set mapping in the complex plane. For mapping (3.25) to be univalent it is required that the domain of the function had a horizontal stripe with the width of $2\pi$. By that stripe $0 < \mathrm{Im}\, Z_n < 2\pi$ appears in the complex plane slit along the ray $[0, +\infty)$ (see Fig.1). Therefore, the range of values of the function $M_n$ will be:

$$0 < \mathrm{Im}\, M_n < 4\pi,$$
$$-\infty < \mathrm{Re}\, M_n < +\infty, \qquad (3.30)$$

or by the definition (3.24)

$$0 < \Phi_n\left(\vec{\xi}_n, t\right) < 4\pi,$$
$$-\infty < S_n\left(\vec{\xi}_n, t\right) < +\infty. \qquad (3.31)$$

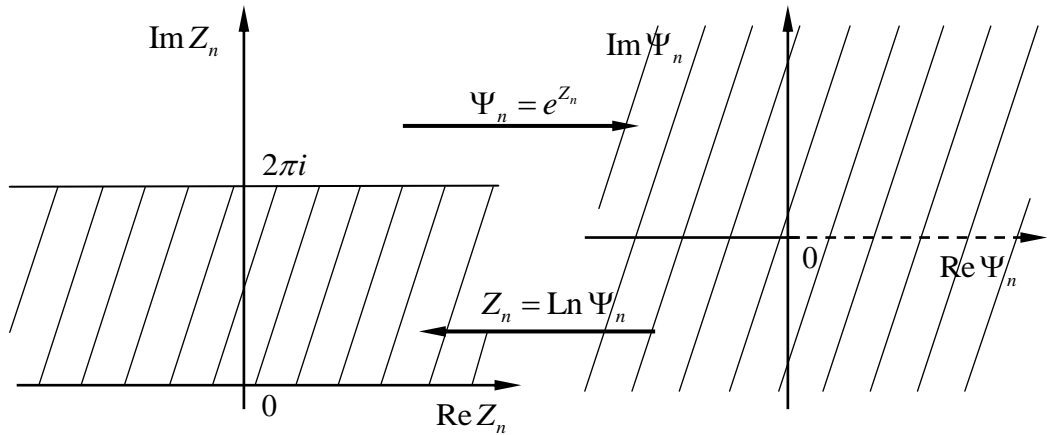

Fig. 1 The Mapping (3.25)

The function $\Phi_n(\vec{\xi}_n, t)$ specifies a scalar velocity potential of the probability flow $\left\langle \overset{(n-1)}{\vec{v}} \right\rangle(\vec{\xi}_n, t)$, and the function $S_n(\vec{\xi}_n, t)$ is connected via the logarithm of a probability density function $f_n(\vec{\xi}_n, t)$. Therefore, the function $\Psi_n(\vec{\xi}_n, t)$ in the moment of time $t$ can be seen not only as a complex function of the real variable $\vec{\xi}_n$ but as a complex function of the complex variable $M_n = 2Z_n$ with domain of the definitions (3.30)-(3.31), that is $\Psi_n(M_n)$.

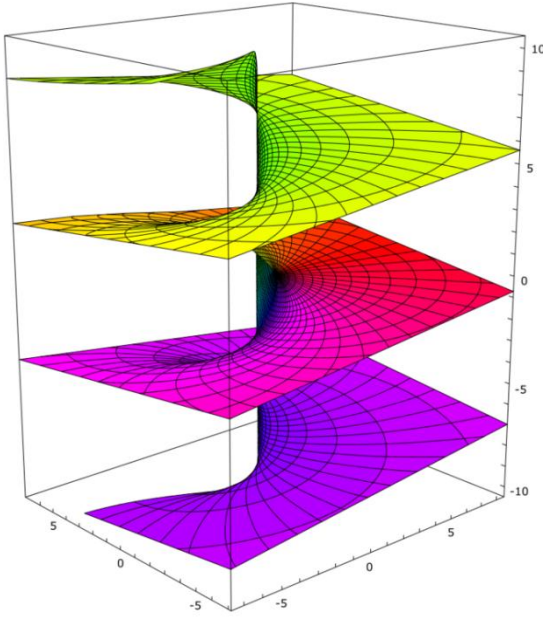

Fig. 2 Riemann surfaces for function $Z_n = \mathrm{Ln}\,\Psi_n$

The range of values of the scalar potential $\Phi_n(\vec{\xi}_n, t)$ in the general case can be wider than (3.31), in this case we can see violation of univalent mapping (3.25). The inverse mapping is a function $\mathrm{Ln}\,\Psi_n$ that is defined on a Riemann surface. The function $\Phi_n(\vec{\xi}_n, t)$ has the physical meaning of an action, the phase of a wave function and the potential of flow probabilities $\left\langle \overset{(n-1)}{\vec{v}} \right\rangle$. Let us consider $\Phi_1(\vec{r}, t)$ as an action $S$, then the limiting transition between classical and quantum mechanics is defined by relation $\frac{S}{\hbar} = \frac{\Phi_1}{2} = \varphi_1$. In classical mechanics $\varphi_1 \gg 1$, that is, the function $\Psi_1$ ($\mathrm{Ln}\,\Psi_1$) is multivalent, as the domain of the function $\Psi_1$ becomes wider than the horizontal stripe $2\pi$ (see Fig.1,2). The value of the function $\Psi_1$ will take several leaves of the Riemann surface.

In quantum mechanics $\varphi_1 \sim 1$, that is, the function $\Psi_1$ ($\mathrm{Ln}\,\Psi_1$) is univalent (see Fig.1,2), and $\varphi_1$ does not exceed $2\pi < 10$. If a violation of univalence, for example, in the case of the two-valence function $0 < \varphi_1 < 4\pi$, value $4\pi > 10 \gg 1$ that is the «movement?» in the direction of classical mechanics.

It is possible, that for higher orders ($n > 1$) the function $\Phi_n$ has the meaning similar to the function $\Phi_1$.

## §4 Generalized motion equation

Based on the expression (3.5), it is possible to obtain motion equations. Let us calculate the gradient of the potential $U_n$. From (2.8) it follows that

$$\nabla_{\underset{v}{(n-2)}}\varphi_n = \frac{1}{2}\nabla_{\underset{v}{(n-2)}}\Phi_n = -\frac{1}{2\alpha_n}\left\langle \overset{(n-1)}{\vec{v}} \right\rangle + \frac{\gamma_n}{2\alpha_n}\vec{A}_n,$$

$$\nabla_{\underset{v}{(n-2)}}\frac{\partial}{\partial t}\varphi_n = -\frac{1}{2\alpha_n}\frac{\partial}{\partial t}\left[\left\langle \overset{(n-1)}{\vec{v}} \right\rangle - \gamma_n\vec{A}_n\right],$$

(4.1)

$$\left|\nabla_{\substack{(n-2)\\v}}\varphi_n\right|^2 = \frac{1}{4\alpha_n^2}\left[\left|\left\langle\vec{v}\right\rangle^{(n-1)}\right|^2 - 2\gamma_n\left(\left\langle\vec{v}\right\rangle^{(n-1)},\vec{A}_n\right) + \gamma_n^2\left|\vec{A}_n\right|^2\right].$$

Taking into account (4.1) and (3.5), the expression $\nabla_{\substack{(n-2)\\v}}U_n$ is as follows:

$$\nabla_{\substack{(n-2)\\v}}U_n = -\frac{1}{\beta_n}\frac{\partial}{\partial t}\nabla_{\substack{(n-2)\\v}}\varphi_n - \nabla_{\substack{(n-2)\\v}}Q_n + \frac{\alpha_n}{\beta_n}\nabla_{\substack{(n-2)\\v}}\left|\nabla_{\substack{(n-2)\\v}}\varphi_n\right|^2 - \frac{1}{\beta_n}\nabla_{\substack{(n-2)\\v}}\left(\gamma_n\vec{A}_n, \nabla_{\substack{(n-2)\\v}}\varphi_n\right) -$$

$$-\frac{1}{\beta_n}\nabla_{\substack{(n-2)\\v}}\left[(\vec{v},\nabla_r\varphi_n) + (\dot{\vec{v}},\nabla_v\varphi_n) + \ldots + \left(\overset{(n-2)}{\vec{v}}, \nabla_{\substack{(n-3)\\v}}\varphi_n\right)\right] =$$

$$= -\frac{1}{\beta_n}\frac{\partial}{\partial t}\left[-\frac{1}{2\alpha_n}\left\langle\vec{v}\right\rangle^{(n-1)} + \frac{\gamma_n}{2\alpha_n}\vec{A}_n\right] + \frac{\alpha_n}{\beta_n}\nabla_{\substack{(n-2)\\v}}\frac{1}{4\alpha_n^2}\left[\left|\left\langle\vec{v}\right\rangle^{(n-1)}\right|^2 - 2\gamma_n\left(\left\langle\vec{v}\right\rangle^{(n-1)},\vec{A}_n\right) + \gamma_n^2\left|\vec{A}_n\right|^2\right] -$$

$$-\frac{1}{\beta_n}\nabla_{\substack{(n-2)\\v}}\left(\gamma_n\vec{A}_n, -\frac{1}{2\alpha_n}\left\langle\vec{v}\right\rangle^{(n-1)} + \frac{\gamma_n}{2\alpha_n}\vec{A}_n\right) - \nabla_{\substack{(n-2)\\v}}Q_n -$$

$$-\frac{1}{\beta_n}\nabla_{\substack{(n-2)\\v}}\left[(\vec{v},\nabla_r\varphi_n) + (\dot{\vec{v}},\nabla_v\varphi_n) + \ldots + \left(\overset{(n-2)}{\vec{v}}, \nabla_{\substack{(n-3)\\v}}\varphi_n\right)\right] =$$

$$= \frac{1}{2\alpha_n\beta_n}\frac{\partial}{\partial t}\left\langle\vec{v}\right\rangle^{(n-1)} - \frac{\gamma_n}{2\alpha_n\beta_n}\frac{\partial}{\partial t}\vec{A}_n + \frac{1}{4\alpha_n\beta_n}\nabla_{\substack{(n-2)\\v}}\left|\left\langle\vec{v}\right\rangle^{(n-1)}\right|^2 - \frac{\gamma_n}{2\alpha_n\beta_n}\nabla_{\substack{(n-2)\\v}}\left(\left\langle\vec{v}\right\rangle^{(n-1)},\vec{A}_n\right) +$$

$$+\frac{\gamma_n^2}{4\alpha_n\beta_n}\nabla_{\substack{(n-2)\\v}}\left|\vec{A}_n\right|^2 + \frac{\gamma_n}{2\alpha_n\beta_n}\nabla_{\substack{(n-2)\\v}}\left(\vec{A}_n,\left\langle\vec{v}\right\rangle^{(n-1)}\right) - \frac{\gamma_n^2}{2\alpha_n\beta_n}\nabla_{\substack{(n-2)\\v}}\left|\vec{A}_n\right|^2 - \nabla_{\substack{(n-2)\\v}}Q_n -$$

$$-\frac{1}{\beta_n}\nabla_{\substack{(n-2)\\v}}\left[(\vec{v},\nabla_r\varphi_n) + (\dot{\vec{v}},\nabla_v\varphi_n) + \ldots + \left(\overset{(n-2)}{\vec{v}}, \nabla_{\substack{(n-3)\\v}}\varphi_n\right)\right],$$

$$\nabla_{\substack{(n-2)\\v}}U_n = \frac{1}{2\alpha_n\beta_n}\frac{\partial}{\partial t}\left\langle\vec{v}\right\rangle^{(n-1)} - \frac{\gamma_n}{2\alpha_n\beta_n}\frac{\partial}{\partial t}\vec{A}_n + \frac{1}{4\alpha_n\beta_n}\nabla_{\substack{(n-2)\\v}}\left|\left\langle\vec{v}\right\rangle^{(n-1)}\right|^2 - \tag{4.2}$$

$$-\frac{\gamma_n^2}{4\alpha_n\beta_n}\nabla_{\substack{(n-2)\\v}}\left|\vec{A}_n\right|^2 - \nabla_{\substack{(n-2)\\v}}Q_n - \frac{1}{\beta_n}\nabla_{\substack{(n-2)\\v}}\left[(\vec{v},\nabla_r\varphi_n) + (\dot{\vec{v}},\nabla_v\varphi_n) + \ldots + \left(\overset{(n-2)}{\vec{v}}, \nabla_{\substack{(n-3)\\v}}\varphi_n\right)\right].$$

The addend $\nabla_{\substack{(n-2)\\v}}\left|\left\langle\vec{v}\right\rangle^{(n-1)}\right|^2$ in the expression (4.2) can be represented in the form

$$\frac{1}{2}\nabla_{\substack{(n-1)\\r}}\left|\left\langle\vec{v}\right\rangle^{(n-1)}\right|^2 = \left(\left\langle\vec{v}\right\rangle^{(n-1)}, \nabla_{\substack{(n-1)\\r}}\right)\left\langle\vec{v}\right\rangle^{(n-1)} + \gamma_n\left[\left\langle\vec{v}\right\rangle^{(n-1)}, {}^{(n)}\vec{B}_n\right], \tag{4.3}$$

where

$${}^{(k)}\vec{B}_n \overset{\text{det}}{=} \operatorname*{rot}_{\substack{(k-1)\\r}}\vec{A}_n = \operatorname*{rot}_{\substack{(k-2)\\v}}\vec{A}_n.$$

Substituting (4.3) into (4.2), we obtain

$$\nabla_{\underset{v}{(n-2)}} U_n = \frac{1}{2\alpha_n \beta_n} \frac{\partial}{\partial t}\left\langle \overset{(n-1)}{\vec{v}} \right\rangle + \frac{1}{2\alpha_n \beta_n}\left(\left\langle \overset{(n-1)}{\vec{v}} \right\rangle, \nabla_{\underset{v}{(n-2)}}\right)\left\langle \overset{(n-1)}{\vec{v}} \right\rangle - \frac{\gamma_n}{2\alpha_n \beta_n}\frac{\partial}{\partial t}\vec{A}_n +$$

$$+ \frac{\gamma_n}{2\alpha_n \beta_n}\left[\left\langle \overset{(n-1)}{\vec{v}} \right\rangle, {}^{(n)}\vec{B}_n\right] - \frac{\gamma_n^2}{4\alpha_n \beta_n}\nabla_{\underset{v}{(n-2)}}\left|\vec{A}_n\right|^2 - \nabla_{\underset{v}{(n-2)}} Q_n - \quad (4.4)$$

$$-\frac{1}{\beta_n}\nabla_{\underset{v}{(n-2)}}\left[(\vec{v},\nabla_r \varphi_n)+(\dot{\vec{v}},\nabla_v \varphi_n)+...+\left(\overset{(n-2)}{\vec{v}},\nabla_{\underset{v}{(n-3)}}\varphi_n\right)\right].$$

Let us transform the following expression

$$\nabla_{\underset{v}{(n-2)}}\left[(\vec{v},\nabla_r \varphi_n)+(\dot{\vec{v}},\nabla_v \varphi_n)+...+\left(\overset{(n-2)}{\vec{v}},\nabla_{\underset{v}{(n-3)}}\varphi_n\right)\right] =$$

$$= \nabla_{\underset{v}{(n-2)}}(\vec{v},\nabla_r \varphi_n) + \nabla_{\underset{v}{(n-2)}}(\dot{\vec{v}},\nabla_v \varphi_n) + ... + \nabla_{\underset{v}{(n-2)}}\left(\overset{(n-2)}{\vec{v}},\nabla_{\underset{v}{(n-3)}}\varphi_n\right). \quad (4.5)$$

If the mixed derivatives of the function $\varphi_n$ are equal, then the following relations are correct

$$\nabla_{\underset{v}{(n-2)}}(\vec{v},\nabla_r \varphi_n) = (\vec{v},\nabla_r)\nabla_{\underset{v}{(n-2)}}\varphi_n,$$

$$\nabla_{\underset{v}{(n-2)}}(\dot{\vec{v}},\nabla_v \varphi_n) = (\dot{\vec{v}},\nabla_v)\nabla_{\underset{v}{(n-2)}}\varphi_n, \quad (4.6)$$

$$...$$

$$\nabla_{\underset{v}{(n-2)}}\left(\overset{(n-2)}{\vec{v}},\nabla_{\underset{v}{(n-3)}}\varphi_n\right) = \nabla_{\underset{v}{(n-3)}}\varphi_n + \left(\overset{(n-2)}{\vec{v}},\nabla_{\underset{v}{(n-3)}}\right)\nabla_{\underset{v}{(n-2)}}\varphi_n.$$

Considering (4.6), the expression (4.5) is as follows

$$\nabla_{\underset{v}{(n-2)}}\left[(\vec{v},\nabla_r \varphi_n)+(\dot{\vec{v}},\nabla_v \varphi_n)+...+\left(\overset{(n-2)}{\vec{v}},\nabla_{\underset{v}{(n-3)}}\varphi_n\right)\right] =$$

$$= \frac{1}{2\alpha_n}\left[(\vec{v},\nabla_r)+(\dot{\vec{v}},\nabla_v)+...+\left(\overset{(n-2)}{\vec{v}},\nabla_{\underset{v}{(n-3)}}\right)\right]\nabla_{\underset{v}{(n-2)}}\alpha_n \Phi_n + \frac{1}{2\alpha_n}\nabla_{\underset{v}{(n-3)}}\alpha_n \Phi_n. \quad (4.7)$$

Substituting (4.7) into (4.4), we obtain

$$-\frac{1}{2\alpha_n\beta_n}\left[\frac{\partial}{\partial t}\left\langle\overset{(n-1)}{\vec{v}}\right\rangle+\left(\left\langle\overset{(n-1)}{\vec{v}}\right\rangle,\nabla_{\underset{v}{(n-2)}}\right)\left\langle\overset{(n-1)}{\vec{v}}\right\rangle\right]=-\nabla_{\underset{v}{(n-2)}}U_n-\nabla_{\underset{v}{(n-2)}}Q_n-\frac{\gamma_n}{2\alpha_n\beta_n}\frac{\partial}{\partial t}\vec{A}_n+$$

$$+\frac{\gamma_n}{2\alpha_n\beta_n}\left[\left\langle\overset{(n-1)}{\vec{v}}\right\rangle,{}^{(n)}\vec{B}_n\right]-\frac{1}{2\alpha_n\beta_n}\nabla_{\underset{v}{(n-3)}}\alpha_n\Phi_n-\frac{\gamma_n^2}{4\alpha_n\beta_n}\nabla_{\underset{v}{(n-2)}}\left|\vec{A}_n\right|^2+$$

$$+\frac{1}{2\alpha_n\beta_n}\left[(\vec{v},\nabla_r)+(\dot{\vec{v}},\nabla_v)+\ldots+\left(\overset{(n-2)}{\vec{v}},\nabla_{\underset{v}{(n-3)}}\right)\right]\left(-\nabla_{\underset{v}{(n-2)}}\alpha_n\Phi_n+\gamma_n\vec{A}_n\right)+$$

$$-\frac{1}{2\alpha_n\beta_n}\left[(\vec{v},\nabla_r)+(\dot{\vec{v}},\nabla_v)+\ldots+\left(\overset{(n-2)}{\vec{v}},\nabla_{\underset{v}{(n-3)}}\right)\right]\gamma_n\vec{A}_n,$$

or

$$-\frac{1}{2\alpha_n\beta_n}\left[\frac{\partial}{\partial t}+(\vec{v},\nabla_r)+(\dot{\vec{v}},\nabla_v)+\ldots+\left(\overset{(n-2)}{\vec{v}},\nabla_{\underset{v}{(n-3)}}\right)+\left(\left\langle\overset{(n-1)}{\vec{v}}\right\rangle,\nabla_{\underset{v}{(n-2)}}\right)\right]\left\langle\overset{(n-1)}{\vec{v}}\right\rangle=$$

$$=-\frac{1}{2\alpha_n\beta_n}\frac{d_n}{dt}\left\langle\overset{(n-1)}{\vec{v}}\right\rangle=-\nabla_{\underset{v}{(n-2)}}U_n-\nabla_{\underset{v}{(n-2)}}Q_n+\frac{\gamma_n}{2\alpha_n\beta_n}\left[\left\langle\overset{(n-1)}{\vec{v}}\right\rangle,{}^{(n)}\vec{B}_n\right]-$$

$$-\frac{1}{2\alpha_n\beta_n}\nabla_{\underset{v}{(n-3)}}\alpha_n\Phi_n-\frac{\gamma_n^2}{4\alpha_n\beta_n}\nabla_{\underset{v}{(n-2)}}\left|\vec{A}_n\right|^2+$$

$$-\frac{\gamma_n}{2\alpha_n\beta_n}\left[\frac{\partial}{\partial t}+(\vec{v},\nabla_r)+(\dot{\vec{v}},\nabla_v)+\ldots+\left(\overset{(n-2)}{\vec{v}},\nabla_{\underset{v}{(n-3)}}\right)\right]\vec{A}_n.$$
(4.8)

Let us transform the expressions

$$(\vec{v},\nabla_r)\vec{A}_n=\nabla_r(\vec{v},\vec{A}_n)-\left[\vec{v},{}^{(1)}\vec{B}_n\right],$$
$$(\dot{\vec{v}},\nabla_v)\vec{A}_n=\nabla_v(\dot{\vec{v}},\vec{A}_n)-\left[\dot{\vec{v}},{}^{(2)}\vec{B}_n\right],$$
$$\ldots$$
$$\left(\overset{(n-2)}{\vec{v}},\nabla_{\underset{v}{(n-3)}}\right)\vec{A}_n=\nabla_{\underset{v}{(n-3)}}\left(\overset{(n-2)}{\vec{v}},\vec{A}_n\right)-\left[\overset{(n-2)}{\vec{v}},{}^{(n-1)}\vec{B}_n\right].$$
(4.9)

Substituting (4.9) into (4.8), we obtain a generalized motion equation

$$-\frac{1}{2\alpha_n\beta_n}\frac{d_n}{dt}\left\langle\overset{(n-1)}{\vec{v}}\right\rangle=-\frac{\gamma_n}{2\alpha_n\beta_n}\frac{\partial}{\partial t}\vec{A}_n-\nabla_{\underset{v}{(n-2)}}U_n-\nabla_{\underset{v}{(n-2)}}Q_n-\frac{\gamma_n^2}{4\alpha_n\beta_n}\nabla_{\underset{v}{(n-2)}}\left|\vec{A}_n\right|^2-$$

$$-\frac{1}{2\alpha_n\beta_n}\left(\nabla_r(\vec{v},\gamma_n\vec{A}_n)+\nabla_v(\dot{\vec{v}},\gamma_n\vec{A}_n)+\ldots+\nabla_{\underset{v}{(n-3)}}\left(\overset{(n-2)}{\vec{v}},\gamma_n\vec{A}_n\right)+\alpha_n\nabla_{\underset{v}{(n-3)}}\Phi_n\right)+$$
(4.10)

$$+\frac{\gamma_n}{2\alpha_n\beta_n}\left(\left[\vec{v},{}^{(1)}\vec{B}_n\right]+\left[\dot{\vec{v}},{}^{(2)}\vec{B}_n\right]+\ldots+\left[\overset{(n-2)}{\vec{v}},{}^{(n-1)}\vec{B}_n\right]+\left[\left\langle\overset{(n-1)}{\vec{v}}\right\rangle,{}^{(n)}\vec{B}_n\right]\right).$$

The equation (4.10) can be rewritten in a compact form, considering the formalism of the $\Psi$-algebra (see §1). Let us introduce the following notations

$$_n\underline{\chi} = \begin{pmatrix} {}^{(1)}\chi_n \\ {}^{(2)}\chi_n \\ ... \\ {}^{(n-1)}\chi_n \\ {}^{(n)}\chi_n \end{pmatrix} = \begin{pmatrix} (\vec{v},\vec{A}_n) + {}^{(1)}c_n \\ (\dot{\vec{v}},\vec{A}_n) + {}^{(2)}c_n \\ ... \\ \left(\overset{(n-2)}{\vec{v}},\vec{A}_n\right) + \dfrac{\alpha_n}{\gamma_n}\Phi_n + {}^{(n-1)}c_n \\ \dfrac{2\alpha_n\beta_n}{\gamma_n}\left(U_n + Q_n + \dfrac{1}{4\alpha_n\beta_n}|\gamma_n\vec{A}_n|^2\right) + {}^{(n)}c_n \end{pmatrix}, \quad _n\vec{\underline{B}} = \begin{pmatrix} {}^{(1)}\vec{B}_n \\ {}^{(2)}\vec{B}_n \\ ... \\ {}^{(n-1)}\vec{B}_n \\ {}^{(n)}\vec{B}_n \end{pmatrix}. \quad (4.11)$$

where ${}^{(k)}c_n$ are some functions having the properties:

$$^{(1)}c_n = {}^{(1)}c_n\left(\vec{v},\dot{\vec{v}},...,\overset{(n-2)}{\vec{v}},t\right),$$

$$^{(2)}c_n = {}^{(2)}c_n\left(\vec{r},\dot{\vec{v}},...,\overset{(n-2)}{\vec{v}},t\right),$$

...

$$^{(n-1)}c_n = {}^{(n-1)}c_n\left(\vec{r},\vec{v},...,\overset{(n-4)}{\vec{v}},\overset{(n-2)}{\vec{v}},t\right),$$

$$^{(n)}c_n = {}^{(n)}c_n\left(\vec{r},\vec{v},...,\overset{(n-3)}{\vec{v}},t\right).$$

As a result, we obtain

$$-\dfrac{d_n}{dt}\left\langle \overset{(n-1)}{\vec{v}} \right\rangle = -\gamma_n \dfrac{\partial}{\partial t}\vec{A}_n - \gamma_n\left(_n\nabla_n\underline{\chi}\right)_\xi + \gamma_n\left[_n\vec{\underline{u}},_n\vec{\underline{B}}\right]_\xi,$$

$$\dfrac{d_n}{dt}\left\langle \overset{(n-1)}{\vec{v}} \right\rangle = -\gamma_n\left(\vec{E}_n + \left[_n\vec{\underline{u}},_n\vec{\underline{B}}\right]_\xi\right), \quad (4.12)$$

where

$$\vec{E}_n \overset{\text{det}}{=} -\dfrac{\partial}{\partial t}\vec{A}_n - \left(_n\nabla_n\underline{\chi}\right)_\xi.$$

The motion equations (4.10) can also be written in other form

$$-\dfrac{1}{2\alpha_n\beta_n}\dfrac{d_n}{dt}\left\langle \overset{(n-1)}{\vec{v}} \right\rangle = -\nabla_{\underset{v}{(n-2)}}U_n - \nabla_{\underset{v}{(n-2)}}Q_n + \dfrac{\gamma_n}{2\alpha_n\beta_n}\left[\left\langle \overset{(n-1)}{\vec{v}} \right\rangle, {}^{(n)}\vec{B}_n\right] -$$

$$-\dfrac{1}{2\alpha_n\beta_n}\nabla_{\underset{v}{(n-3)}}\alpha_n\Phi_n - \dfrac{1}{2\alpha_n\beta_n}\left[\dfrac{1}{2}\nabla_{\underset{v}{(n-2)}}|\gamma_n\vec{A}_n|^2 - \left(\left\langle \overset{(n-1)}{\vec{v}} \right\rangle, \nabla_{\underset{v}{(n-2)}}\right)\gamma_n\vec{A}_n\right] - \dfrac{\gamma_n}{2\alpha_n\beta_n}\dfrac{d_n}{dt}\vec{A}_n,$$

or

$$\frac{d_n}{dt}\left\langle \vec{v}^{(n-1)}\right\rangle = -\gamma_n\left(\vec{E}_n + \left[\left\langle \vec{v}^{(n-1)}\right\rangle, {}^{(n)}\vec{B}_n\right]\right), \tag{4.13}$$

where

$$^{(n)}\chi_n = \frac{2\alpha_n \beta_n}{\gamma_n}\left(U_n + Q_n + \frac{1}{2}\left|\gamma_n \vec{A}_n\right|^2\right),$$

$$\vec{E}_n = -\frac{d_n}{dt}\vec{A}_n - \nabla_{v}^{(n-2)}{}^{(n)}\chi_n - \frac{\alpha_n}{\gamma_n}\nabla_{v}^{(n-3)}\Phi_n + \left(\left\langle \vec{v}^{(n-1)}\right\rangle, \nabla_{v}^{(n-2)}\right)\vec{A}_n.$$

The representations (4.12) and (4.13) of motion equations are equivalents and differ only in the form they are written.

In the case $n=1$, the equations (4.12), (4.13) are of the classical form [10]

$$\frac{d}{dt}\langle \vec{v}\rangle = -\gamma_1\left(\vec{E}_1 + \left[\langle \vec{v}\rangle, {}^{(1)}\vec{B}_1\right]\right), \tag{4.14}$$

where

$$\vec{E}_1 = -\frac{\partial}{\partial t}\vec{A}_1 - \nabla_r {}^{(1)}\chi_1, \quad {}^{(1)}\vec{B}_1 = \operatorname{rot}_r \vec{A}_1,$$

$$e^{(1)}\chi_1 = U_1 + Q_1 + \frac{e^2}{2m^2}\left|\vec{A}_1\right|^2.$$

At $n=2$, the equation (4.13) is of the form

$$\frac{d_2}{dt}\langle \dot{\vec{v}}\rangle = -\gamma_2\left(\vec{E}_2 + \left[\langle \dot{\vec{v}}\rangle, {}^{(2)}\vec{B}_2\right]\right), \tag{4.15}$$

where

$$\vec{E}_2 = -\frac{\partial \vec{A}_2}{\partial t} - \nabla_v {}^{(2)}\chi_2 - \frac{\alpha_2}{\gamma_2}\nabla_r \Phi_2 - (\vec{v}, \nabla_r)\vec{A}_2,$$

$$^{(2)}\chi_2 = \frac{2\alpha_2 \beta_2}{\gamma_2}\left(U_2 + Q_2 + \frac{1}{2}\left|\gamma_2 \vec{A}_2\right|^2\right).$$

The equation (4.12)

$$\frac{d_2}{dt}\langle \dot{\vec{v}}\rangle = -\gamma_2\left(\vec{E}_2 + \left[\vec{v}, {}^{(1)}\vec{B}_2\right] + \left[\langle \dot{\vec{v}}\rangle, {}^{(2)}\vec{B}_2\right]\right), \tag{4.16}$$

where

$$\vec{E}_2 = -\frac{\partial \vec{A}_2}{\partial t} - \nabla_v {}^{(2)}\chi_2 - \nabla_r {}^{(1)}\chi_2,$$

$$^{(1)}\chi_2 = \frac{\alpha_2}{\gamma_2}\Phi_2 + (\vec{v}, \vec{A}_2).$$

*Hydrodynamic representation*

In paper [14] we obtained a chain of relations for the derivatives and average values of kinematic characteristics of the for

$$f_0 \langle\langle \vec{v} \rangle\rangle = f_0 \frac{d_0}{dt}\langle \vec{r} \rangle + \langle \vec{r} \rangle \frac{d_0}{dt} f_0, \tag{4.17}$$

$$f_1 \langle\langle \dot{\vec{v}} \rangle\rangle = f_1 \frac{d_1}{dt}\langle \vec{v} \rangle + \int_{(\infty)} (\vec{v} - \langle \vec{v} \rangle)(\vec{v} - \langle \vec{v} \rangle, \nabla_r f_2) d^3v,$$

$$f_2 \langle\langle \ddot{\vec{v}} \rangle\rangle = f_2 \frac{d_2}{dt}\langle \dot{\vec{v}} \rangle + \int_{(\infty)} (\dot{\vec{v}} - \langle \dot{\vec{v}} \rangle)(\dot{\vec{v}} - \langle \dot{\vec{v}} \rangle, \nabla_v f_3) d^3\dot{v},$$

...

The relations (4.17) can be rewritten in terms of the hydrodynamic stress-elasticity tensor [5, 6]

$$^{(1)}P_{\mu\lambda} = \int_{(\infty)} f_2(\vec{r}, \vec{v}, t)(v_\mu - \langle v_\mu \rangle)(v_\lambda - \langle v_\lambda \rangle) d^3v,$$

$$^{(2)}P_{\mu\lambda} = \int_{(\infty)} f_3(\vec{r}, \vec{v}, \dot{\vec{v}}, t)(\dot{v}_\mu - \langle \dot{v}_\mu \rangle)(\dot{v}_\lambda - \langle \dot{v}_\lambda \rangle) d^3\dot{v}, \tag{4.18}$$

....

Let us calculate the derivative $\dfrac{\partial {}^{(1)}P_{\mu\lambda}}{\partial x^\lambda}$, we obtain

$$\frac{\partial {}^{(1)}P_{\mu\lambda}}{\partial x^\lambda} = \frac{\partial}{\partial x^\lambda} \int_{(\infty)} (v_\mu - \langle v_\mu \rangle)(v_\lambda - \langle v_\lambda \rangle) f_2 d^3v = \int_{(\infty)} (v_\mu - \langle v_\mu \rangle)(v_\lambda - \langle v_\lambda \rangle) \frac{\partial f_2}{\partial x^\lambda} d^3v -$$

$$- \frac{\partial \langle v_\mu \rangle}{\partial x^\lambda} \int_{(\infty)} (v_\lambda - \langle v_\lambda \rangle) f_2 d^3v - \frac{\partial \langle v_\lambda \rangle}{\partial x^\lambda} \int_{(\infty)} (v_\mu - \langle v_\mu \rangle) f_2 d^3v,$$

$$\frac{\partial {}^{(1)}P_{\mu\lambda}}{\partial x^\lambda} = \int_{(\infty)} (v_\mu - \langle v_\mu \rangle)(v_\lambda - \langle v_\lambda \rangle) \frac{\partial f_2}{\partial x^\lambda} d^3v. \tag{4.19}$$

By analogy, for $\dfrac{\partial {}^{(2)}P_{\mu\lambda}}{\partial v^\lambda}, \dfrac{\partial {}^{(3)}P_{\mu\lambda}}{\partial \dot{v}^\lambda},\ldots$ we obtain

$$\frac{\partial {}^{(2)}P_{\mu\lambda}}{\partial v^\lambda} = \int_{(\infty)} (\dot{v}_\mu - \langle \dot{v}_\mu \rangle)(\dot{v}_\lambda - \langle \dot{v}_\lambda \rangle) \frac{\partial f_3}{\partial v^\lambda} d^3\dot{v},$$

$$\frac{\partial {}^{(3)}P_{\mu\lambda}}{\partial \dot{v}^\lambda} = \int_{(\infty)} (\ddot{v}_\mu - \langle \ddot{v}_\mu \rangle)(\ddot{v}_\lambda - \langle \ddot{v}_\lambda \rangle) \frac{\partial f_4}{\partial \dot{v}^\lambda} d^3\ddot{v},$$

...

Substituting (4.19) into (4.17), we obtain a set of motion equations

$$\frac{d_1}{dt}\langle v_\mu \rangle = -\frac{1}{f_1}\frac{\partial^{(1)} P_{\mu\lambda}}{\partial x^\lambda} + \langle\langle \dot{v}_\mu \rangle\rangle, \qquad (4.20)$$

$$\frac{d_2}{dt}\langle \dot{v}_\mu \rangle = -\frac{1}{f_2}\frac{\partial^{(2)} P_{\mu\lambda}}{\partial v^\lambda} + \langle\langle \ddot{v}_\mu \rangle\rangle,$$

...

The motion equations (4.20) are related to the obtained generalized motion equations (4.12), (4.13). The mean values of $\langle\langle \dot{v}_\mu \rangle\rangle$, $\langle\langle \ddot{v}_\mu \rangle\rangle$,... in the hydrodynamic representation correspond to external interactions. The derivatives $\frac{\partial^{(1)} P_{\mu\lambda}}{\partial x^\lambda}$, $\frac{\partial^{(2)} P_{\mu\lambda}}{\partial v^\lambda}$,... in the hydrodynamic representation are associated with «forces of pressure».

When considering quantum mechanics, the generalized quantum potential $Q_n$ leads to the generalized quantum pressure $^{(n)}P_{\mu\lambda}^{(q)}$ of the form

$$^{(n)}P_{\mu\lambda}^{(q)} = -\alpha_n^2 f_n \frac{\partial^2 S_n}{\partial^{(n)} x^\mu \partial^{(n)} x^\lambda}, \qquad (4.21)$$

where $^{(n)}x^\lambda \stackrel{\text{det}}{=} \left[ x^\lambda \right]^{(n-1)}$, that is $^{(1)}x^\lambda = x^\lambda$, $^{(2)}x^\lambda = v^\lambda$, $^{(3)}x^\lambda = \dot{v}^\lambda$,.... Indeed, calculating the "force" of the generalized quantum pressure $-\frac{1}{f_n}\frac{\partial^{(n)} P_{\mu\lambda}^{(q)}}{\partial^{(n)} x^\lambda}$, we obtain

$$\frac{\partial}{\partial^{(n)} x^\lambda} {}^{(n)}P_{\mu\lambda}^{(q)} = -\alpha_n^2 \frac{\partial}{\partial^{(n)} x_n^\lambda}\left( f_n \frac{\partial^2 \ln f_n}{\partial^{(n)} x^\mu \partial^{(n)} x^\lambda} \right) = -\alpha_n^2 \frac{\partial f_n}{\partial^{(n)} x^\lambda} \frac{\partial^2 \ln f_n}{\partial^{(n)} x^\mu \partial^{(n)} x^\lambda} - \qquad (4.22)$$

$$-\alpha_n^2 f_n \frac{\partial}{\partial^{(n)} x^\mu}\left( \frac{\partial^2 \ln f_1}{\partial^{(n)} x^\lambda \partial^{(n)} x^\lambda} \right).$$

We take into consideration, that

$$\frac{\partial^2 \ln f_n}{\partial^{(n)} x^\lambda \partial^{(n)} x^\lambda} = \frac{2}{\sqrt{f_n}}\frac{\partial^2 \sqrt{f_n}}{\partial^{(n)} x^\lambda \partial^{(n)} x^\lambda} - \frac{1}{2}\frac{\partial \ln f_n}{\partial^{(n)} x^\lambda}\frac{\partial \ln f_n}{\partial^{(n)} x^\lambda}.$$

then (4.22) is as follows

$$\frac{\partial}{\partial^{(n)}x^{\lambda}}{}^{(n)}P^{(q)}_{\mu\lambda} = -\alpha_n^2 \frac{\partial f_n}{\partial^{(n)}x^{\lambda}} \frac{\partial^2 \ln f_n}{\partial^{(n)}x^{\mu}\partial^{(n)}x^{\lambda}} - 2\alpha_n^2 f_n \frac{\partial}{\partial^{(n)}x^{\mu}}\left(\frac{1}{\sqrt{f_n}} \frac{\partial^2 \sqrt{f_n}}{\partial^{(n)}x^{\lambda}\partial^{(n)}x^{\lambda}}\right) +$$

$$+\alpha_n^2 \frac{1}{2} f_n \frac{\partial}{\partial^{(n)}x^{\mu}}\left(\frac{\partial \ln f_n}{\partial^{(n)}x^{\lambda}} \frac{\partial \ln f_n}{\partial^{(n)}x^{\lambda}}\right) = -2\alpha_n^2 f_n \frac{\partial}{\partial^{(n)}x^{\mu}}\left(\frac{1}{\sqrt{f_n}} \frac{\partial^2 \sqrt{f_n}}{\partial^{(n)}x^{\lambda}\partial^{(n)}x^{\lambda}}\right) -$$

$$-\alpha_n^2 f_n \left(\frac{1}{f_n} \frac{\partial f_n}{\partial^{(n)}x^{\lambda}} \frac{\partial^2 \ln f_n}{\partial^{(n)}x^{\mu}\partial^{(n)}x^{\lambda}} - \frac{\partial \ln f_n}{\partial^{(n)}x^{\lambda}} \frac{\partial^2 \ln f_n}{\partial^{(n)}x^{\lambda}\partial^{(n)}x^{\mu}}\right) = -2\alpha_n^2 f_n \frac{\partial}{\partial^{(n)}x^{\mu}}\left(\frac{1}{\sqrt{f_n}} \frac{\partial^2 \sqrt{f_n}}{\partial^{(n)}x^{\lambda}\partial^{(n)}x^{\lambda}}\right) -$$

$$-\alpha_n^2 f_n \left(\frac{\partial \ln f_n}{\partial^{(n)}x^{\lambda}} \frac{\partial^2 \ln f_n}{\partial^{(n)}x^{\mu}\partial^{(n)}x^{\lambda}} - \frac{\partial \ln f_n}{\partial^{(n)}x^{\lambda}} \frac{\partial^2 \ln f_n}{\partial^{(n)}x^{\lambda}\partial^{(n)}x^{\mu}}\right) = -2\alpha_n^2 f_n \frac{\partial}{\partial^{(n)}x^{\mu}}\left(\frac{1}{\sqrt{f_n}} \frac{\partial^2 \sqrt{f_n}}{\partial^{(n)}x^{\lambda}\partial^{(n)}x^{\lambda}}\right),$$

from this we obtain

$$-\frac{1}{f_n} \frac{\partial^{(n)} P^{(q)}_{\mu\lambda}}{\partial^{(n)}x^{\lambda}} = 2\alpha_n^2 \frac{\partial}{\partial^{(n)}x^{\mu}}\left(\frac{1}{\sqrt{f_n}} \frac{\partial^2 \sqrt{f_n}}{\partial^{(n)}x^{\lambda}\partial^{(n)}x^{\lambda}}\right) = 2\alpha_n \beta_n \frac{\partial Q_n}{\partial^{(n)}x^{\mu}}. \tag{4.23}$$

The addend (4.23) is included into the generalized motion equations (4.12), (4.13). According to the hydrodynamic representation (4.20), the addend (4.23) is associated with the «force» of the generalized quantum pressure (4.21).

§5 Equations of the "electromagnetic" field

Based on the generalized motion equations (4.12), (4.13), we can obtain equations foe the fields $\vec{E}$, $\vec{B}$, $\vec{D}$, $\vec{H}$.

From (4.3) it follows that

$$\operatorname{div}_r {}^{(1)}\vec{B}_n = 0, \quad \operatorname{div}_v {}^{(2)}\vec{B}_n = 0, \ldots, \tag{5.1}$$

or in the form of $\Psi$-vectors (see §1)

$$_n\vec{\underline{B}} = \left[{}_n\underline{\nabla}, {}_n\vec{\underline{A}}\right], \quad _n\vec{\underline{A}} \stackrel{\text{det}}{=} \begin{pmatrix} \vec{A}_n \\ \ldots \\ 0 \end{pmatrix}, \quad \left({}_n\underline{\nabla}, {}_n\vec{\underline{B}}\right) = \begin{pmatrix} 0 \\ \ldots \\ 0 \end{pmatrix}, \tag{5.2}$$

$$\left({}_n\underline{\nabla}, {}_n\vec{\underline{B}}\right) = 0. \tag{5.3}$$

From the equations (5.1) and (5.3) it follows that the divergence and $\Psi$-divergence of the fields $\vec{B}$ equals zero. At $n=1$, the equations (5.1) and (5.3) become a single equation

$$\operatorname{div}_r {}^{(1)}\vec{B}_1 = 0. \tag{5.4}$$

The equation (5.4) is included into the Maxwell's equation system.

Let us calculate the rotation of the expression (4.12), $\operatorname{rot}_{(k) \atop r} \vec{E}_n$ for $k=0,\ldots,n$ we obtain

$$-\frac{\partial^{(1)}\vec{B}_n}{\partial t} = \text{rot}_r\,\vec{E}_n + \text{rot}_r\left(\nabla_v{}^{(2)}\chi + \ldots + \nabla_{(n-2)\atop v}{}^{(n)}\chi\right),$$

$$-\frac{\partial^{(2)}\vec{B}_n}{\partial t} = \text{rot}_v\,\vec{E}_n + \text{rot}_v\left(\nabla_r{}^{(1)}\chi + \nabla_{\dot{v}}{}^{(3)}\chi + \ldots + \nabla_{(n-2)\atop v}{}^{(n)}\chi\right), \quad (5.5)$$

...

$$-\frac{\partial^{(n)}\vec{B}_n}{\partial t} = \text{rot}_{(n-2)\atop v}\vec{E}_n + \text{rot}_v\left(\nabla_r{}^{(1)}\chi + \nabla_v{}^{(2)}\chi + \ldots + \nabla_{(n-3)\atop v}{}^{(n-1)}\chi\right).$$

At $n=1$, the equations (5.5) become a single equation

$$-\frac{\partial^{(1)}\vec{B}_1(\vec{r},t)}{\partial t} = \text{rot}_r\,\vec{E}_1(\vec{r},t). \qquad (5.6)$$

The equation (5.6) is also included in the Maxwell's equation system. Before considering the system of equations (5.5) for the cases $n>1$, let us define the following relations, making relations between the fields $\vec{E}_n$ and $\vec{E}_{n+1}$, $\vec{A}_n$ and $\vec{A}_{n+1}$ as well as between the potentials ${}^{(n)}\chi$ and ${}^{(n+1)}\chi$.

$$\vec{E}_n\left(\vec{r},\vec{v},\ldots,\overset{(n-2)}{\vec{v}},t\right) = \int_{(\infty)} \vec{E}_{n+1}\left(\vec{r},\vec{v},\ldots,\overset{(n-1)}{\vec{v}},t\right) d^3\overset{(n-1)}{v}, \qquad (5.7)$$

$$\vec{A}_n\left(\vec{r},\vec{v},\ldots,\overset{(n-2)}{\vec{v}},t\right) = \int_{(\infty)} \vec{A}_{n+1}\left(\vec{r},\vec{v},\ldots,\overset{(n-1)}{\vec{v}},t\right) d^3\overset{(n-1)}{v}, \qquad (5.8)$$

$${}^{(k)}\chi_n\left(\vec{r},\vec{v},\ldots,\overset{(n-2)}{\vec{v}},t\right) = \int_{(\infty)} {}^{(k)}\chi_{n+1}\left(\vec{r},\vec{v},\ldots,\overset{(n-1)}{\vec{v}},t\right) d^3\overset{(n-1)}{v}, \quad k=1,\ldots,n. \qquad (5.9)$$

The correctness of the applied conditions (5.7)-(5.9) should be seen from the further constructions.

For all $k=1,\ldots,n-1$ the condition (5.9) is fulfilled automatically. In the case $k=n$, the condition (5.9) is equivalent to the expression

$$U_n + Q_n + \frac{1}{4\alpha_n\beta_n}\left|\gamma_n\vec{A}_n\right|^2 + {}^{(n)}c_n = \frac{\gamma_n}{2\alpha_n\beta_n}\int_{(\infty)}\left[\left(\overset{(n-1)}{\vec{v}},\vec{A}_{n+1}\right) + \frac{\alpha_{n+1}}{\gamma_{n+1}}\Phi_{n+1} + {}^{(n)}c_{n+1}\right]d^3\overset{(n-1)}{v}. \qquad (5.10)$$

If the vortex component $\vec{A}_n$ of the field is absent, the expression (5.10) is as follows

$$U_n + Q_n + {}^{(n)}c_n = \frac{\gamma_n}{2\alpha_n\beta_n}\int_{(\infty)}\left(\frac{\alpha_{n+1}}{\gamma_{n+1}}\Phi_{n+1} + {}^{(n)}c_{n+1}\right)d^3\overset{(n-1)}{v}.$$

Note that it follows from the condition (5.8) the field property ${}^{(k)}\vec{B}_n = \text{rot}_{(k-2)\atop v}\vec{A}_n$

$$^{(k)}\vec{B}_n\left(\vec{r},\vec{v},..,\overset{(n-2)}{\vec{v}},t\right) = \operatorname{rot}_{(k-1)\atop r}\vec{A}_n = \int\limits_{(\infty)}\operatorname{rot}_{(k-1)\atop r}\vec{A}_{n+1}\left(\vec{r},\vec{v},..,\overset{(n-1)}{\vec{v}},t\right)d^3\overset{(n-1)}{v} =$$
$$= \int\limits_{(\infty)}{}^{(k)}\vec{B}_{n+1}\left(\vec{r},\vec{v},..,\overset{(n-1)}{\vec{v}},t\right)d^3\overset{(n-1)}{v}. \tag{5.11}$$

So, let us consider the case $n=2$. We obtain from (5.5) two new equations

$$-\frac{\partial^{(1)}\vec{B}_2(\vec{r},\vec{v},t)}{\partial t} = \operatorname{rot}_r \vec{E}_2(\vec{r},\vec{v},t) + \operatorname{rot}_r \nabla_v{}^{(2)}\chi_2(\vec{r},\vec{v},t),$$
$$-\frac{\partial^{(2)}\vec{B}_2(\vec{r},\vec{v},t)}{\partial t} = \operatorname{rot}_v \vec{E}_2(\vec{r},\vec{v},t) + \operatorname{rot}_v \nabla_r{}^{(1)}\chi_2(\vec{r},\vec{v},t). \tag{5.12}$$

Let us integrate the equations (5.12) over the velocity space using of the relations (5.7), (5.11). For the first equation from (5.12), taking into consideration (1.21), we obtain

$$-\frac{\partial}{\partial t}\int\limits_{(\infty)}{}^{(1)}\vec{B}_2 d^3v = \operatorname{rot}_r \int\limits_{(\infty)}\vec{E}_2 d^3v - \int\limits_{(\infty)}\operatorname{rot}_v \nabla_r{}^{(2)}\chi_2 d^3v,$$
$$-\frac{\partial^{(1)}\vec{B}_1(\vec{r},t)}{\partial t} = \operatorname{rot}_r \vec{E}_1(\vec{r},t). \tag{5.13}$$

When obtaining (5.13), it was supposed that the field $\nabla_r{}^{(2)}\chi_2$ vanishes rapidly «enough» at $v\to\infty$, that is

$$\int\limits_{(\infty)}\operatorname{rot}_v \nabla_r{}^{(2)}\chi_2 d^3v = \int\limits_{\Sigma_\infty}\left[d\vec{\sigma}_v, \nabla_r{}^{(2)}\chi_2\right] = 0.$$

The equation (5.13) completely coincides with the equation (5.6). Integration of the second equation from (5.12) gives a trivial result «0=0».

Let us calculate the divergence from $\vec{E}_n$ (4.12) and take into account (2.7), we obtain

$$-\frac{\partial}{\partial t}\operatorname{div}_r \vec{A}_n = \operatorname{div}_r \vec{E}_n + \Delta_r{}^{(1)}\chi_n + \operatorname{div}_r\left(\nabla_v{}^{(2)}\chi_n + ... + \nabla_{(n-2)\atop v}{}^{(n)}\chi_n\right),$$
$$-\frac{\partial}{\partial t}\operatorname{div}_v \vec{A}_n = \operatorname{div}_v \vec{E}_n + \Delta_v{}^{(2)}\chi_n + \operatorname{div}_v\left(\nabla_r{}^{(1)}\chi_n + \nabla_{\dot v}{}^{(3)}\chi_n + ... + \nabla_{(n-2)\atop v}{}^{(n)}\chi_n\right), \tag{5.14}$$
...
$$-\frac{\partial}{\partial t}\operatorname{div}_{(n-2)\atop v}\vec{A}_n = 0 = \operatorname{div}_{(n-2)\atop v}\vec{E}_n + \Delta_{(n-2)\atop v}{}^{(n)}\chi_n + \operatorname{div}_{(n-2)\atop v}\left(\nabla_r{}^{(1)}\chi_n + \nabla_v{}^{(2)}\chi_n + ... + \nabla_{(n-3)\atop v}{}^{(n-1)}\chi_n\right).$$

In the case $n=1$, the equations (5.14) become a single equation

$$\operatorname{div}_r \vec{E}_1 = -\Delta_r{}^{(1)}\chi_1, \tag{5.15}$$

where $e^{(1)}\chi_1 = U_1 + Q_1 + \dfrac{e^2}{2m^2}|\vec{A}_1|^2$. When considering the model of «strongly self-consistent field» [10], the right side of the equation (5.15) corresponds the density $f_1(\vec{r},t)$.

At $n = 2$, the system (5.14) consists of two equations

$$-\frac{\partial}{\partial t}\text{div}_r\,\vec{A}_2 = \text{div}_r\,\vec{E}_2 + \Delta_r{}^{(1)}\chi_2 + \text{div}_r\,\nabla_v{}^{(2)}\chi_2,$$
$$-\frac{\partial}{\partial t}\text{div}_v\,\vec{A}_2 = \text{div}_v\,\vec{E}_2 + \Delta_v{}^{(2)}\chi_2 + \text{div}_v\,\nabla_r{}^{(1)}\chi_2. \quad (5.16)$$

Let us integrate the equations (5.16) over the velocity space using the relations (5.7-9) and the property

$$\text{div}_r\,\nabla_v\chi = (\nabla_r, \nabla_v\chi) = (\nabla_r, \nabla_v)\chi = (\nabla_v, \nabla_r)\chi = \text{div}_v\,\nabla_r\chi.$$

For the first equation from (5.16) considering (2.7), we obtain

$$-\frac{\partial}{\partial t}\text{div}_r\int_{(\infty)}\vec{A}_2 d^3v = \text{div}_r\int_{(\infty)}\vec{E}_2 d^3v + \Delta_r\int_{(\infty)}{}^{(1)}\chi_2 d^3v + \int_{(\infty)}\text{div}_v\,\nabla_r{}^{(2)}\chi_2 d^3v,$$

$$-\frac{\partial}{\partial t}\text{div}_r\,\vec{A}_1 = \text{div}_r\,\vec{E}_1 + \Delta_r{}^{(1)}\chi_1,$$

$$\text{div}_r\,\vec{E}_1 = -\Delta_r{}^{(1)}\chi_1. \quad (5.17)$$

The equation (5.17) completely coincides with the equation (5.15). Integrating the second equation over the velocity space gives a trivial result «0=0».

The conditions (5.9-10) on the potential $\chi$ for the case $n = 2$ is as follows

$${}^{(1)}\chi_1(\vec{r},t) = \int_{(\infty)}{}^{(1)}\chi_2(\vec{r},\vec{v},t) d^3v,$$

$$U_1 + Q_1 + \frac{1}{2}|\gamma_1\vec{A}_1|^2 + {}^{(1)}c_1 = \frac{\gamma_1}{2\alpha_1\beta_1}\int_{(\infty)}\left[\frac{\alpha_2}{\gamma_2}\Phi_2 + (\vec{v},\vec{A}_2) + {}^{(1)}c_2\right]d^3v, \quad (5.18)$$

Using the notations from §1, we rewrite the Vlasov equation (2.1)/(2.5) as follows

$$\frac{\partial f_n}{\partial t} + \left({}_n\underline{\nabla}, {}_n\underline{\vec{u}}\,f_n\right)_\xi = 0. \quad (5.19)$$

The function $f_n$ is a scalar function, which can be represented in the form of the divergence of a field ${}_n\vec{D}$, that is

$$f_n \stackrel{\text{det}}{=} \left({}_n\underline{\nabla}, {}_n\underline{\vec{D}}\right)_\xi, \quad (5.20)$$

or

$$f_n = \text{div}_r\,{}^{(1)}\vec{D}_n + \text{div}_v\,{}^{(2)}\vec{D}_n + \ldots + \text{div}_{(n-2)v}\,{}^{(n)}\vec{D}_n.$$

Note that the representation (5.20) agrees with the conditions (2.3). Substituting (5.20) into the equation (5.19), we obtain

$$\left(_n\vec{\nabla}, \frac{\partial\, _n\vec{D}}{\partial t} + {}_n\vec{u}\left(_n\vec{\nabla}, {}_n\vec{D}\right)_\xi\right)_\xi = 0. \tag{5.21}$$

Taking into account the property (1.19) from the expression (5.21), we can obtain the equation

$$\frac{\partial\, _n\vec{D}}{\partial t} + {}_n\vec{J} = \left[_n\vec{\nabla}, {}_n\vec{H}\right], \quad n \in \mathbb{N}_2, \tag{5.22}$$

where $_n\vec{J} \stackrel{\text{det}}{=} {}_n\vec{u}\left(_n\vec{\nabla}, {}_n\vec{D}\right)_\xi$, and $_n\vec{H}$ is a field. At $n=1$ the equation (5.22) becomes formally one of Maxwell's equations

$$\frac{\partial\, ^{(1)}\vec{D}_1(\vec{r},t)}{\partial t} + {}^{(1)}\vec{J}_1(\vec{r},t) = \text{rot}_r\, ^{(1)}\vec{H}_1(\vec{r},t), \tag{5.23}$$

where $^{(1)}\vec{J}_1 = f_1\langle\vec{v}\rangle$, $f_1 = \text{div}_r\, ^{(1)}\vec{D}_1$. In the case $n=2$ we obtain two equations

$$\begin{pmatrix} \dfrac{\partial\, ^{(1)}\vec{D}_2}{\partial t} \\ \dfrac{\partial\, ^{(2)}\vec{D}_2}{\partial t} \end{pmatrix} + \begin{pmatrix} \vec{v}f_2 \\ \langle\dot{\vec{v}}\rangle f_2 \end{pmatrix} = \begin{pmatrix} \text{rot}_r\, ^{(1)}\vec{H}_2 + \text{rot}_v\, ^{(2)}\vec{H}_2 \\ \text{rot}_r\, ^{(2)}\vec{H}_2 + \text{rot}_v\, ^{(1)}\vec{H}_2 \end{pmatrix},$$

or

$$\frac{\partial\, ^{(1)}\vec{D}_2(\vec{r},\vec{v},t)}{\partial t} + \vec{v}f_2(\vec{r},\vec{v},t) = \text{rot}_r\, ^{(1)}\vec{H}_2(\vec{r},\vec{v},t) + \text{rot}_v\, ^{(2)}\vec{H}_2(\vec{r},\vec{v},t),$$
$$\frac{\partial\, ^{(2)}\vec{D}_2(\vec{r},\vec{v},t)}{\partial t} + \langle\dot{\vec{v}}\rangle(\vec{r},\vec{v},t) f_2(\vec{r},\vec{v},t) = \text{rot}_r\, ^{(2)}\vec{H}_2(\vec{r},\vec{v},t) + \text{rot}_v\, ^{(1)}\vec{H}_2(\vec{r},\vec{v},t). \tag{5.24}$$

As before (5.7-9), we impose analog conditions on the fields $\vec{D}$ and $\vec{H}$

$$\vec{D}_n\left(\vec{r},\vec{v},...,\overset{(n-2)}{\vec{v}},t\right) = \int\limits_{(\infty)} \vec{D}_{n+1}\left(\vec{r},\vec{v},...,\overset{(n-1)}{\vec{v}},t\right) d^3\overset{(n-1)}{\vec{v}}, \tag{5.25}$$

$$\vec{H}_n\left(\vec{r},\vec{v},...,\overset{(n-2)}{\vec{v}},t\right) = \int\limits_{(\infty)} \vec{H}_{n+1}\left(\vec{r},\vec{v},...,\overset{(n-1)}{\vec{v}},t\right) d^3\overset{(n-1)}{\vec{v}}. \tag{5.26}$$

Let us integrate the first equation from (5.24) over the velocity space taking into account (2.4), (5.25-26), we obtain

$$\frac{\partial}{\partial t}\int\limits_{(\infty)} {}^{(1)}\vec{D}_2(\vec{r},\vec{v},t)d^3v + \int\limits_{(\infty)} \vec{v}f_2(\vec{r},\vec{v},t)d^3v = \text{rot}_r\int\limits_{(\infty)} {}^{(1)}\vec{H}_2(\vec{r},\vec{v},t)d^3v + \int\limits_{(\infty)} \text{rot}_v\, ^{(2)}\vec{H}_2(\vec{r},\vec{v},t)d^3v,$$

$$\frac{\partial {}^{(1)}\vec{D}_1(\vec{r},t)}{\partial t} + {}^{(1)}\vec{J}_1 = \operatorname{rot}_r {}^{(1)}\vec{H}_1(\vec{r},t), \tag{5.27}$$

The obtained equation (5.27) completely coincides with the Maxwell equation (5.23), obtained at $n=1$. Integrating the second equation in (5.24) over the velocity space gives a new equation

$$\frac{\partial}{\partial t}\int_{(\infty)} {}^{(2)}\vec{D}_2 d^3v + \int_{(\infty)} \langle \dot{\vec{v}} \rangle f_2(\vec{r},\vec{v},t) d^3v = \operatorname{rot}_r \int_{(\infty)} {}^{(2)}\vec{H}_2 d^3v + \int_{(\infty)} \operatorname{rot}_v {}^{(1)}\vec{H}_2 d^3v,$$

$$\frac{\partial {}^{(2)}\vec{D}_1(\vec{r},t)}{\partial t} + f_1 \langle\langle \dot{\vec{v}} \rangle\rangle = \operatorname{rot}_r {}^{(2)}\vec{H}_1(\vec{r},t), \tag{5.28}$$

The equation (5.28) significantly differs from the Maxwell equation (5.23), as it contains new «current» density $f_1 \langle\langle \dot{\vec{v}} \rangle\rangle$ or rather the current density of an external force (4.17) or (4.20). According to the physical content, the equation (5.28) corresponds to the motion equation in the hydrodynamic representation (4.20).

Thus, a system of the generalized field equations (5.3), (5.5), (5.14), (5.22) is obtained. In the particular case, at $n=1$ or when integrating over the corresponding kinematic spaces $d^3v d^3\dot{v}...$, the obtained system becomes the known system of Maxwell's equations.

### §6 Example of the quantum harmonic oscillator in the generalized phase space

Let us consider a model of the one-dimensional harmonic oscillator. The oscillation equation is of the form

$$\ddot{x} + \omega_1^2 x = 0, \tag{6.1}$$

where $\omega_1$ is the oscillation frequency. In follows from the equation (6.1) that the following relations are true

$$\dot{v} = -\omega_1^2 x, \quad \ddot{v} = -\omega_1^2 v, \quad \dddot{v} = -\omega_1^2 \dot{v}, \quad \ddddot{v} = -\omega_1^2 \ddot{v}, ...., \tag{6.2}$$

From (6.1) according to (2.2) and (2.4), we obtain

$$\langle v \rangle = 0, \quad \langle \dot{v} \rangle = -\omega_1^2 x, \quad \langle \ddot{v} \rangle = -\omega_1^2 v, \quad \langle \dddot{v} \rangle = -\omega_1^2 \dot{v}, \quad \langle \ddddot{v} \rangle = -\omega_1^2 \ddot{v}, ...., \tag{6.3}$$

$$Q_1 = \operatorname{div}_r \langle v \rangle = 0, \quad Q_2 = \operatorname{div}_v \langle \dot{v} \rangle = 0, \quad Q_3 = \operatorname{div}_{\dot{v}} \langle \ddot{v} \rangle = 0, \quad Q_4 = \operatorname{div}_{\ddot{v}} \langle \dddot{v} \rangle = 0, ....$$

The chain of Vlasov equations (2.1)/(2.5) for the harmonic oscillator in the one-dimensional stationary case is as follows

$$\langle v \rangle \frac{\partial f_1}{\partial x} = 0, \tag{6.4}$$

$$v \frac{\partial f_2}{\partial x} + \langle \dot{v} \rangle \frac{\partial f_2}{\partial v} = 0,$$

$$v\frac{\partial f_3}{\partial x}+\dot v\frac{\partial f_3}{\partial v}+\langle\ddot v\rangle\frac{\partial f_3}{\partial \dot v}=0,$$

$$v\frac{\partial f_4}{\partial x}+\dot v\frac{\partial f_4}{\partial v}+\ddot v\frac{\partial f_4}{\partial \dot v}+\langle\dddot v\rangle\frac{\partial f_4}{\partial \ddot v}=0,$$

...

The characteristic equations for (6.4) are of the form

$$\frac{dx}{\langle v\rangle}=0,$$

$$\frac{dx}{v}=\frac{dv}{\langle \dot v\rangle},$$

$$\frac{dx}{v}=\frac{dv}{\dot v}=\frac{d\dot v}{\langle\ddot v\rangle}, \quad (6.5)$$

$$\frac{dx}{v}=\frac{dv}{\dot v}=\frac{d\dot v}{\ddot v}=\frac{d\ddot v}{\langle\dddot v\rangle},$$

...

From (6.5) we obtain the following first integrals

$$\xi_1 = x^2,$$
$$\xi_2 = v^2 + \omega_1^2 x^2,$$
$$\xi_3 = \dot v^2 + \omega_1^2 v^2, \quad \eta_3 = \dot v + \omega_1^2 x, \quad (6.6)$$
$$\xi_4 = \ddot v^2 + \omega_1^2 \dot v^2, \ \eta_4 = \ddot v + \omega_1^2 v, \ \varsigma_4 = \omega_1\left(\dot v + \omega_1^2 x\right) + \eta_4 \arcsin\frac{\ddot v}{\sqrt{\xi_4}},$$

...

As a result, the solution of the equation chain (6.4) can be represented as follows

$$f_{1,n}(x) = F_{1,n}(\xi_1(x)),$$
$$f_{2,n}(x,v) = F_{2,n}(\xi_2(x,v)),$$
$$f_{3,n}(x,v,\dot v) = F_{3,n}(\xi_3(v,\dot v),\eta_3(x,\dot v)), \quad (6.6)$$
$$f_{4,n}(x,v,\dot v,\ddot v) = F_{4,n}(\xi_4(\dot v,\ddot v),\eta_4(v,\ddot v),\varsigma_4(x,v,\dot v,\ddot v)),$$

...

where $F$ is a functions, which can be defined from the initial-boundary conditions. The form of the functions $F_{1,n}$ and $F_{2,n}$ for the harmonic oscillator is known. The function $F_{1,n}$ is obtained from the solution of the Schrödinger equation for the harmonic oscillator

$$F_{1,n} = |\psi_n(x)|^2,$$

$$E_n = \hbar\omega_1\left(n+\frac{1}{2}\right), \ n = 0,1,2,...$$

$$\psi_n(x) = \frac{1}{\sqrt{2^n n!}} \left(\frac{m\omega_1}{\pi\hbar}\right)^{\frac{1}{4}} e^{-\frac{m\omega_1 x^2}{2\hbar}} H_n\left(\sqrt{\frac{m\omega_1}{\hbar}} x\right), \tag{6.7}$$

where $H_n$ is the Hermite polynomials.

The function $F_{2,n}$ is the Wigner function for the harmonic oscillator. The form of the functions $F_{k,n}$, $k > 2$ can be constructed from the characteristic equations as follows

$$F_{1,n} = c_1 e^{-\frac{\xi_1}{2\sigma_r^2}} H_n^2\left(\sqrt{\frac{\xi_1}{2\sigma_r^2}}\right), \tag{6.8}$$

$$F_{2,n} = c_2 e^{-\frac{\xi_2}{2\sigma_v^2}} L_n\left(\frac{\xi_2}{\sigma_v^2}\right), \tag{6.9}$$

$$F_{3,n}(\xi_3, \eta_3) = c_3 e^{-\frac{\xi_3}{2\sigma_{\dot v}^2}} L_n\left(\frac{\xi_3}{\sigma_{\dot v}^2}\right) \delta(\eta_3), \tag{6.10}$$

$$F_{4,n}(\xi_4, \eta_4, \varsigma_4) = c_4 e^{-\frac{\xi_4}{2\sigma_{\ddot v}^2}} L_n\left(\frac{\xi_4}{\sigma_{\ddot v}^2}\right) \delta(\eta_4) \delta(\varsigma_4), \tag{6.11}$$

...

where $L_n$ is polynomials Laguerre, and $\delta$ is the Dirac delta function. As a result, the solution of the Vlasov equation chain (6.4) for the harmonic oscillator is of the form

$$f_{1,n}(x) = c_1 e^{-\frac{x^2}{2\sigma_r^2}} H_n^2\left(\frac{x}{\sqrt{2}\sigma_r}\right), \tag{6.12}$$

$$f_{2,n}(x,v) = c_2 e^{-\frac{v^2 + \omega_1^2 x^2}{2\sigma_v^2}} L_n\left(\frac{v^2 + \omega_1^2 x^2}{\sigma_v^2}\right), \tag{6.13}$$

$$f_{3,n}(x,v,\dot v) = c_3 e^{-\frac{\dot v^2 + \omega_1^2 v^2}{2\sigma_{\dot v}^2}} L_n\left(\frac{\dot v^2 + \omega_1^2 v^2}{\sigma_{\dot v}^2}\right) \delta(\dot v + \omega_1^2 x), \tag{6.14}$$

$$f_{4,n}(x,v,\dot v,\ddot v) = c_4 e^{-\frac{\ddot v^2 + \omega_1^2 \dot v^2}{2\sigma_{\ddot v}^2}} L_n\left(\frac{\ddot v^2 + \omega_1^2 \dot v^2}{\sigma_{\ddot v}^2}\right) \delta(\ddot v + \omega_1^2 v) \times$$
$$\times \delta\left(\omega_1(\dot v + \omega_1^2 x) + (\ddot v + \omega_1^2 v) \arcsin\frac{\ddot v}{\sqrt{\ddot v^2 + \omega_1^2 \dot v^2}}\right), \tag{6.15}$$

...

**Remark**

As it is seen from the solutions (6.12)-(6.15), starting from the third function $f_{3,n}, f_{4,n}, ...$ and further all the solutions are obtained from the previous solutions by multiplying by the Dirac delta function of the following characteristic. Such a solution structure is because only the first two solutions $f_{1,n}$ and $f_{2,n}$ are «independent». For the harmonic oscillator all the kinematic characteristic of higher orders $\dot v, \ddot v, \dddot v, ...$ according to (6.2), are expressed only in terms of coordinate and velocity, indeed

$$\dot{v}=-\omega_1^2 x, \quad \ddot{v}=-\omega_1^2 v, \quad \dddot{v}=\omega_1^4 x, \quad \ddddot{v}=\omega_1^4 v,\ldots \tag{6.16}$$

The coefficients $c_k$ in (6.12)-(6.15) can be defined from the normalization condition of the probability density function (2.3). Indeed, for the function $f_{3,n}(x,v,\dot{v})$ we obtain

$$f_{3,n}(x,v,\dot{v}) = \int_{-\infty}^{+\infty} f_{4,n}(x,v,\dot{v},\ddot{v})d\ddot{v} =$$

$$= c_4 \int_{-\infty}^{+\infty} e^{-\frac{\ddot{v}^2+\omega_1^2\dot{v}^2}{2\sigma_{\ddot{v}}^2}} L_n\left(\frac{\ddot{v}^2+\omega_1^2\dot{v}^2}{\sigma_{\ddot{v}}^2}\right)\delta(\ddot{v}+\omega_1^2 v)\delta\left(\omega_1(\dot{v}+\omega_1^2 x)+(\ddot{v}+\omega_1^2 v)\arcsin\frac{\ddot{v}}{\sqrt{\ddot{v}^2+\omega_1^2\dot{v}^2}}\right)d\ddot{v} =$$

$$= \frac{c_4}{\omega_1} e^{-\frac{\omega_1^2(\omega_1^2 v^2+\dot{v}^2)}{2\sigma_{\ddot{v}}^2}} L_n\left(\frac{\omega_1^2(\omega_1^2 v^2+\dot{v}^2)}{\sigma_{\ddot{v}}^2}\right)\delta(\dot{v}+\omega_1^2 x),$$

It follows that

$$c_3 = \frac{c_4}{\omega_1}, \quad \omega_1 = \frac{\sigma_{\ddot{v}}}{\sigma_{\dot{v}}}. \tag{6.17}$$

By analogy, for the function $f_{2,n}(x,v)$

$$f_{2,n}(x,v) = \int_{-\infty}^{+\infty} f_{3,n}(x,v,\dot{v})d\dot{v} = c_3 \int_{-\infty}^{+\infty} e^{-\frac{\omega_1^2 v^2+\dot{v}^2}{2\sigma_{\dot{v}}^2}} L_n\left(\frac{\omega_1^2 v^2+\dot{v}^2}{\sigma_{\dot{v}}^2}\right)\delta(\dot{v}+\omega_1^2 x)d\dot{v} =$$

$$= c_3 e^{-\frac{\omega_1^2(v^2+\omega_1^2 x^2)}{2\sigma_{\dot{v}}^2}} L_n\left(\frac{\omega_1^2(v^2+\omega_1^2 x^2)}{\sigma_{\dot{v}}^2}\right),$$

from this we obtain

$$c_2 = c_3 = \frac{c_4}{\omega_1}, \quad \omega_1 = \frac{\sigma_{\dot{v}}}{\sigma_v} = \frac{\sigma_{\ddot{v}}}{\sigma_{\dot{v}}}. \tag{6.18}$$

As a result, the function $f_{2,n}(x,v)$ is of the form

$$f_{2,n}(x,v) = c_2 e^{-\frac{(v^2+\omega_1^2 x^2)}{2\sigma_v^2}} L_n\left(\frac{v^2+\omega_1^2 x^2}{\sigma_v^2}\right) \tag{6.19}$$

Since the function $f_{2,n}(x,v)$ is the Wigner function

$$W_n(x,p) = \frac{(-1)^n}{\pi\hbar} e^{-2\varepsilon(x,p)} L_n(4\varepsilon(x,p)), \quad \varepsilon(x,p) = \frac{1}{\hbar\omega_1}\left(\frac{p^2}{2m}+\frac{m\omega_1^2 x^2}{2}\right), \tag{6.20}$$

or

$$W_n(x,v) \stackrel{det}{=} mW_n(x,mv) = \frac{(-1)^n m}{\pi \hbar} e^{-m\frac{v^2+\omega_1^2 x^2}{\hbar \omega_1}} L_n\left(2m\frac{v^2+\omega_1^2 x^2}{\hbar \omega_1}\right),$$

then comparing $W_n(x,v)$ (6.20) and $f_{2,n}(x,v)$ (6.19), we obtain

$$\sigma_v^2 = \frac{\hbar \omega_1}{2m}, \quad c_2 = c_3 = \frac{(-1)^n m}{\pi \hbar}, \quad c_4 = \frac{(-1)^n m\omega_1}{\pi \hbar}. \tag{6.21}$$

Taking into account the uncertainty relation for the harmonic oscillator $\sigma_v \sigma_r = \frac{\hbar}{2m}$, we obtain

$$\omega_1 = \frac{\sigma_v}{\sigma_r} = \frac{\sigma_{\dot{v}}}{\sigma_v} = \frac{\sigma_{\ddot{v}}}{\sigma_{\dot{v}}}. \tag{6.22}$$

Thus, the solutions (6.12)-(6.15) of the Vlasov chain are of the form

$$f_{1,n}(x) = \frac{1}{2^n n!} \frac{1}{\sqrt{2\pi}\sigma_r} e^{-\frac{x^2}{2\sigma_r^2}} H_n^2\left(\frac{x}{\sqrt{2}\sigma_r}\right), \tag{6.23}$$

$$f_{2,n}(x,v) = \frac{(-1)^n}{2\pi\sigma_v\sigma_r} e^{-\frac{v^2}{2\sigma_v^2} - \frac{x^2}{2\sigma_r^2}} L_n\left(2\left(\frac{v^2}{2\sigma_v^2} + \frac{x^2}{2\sigma_r^2}\right)\right), \tag{6.24}$$

$$f_{3,n}(x,v,\dot{v}) = \frac{(-1)^n}{2\pi\sigma_v\sigma_r} e^{-\frac{\dot{v}^2}{2\sigma_{\dot{v}}^2} - \frac{v^2}{2\sigma_v^2}} L_n\left(2\left(\frac{\dot{v}^2}{2\sigma_{\dot{v}}^2} + \frac{v^2}{2\sigma_v^2}\right)\right) \delta(\dot{v} + \omega_1^2 x), \tag{6.25}$$

$$f_{4,n}(x,v,\dot{v},\ddot{v}) = \frac{(-1)^n}{2\pi\sigma_v\sigma_r} e^{-\frac{\ddot{v}^2}{2\sigma_{\ddot{v}}^2} - \frac{\dot{v}^2}{2\sigma_{\dot{v}}^2}} L_n\left(2\left(\frac{\ddot{v}^2}{2\sigma_{\ddot{v}}^2} + \frac{\dot{v}^2}{2\sigma_{\dot{v}}^2}\right)\right) \delta(\ddot{v} + \omega_1^2 v) \times$$

$$\times \delta\left(\dot{v} + \omega_1^2 x + \frac{\ddot{v} + \omega_1^2 v}{\omega_1} \arcsin \frac{\ddot{v}}{\sqrt{\ddot{v}^2 + \omega_1^2 \dot{v}^2}}\right), \tag{6.26}$$

…

The functions (6.23) and (6.24) are known functions. The functions (6.25), (6.26),.. are new solutions and have never been obtained before.

The mean values of the kinematic variables (2.4) are of the form

$$\langle v \rangle = \frac{1}{f_{1,n}(x)} \int_{-\infty}^{+\infty} f_{2,n}(x,v) v \, dv = 0, \quad \langle\langle v \rangle\rangle = 0,$$

$$\langle \dot{v} \rangle = \frac{1}{f_{2,n}(x,v)} \int_{-\infty}^{+\infty} f_{3,n}(x,v,\dot{v}) \dot{v} \, d\dot{v} = -\omega_1^2 x, \quad \langle\langle \dot{v} \rangle\rangle = -\omega_1^2 x, \quad \langle\langle\langle \dot{v} \rangle\rangle\rangle = 0, \quad (6.27)$$

$$\langle \ddot{v} \rangle = \frac{1}{f_{3,n}(x,v,\dot{v})} \int_{-\infty}^{+\infty} f_{4,n}(x,v,\dot{v},\ddot{v}) \ddot{v} \, d\ddot{v} = -\omega_1^2 v, \quad \langle\langle \ddot{v} \rangle\rangle = -\omega_1^2 v, \quad \langle\langle\langle \ddot{v} \rangle\rangle\rangle = 0, \quad \langle\langle\langle\langle \ddot{v} \rangle\rangle\rangle\rangle = 0,$$

...

The obtained expressions (6.27) correspond to (6.3).

Let us calculate standard deviations $\Sigma^2$ of the kinematic variables. Note that in the general case, the variables introduced earlier $\sigma_r^2, \sigma_v^2, \sigma_{\dot{v}}^2, \sigma_{\ddot{v}}^2, ...$ do not have to coincide with the standard deviations $\Sigma^2$ of the corresponding random variables $X, V, \dot{V}, \ddot{V}, ...$. As an example, let us calculate $\Sigma_{\ddot{V},0}^2$.

$$\Sigma_{\ddot{V},n}^2 = \int_{-\infty}^{+\infty}\int_{-\infty}^{+\infty}\int_{-\infty}^{+\infty}\int_{-\infty}^{+\infty} f_{4,n}(x,v,\dot{v},\ddot{v}) \left(\ddot{v} - \langle\langle\langle\langle \ddot{v} \rangle\rangle\rangle\rangle\right)^2 dx\, dv\, d\dot{v}\, d\ddot{v} =$$

$$= \frac{(-1)^n \omega_1^4}{2\pi\sigma_v \sigma_r} \int_{-\infty}^{+\infty}\int_{-\infty}^{+\infty}\int_{-\infty}^{+\infty} e^{-\frac{\omega_1^2 v^2}{2\sigma_{\dot{v}}^2} - \frac{\dot{v}^2}{2\sigma_{\ddot{v}}^2}} L_n\left(2\left(\frac{\omega_1^2 v^2}{2\sigma_{\dot{v}}^2} + \frac{\dot{v}^2}{2\sigma_{\ddot{v}}^2}\right)\right) \delta(\dot{v} + \omega_1^2 x) v^2 \, dx\, dv\, d\dot{v} =$$

$$= \frac{(-1)^n \omega_1^4}{2\pi\sigma_v \sigma_r} \int_{-\infty}^{+\infty}\int_{-\infty}^{+\infty} e^{-\frac{v^2}{2\sigma_v^2} - \frac{\omega_1^2 x^2}{2\sigma_v^2}} L_n\left(2\left(\frac{v^2}{2\sigma_v^2} + \frac{\omega_1^2 x^2}{2\sigma_v^2}\right)\right) v^2 \, dx\, dv,$$

it follows that

$$\Sigma_{\ddot{V},0}^2 = \frac{\omega_1^4}{2\pi\sigma_v \sigma_r} \int_{-\infty}^{+\infty}\int_{-\infty}^{+\infty} e^{-\frac{v^2}{2\sigma_v^2} - \frac{\omega_1^2 x^2}{2\sigma_v^2}} v^2 \, dx\, dv = \frac{4\sigma_v^4 \omega_1^3}{2\pi\sigma_v \sigma_r} \int_{-\infty}^{+\infty} e^{-\frac{\omega_1^2 x^2}{2\sigma_v^2}} d\frac{\omega_1 x}{\sqrt{2}\sigma_v} \int_{-\infty}^{+\infty} e^{-\frac{v^2}{2\sigma_v^2}} \frac{v^2}{2\sigma_v^2} d\frac{v}{\sqrt{2}\sigma_v} =$$

$$= \frac{2\sigma_v^3 \omega_1^3}{\pi\sigma_r} \int_{-\infty}^{+\infty} e^{-\mu^2} d\mu \int_{-\infty}^{+\infty} e^{-\lambda^2} \lambda^2 d\lambda = \frac{2\sigma_v^3 \omega_1^3}{\pi\sigma_r} \sqrt{\pi} \frac{\sqrt{\pi}}{2} = \frac{\sigma_v^3 \omega_1^4}{\sigma_r \omega_1} = \frac{\sigma_v^3 \sigma_{\ddot{v}}^2}{\sigma_r \sigma_{\dot{v}}^2 \omega_1} \omega_1^2 = \frac{\sigma_v^3 \sigma_{\ddot{v}}^2 \sigma_{\dot{v}}^2}{\sigma_r \sigma_{\dot{v}}^2 \sigma_v^2 \omega_1} =$$

$$= \frac{\sigma_v \sigma_{\ddot{v}}^2}{\sigma_r \omega_1} = \frac{\omega_1}{\omega_1} \sigma_{\ddot{v}}^2,$$

$$\Sigma_{\ddot{V},0}^2 = \sigma_{\ddot{v}}^2. \quad (6.28)$$

**Remark**

Note that in deriving the Vlasov equation chain [5,6,8] (2.1)/(2.5) the condition of the positivity of the probability density function is not imposed anywhere. Therefore, the presence of negative quantities for the Wigner function (6.24), for example, or for subsequent functions (6.25), (6.26) is consistent.

The question arises if there are positive probability density functions in the phase space $(x,v)$ and further in the generalized phase space $(x,v,\dot{v},\ddot{v},...)$. The answer to this question can be obtained on the base of the chain of Vlasov equations (2.1).

Let us suppose there is a solution $f_{1,n}(x)$ (6.23), obtained from the Schrödinger equation for the quantum harmonic oscillator. The function $f_{2,n}(x)$ is related to the functions $f_{1,n}(x)$ by the relation (2.3)

$$f_{1,n}(x) = \int_{-\infty}^{+\infty} f_{2,n}(x,v)\,dv. \tag{6.29}$$

The solution of the equation chain (2.1) is obtained by the method of characteristics. Along the characteristics (6.6), the probability density remains constant. For the function $f_{2,n}(x)$ according to (6.6), circles (ellipses) are the characteristics

$$v^2 + \omega_1^2 x^2 = const. \tag{6.30}$$

If the function $f_{1,n}(x)$ has zeros, that is $\exists x_k, k=1,\ldots,n : f_{1,n}(x_k)=0$, then it follows from (6.29) that

$$\int_{-\infty}^{+\infty} f_{2,n}(x_k, v)\,dv = 0. \tag{6.31}$$

Let the probability density function satisfies the condition $f_{2,n}(x,v) \geq 0$, then it follows from (6.31) that $f_{2,n}(x_k, v) = 0$ almost on the whole interval $v \in (-\infty, +\infty)$. Consequently, on the whole circle (6.30) passing through the point $(x_k, 0)$, the function $f_{2,n}$ also equals zero (see Fig.3).

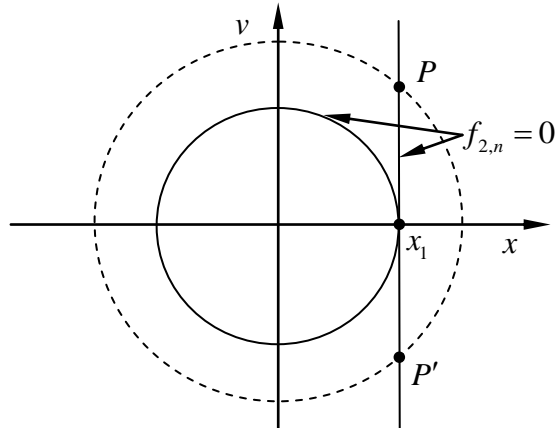

Fig. 3 Values of the function $f_{2,n}$

Fig. 3 shows the position of the first zero $x_1$ of the function $f_{1,n}$. All circles (6.30) of a larger radius intersect the vertical line $x = x_1$, for example, in points $P$ and $P'$. Consequently, on all such circles the value of the function $f_{2,n}$ equals zero. That is the function $f_{2,n}$ differs from zero only inside a circle of the radius $x_1$.

Thus, of all functions $f_{2,n}$, only the functions for which $f_{1,n}$ does not have zeros are positive in the whole domain. Only one function in (6.23) does not have zeros – it is $f_{1,0}$. Therefore of all functions in (6.24), only $f_{2,0}$ is positive in the whole domain. Analog statements are true for the rest functions in the chain.

Positive solutions can be obtained from the solution of the wave equations (2.31), (2.32). As an example, we consider the equation (2.37) for the wave function $\Psi_2(\vec{r},\vec{v},t)$. For the one-dimensional harmonic oscillator, the equation (2.37) is as follows

$$\frac{i}{\beta_2}\frac{\partial \Psi_2}{\partial t} = \frac{\alpha_2}{\beta_2}\frac{\partial^2}{\partial v^2}\Psi_2 - \frac{i}{\beta_2}v\frac{\partial}{\partial x}\Psi_2 + U_2\Psi_2. \qquad (6.32)$$

Let us represent the solution in the form

$$\Psi_2(\vec{r},\vec{v},t) = \psi(\vec{r},\vec{v})e^{-i\beta_2 \mathcal{E}_2 t}, \qquad (6.33)$$

where $\mathcal{E}_2$ is a constant value. Substituting (6.33) into (6.32), we obtain

$$\alpha_2 \psi_{vv} - iv\psi_x + \beta_2(U_2 - \mathcal{E}_2)\psi = 0. \qquad (6.34)$$

It follows from (6.24) and from Fig.3 that the solution (6.34) can be obtained on the form

$$\psi = c e^{-\frac{v^2}{4\sigma_v^2} - \frac{\omega_1^2 x^2}{4\sigma_v^2} + i\lambda xv}, \qquad (6.35)$$

where $\lambda$ is some constant variable. Substituting (6.35) into (6.34), we obtain

$$\psi_x = c e^{-\frac{v^2}{4\sigma_v^2} - \frac{\omega_1^2 x^2}{4\sigma_v^2} + i\lambda xv}\left(-\frac{\omega_1^2 x}{2\sigma_v^2} + i\lambda v\right) = \left(-\frac{\omega_1^2 x}{2\sigma_v^2} + i\lambda v\right)\psi,$$

$$\psi_v = c e^{-\frac{v^2}{4\sigma_v^2} - \frac{\omega_1^2 x^2}{4\sigma_v^2} + i\lambda xv}\left(-\frac{v}{2\sigma_v^2} + i\lambda x\right) = \left(-\frac{v}{2\sigma_v^2} + i\lambda x\right)\psi,$$

$$\psi_{vv} = \left(-\frac{v}{2\sigma_v^2} + i\lambda x\right)\psi_v - \frac{1}{2\sigma_v^2}\psi = \left(-\frac{v}{2\sigma_v^2} + i\lambda x\right)^2 \psi - \frac{1}{2\sigma_v^2}\psi =$$

$$= \left(\frac{v^2}{4\sigma_v^4} - \lambda^2 x^2 - \frac{1}{2\sigma_v^2} - i\frac{2\lambda}{2\sigma_v^2}vx\right)\psi = \frac{1}{\sigma_v^2}\left(\frac{v^2}{4\sigma_v^2} - \lambda^2 \sigma_v^2 x^2 - \frac{1}{2} - i\lambda vx\right)\psi,$$

as a result

$$\alpha_2 \frac{1}{\sigma_v^2}\left(\frac{v^2}{4\sigma_v^2} - \lambda^2 \sigma_v^2 x^2 - \frac{1}{2} - i\lambda vx\right)\psi - iv\left(-\frac{\omega_1^2 x}{2\sigma_v^2} + i\lambda v\right)\psi + \beta_2(U_2 - \mathcal{E}_2)\psi = 0,$$

$$\frac{\alpha_2}{\sigma_v^2}\left(\frac{v^2}{4\sigma_v^2} - \lambda^2 \sigma_v^2 x^2 - \frac{1}{2} - i\lambda vx\right) - iv\left(-\frac{\omega_1^2 x}{2\sigma_v^2} + i\lambda v\right) + \beta_2(U_2 - \mathcal{E}_2) = 0,$$

$$v^2 \frac{\alpha_2}{4\sigma_v^4} + \lambda v^2 - \alpha_2 \lambda^2 x^2 + \beta_2 (U_2 - \mathcal{E}_2) - \frac{\alpha_2}{2\sigma_v^2} + i(\omega_1^2 - 2\alpha_2 \lambda)\frac{vx}{2\sigma_v^2} = 0. \tag{6.36}$$

The expression (6.36) is complex, therefore it is equivalent to two conditions

$$v^2 \frac{\alpha_2}{4\sigma_v^4} + \lambda v^2 - \alpha_2 \lambda^2 x^2 + \beta_2 (U_2 - \mathcal{E}_2) - \frac{\alpha_2}{2\sigma_v^2} = 0,$$
$$\omega_1^2 - 2\alpha_2 \lambda = 0. \tag{6.37}$$

It follows from the second condition (6.37) that

$$\lambda = \frac{\omega_1^2}{2\alpha_2}. \tag{6.38}$$

From the first condition (6.37) taking into consideration (6.38), we obtain

$$v^2 \frac{\alpha_2}{4\sigma_v^4} + \frac{\omega_1^2}{2\alpha_2} v^2 - \frac{\omega_1^4}{4\alpha_2} x^2 + \beta_2 U_2 - \beta_2 \mathcal{E}_2 - \frac{\alpha_2}{2\sigma_v^2} = 0. \tag{6.39}$$

The variable $\mathcal{E}_2$ by the definition (6.33) is constant, therefore we suppose

$$\mathcal{E}_2 = -\frac{\alpha_2}{2\beta_2 \sigma_v^2} = \frac{\sigma_{\dot{v}}}{2\beta_2 \sigma_v} = \frac{\omega_2}{2\beta_2} = \frac{\hbar_2 \omega_2}{2}, \tag{6.40}$$

where for the harmonic oscillator the conditions (2.33), (2.35) are taken into consideration

$$\sigma_v \sigma_{\dot{v}} = -\alpha_2 = \frac{\hbar_2}{2m}, \quad \omega_2 = \frac{\sigma_{\dot{v}}}{\sigma_v}. \tag{6.41}$$

As a result, from (6.39) with consideration of (6.40) for the potential $U_2$ we obtain the expression

$$U_2 = -\frac{1}{\beta_2}\left(\frac{\alpha_2}{4\sigma_v^4} + \frac{\omega_1^2}{2\alpha_2}\right)v^2 + \frac{\omega_1^4}{4\alpha_2 \beta_2} x^2,$$

$$U_2 = -\frac{\omega_1^2}{2\alpha_2 \beta_2}\left[\left(1 + \frac{\alpha_2^2}{2\sigma_v^4 \omega_1^2}\right)v^2 - \frac{\omega_1^2 x^2}{2}\right],$$

or, taking into consideration (6.41),

$$U_2(x,v) = m\omega_1^2\left[\left(1 + \frac{\omega_2^2}{2\omega_1^2}\right)v^2 - \frac{\omega_1^2 x^2}{2}\right]. \tag{6.42}$$

The probability density function $f_{2,0}$ is of the form

$$f_{2,0}(x,v) = c^2 e^{-\frac{v^2}{2\sigma_v^2} - \frac{\omega_1^2 x^2}{2\sigma_v^2}} = c^2 e^{-\frac{v^2}{2\sigma_v^2} - \frac{x^2}{2\sigma_r^2}}, \tag{6.43}$$

which responds to (6.23) $f_{2,0}(x,v)$. The expression for the phase of the wave function $\Psi_2$ is of the form (6.33), (6.35)

$$\varphi_2 = -\frac{m\omega_1^2}{\hbar_2} xv - \frac{\mathcal{E}_2}{\hbar_2} t = -\frac{m\omega_1^2}{\hbar_2} xv - \frac{\omega_2}{2} t. \tag{6.44}$$

It follows from (6.44) and (2.4) that the mean accelerations of the form

$$\langle \dot{v} \rangle = -2\alpha_2 \frac{\partial \varphi_2}{\partial v} = 2\frac{\hbar_2}{2m} \frac{\partial \varphi_2}{\partial v} = -\frac{\hbar_2}{m} \frac{m\omega_1^2 x}{\hbar_2} = -\omega_1^2 x. \tag{6.45}$$

The expression (6.45) coincides with the Vlasov approximation and with (6.3), (6.27)

$$\langle \dot{v} \rangle = -\frac{1}{m} \frac{\partial U_1}{\partial x} = -\frac{1}{m} \frac{\partial}{\partial x} \frac{m\omega_1^2 x^2}{2} = -\omega_1^2 x. \tag{6.46}$$

The expression (6.42) for the potential $U_2$ can be obtained also by the general formula (3.5), (3.6), (3.9)

$$U_2(x,v) = -\frac{1}{\beta_2} \left\{ -\beta_2 \mathcal{E}_2 + \alpha_2 \left[ \frac{1}{2} \left( \Delta_v S_{2,0} + \frac{|\nabla_v S_{2,0}|^2}{2} \right) - \frac{m^2 \omega_1^4 x^2}{\hbar_2^2} \right] - \frac{m\omega_1^2 v^2}{\hbar_2} \right\},$$

where $S_{2,0} = \ln f_{2,0}$

$$S_{2,0}(x,v) = -\frac{v^2}{2\sigma_v^2} - \frac{\omega_1^2 x^2}{2\sigma_v^2} + const.$$

$$\frac{\partial S_{2,0}}{\partial v} = -\frac{v}{\sigma_v^2}, \quad \frac{\partial^2 S_{2,0}}{\partial v^2} = -\frac{1}{\sigma_v^2}. \tag{6.47}$$

As a result,

$$U_2(x,v) = \mathcal{E}_2 + \frac{\hbar_2^2}{4m} \left( -\frac{1}{\sigma_v^2} + \frac{v^2}{2\sigma_v^4} \right) - \frac{m\omega_1^4 x^2}{2} + m\omega_1^2 v^2 =$$

$$= \mathcal{E}_2 + \frac{\hbar_2^2}{4\sigma_v^2 m} \left( \frac{v^2}{2\sigma_v^2} - 1 \right) - \frac{m\omega_1^4 x^2}{2} + m\omega_1^2 v^2 =$$

$$= v^2 \left( m\omega_1^2 + \frac{\hbar_2^2}{8\sigma_v^4 m} \right) - \frac{m\omega_1^4 x^2}{2} + \mathcal{E}_2 - \frac{\hbar_2 \sigma_v \sigma_{\dot{v}}}{2\sigma_v^2} = \tag{6.48}$$

$$= \omega_1^2 \left[ mv^2 \left( 1 + \frac{\hbar_2^2}{8\sigma_v^4 \omega_1^2 m^2} \right) - \frac{m\omega_1^2 x^2}{2} \right] + \mathcal{E}_2 - \frac{\hbar_2 \omega_2}{2} = m\omega_1^2 \left[ v^2 \left( 1 + \frac{\omega_2^2}{2\omega_1^2} \right) - \frac{\omega_1^2 x^2}{2} \right].$$

The expression (6.48) completely coincides with the expression (6.42), which was to be proved.

Let us write the motion equations (see §4). For the representation (4.16) ($n=2$), we obtain

$$\langle \dot{\vec{v}} \rangle = -\omega_1^2 x \vec{e}_x,$$

$$^{(2)}\chi_2 = \frac{2\alpha_2 \beta_2}{\gamma_2}(Q_2 + U_2) + {}^{(2)}c_2 = \frac{2\alpha_2 \beta_2}{\gamma_2}\left(\mathcal{E}_2 + m\omega_1^2\left(v^2 - \frac{\omega_1^2 x^2}{2}\right)\right) + {}^{(2)}c_2, \qquad (6.49)$$

$$^{(1)}\chi_2 = \frac{\alpha_2}{\gamma_2}\Phi_2 + {}^{(1)}c_2 = -\frac{2\alpha_2}{\gamma_2}\left(\frac{m\omega_1^2}{\hbar_2}xv + \frac{\mathcal{E}_2}{\hbar_2}t\right) + {}^{(1)}c_2,$$

$$\vec{E}_2 = -\nabla_v{}^{(2)}\chi_2 - \nabla_r{}^{(1)}\chi_2 = \vec{e}_x \frac{2\alpha_2}{\gamma_2}\left(\frac{m\omega_1^2}{\hbar_2}v - \frac{2m\omega_1^2}{\hbar_2}v\right) = -\frac{2\alpha_2}{\gamma_2}\frac{m\omega_1^2}{\hbar_2}v\vec{e}_x = \frac{\omega_1^2 v}{\gamma_2}\vec{e}_x,$$

$$^{(1)}\vec{B}_2 = \vec{0}, \quad {}^{(2)}\vec{B}_2 = \vec{0}.$$

As a result, considering (6.49), the motion equation (4.16) is as follows

$$\frac{d_2}{dt}\langle \dot{\vec{v}} \rangle = -\gamma_2 \vec{E}_2, \qquad (6.50)$$

which is a valid identical equation as

$$-\omega_1^2\left(\frac{\partial}{\partial t} + v\frac{\partial}{\partial x} + \langle \dot{v} \rangle \frac{\partial}{\partial v}\right)x = -\omega_1^2 v = -\gamma_2 \vec{E}_2 = -\omega_1^2 v,$$

which was to be proved.

The Hamilton-Jacobi equation (3.17) with (6.44), (6.49) gives a valid identical equation

$$-\hbar_2 \frac{\partial \varphi_2}{\partial t} = \frac{m}{2}\left|\langle \dot{\vec{v}} \rangle\right|^2 + e\chi_2 = H_2,$$

$$\frac{\hbar_2 \omega_2}{2} = \frac{m\omega_1^4 x^2}{2} + \frac{\hbar_2 \omega_2}{2} - \frac{m\omega_1^4 x^2}{2} = H_2, \qquad (6.51)$$

where

$$e\chi_2 = U_2 + Q_2 + \hbar_2 v \frac{\partial \varphi_2}{\partial x} = \frac{\hbar_2 \omega_2}{2} + m\omega_1^2\left(v^2 - \frac{\omega_1^2 x^2}{2}\right) - m\omega_1^2 v^2 = \frac{\hbar_2 \omega_2}{2} - \frac{m\omega_1^4 x^2}{2}.$$

For the quantum pressure (4.21) and the quantum pressure force (4.23), we obtain

$$^{(2)}P^{(q)} = -\alpha_2^2 f_{2,0}\frac{\partial^2 S_{2,0}}{\partial v^2} = \frac{\sigma_v^2}{2\pi\sigma_v\sigma_r}e^{-\frac{v^2}{2\sigma_v^2} - \frac{x^2}{2\sigma_r^2}},$$

$$-\frac{1}{m}\frac{\partial Q_2}{\partial v} = -\frac{1}{f_{2,0}}\frac{\partial{}^{(2)}P^{(q)}}{\partial v} = \omega_2^2 v.$$

The generalized Lagrangian (3.11) with (6.44), (6.49) is of the form

$$-\frac{1}{\beta_2}\frac{d_2\varphi_2}{dt} = \frac{1}{2\alpha_2\beta_2}\frac{1}{2}\left|\langle\dot{\vec{v}}_p\rangle\right|^2 + U_2 + Q_2 = -L_2,$$

$$-\hbar_2\left(\frac{\partial}{\partial t} + v\frac{\partial}{\partial x} + \langle\dot{v}\rangle\frac{\partial}{\partial v}\right)\varphi_2 = -\frac{m}{2}\omega_1^4 x^2 + \frac{\hbar_2\omega_2}{2} + m\omega_1^2\left(v^2 - \frac{\omega_1^2 x^2}{2}\right) = -L_2,$$

which gives a valid identical equation

$$\frac{\hbar_2\omega_2}{2} + m\omega_1^2 v^2 - m\omega_1^4 x^2 = \frac{\hbar_2\omega_2}{2} + m\omega_1^2 v^2 - m\omega_1^4 x^2 = -L_2, \tag{6.52}$$

The Legendre transformation (3.21) for $H_2$ and $L_2$ taking (6.51), (6.52), (6.49), (6.44) into account also gives a valid identical equation

$$L_2 + H_2 = \hbar_2\left(\vec{v}, \nabla_r\varphi_2\right) + m\left(\langle\dot{\vec{v}}\rangle, \langle\dot{\vec{v}}_p\rangle\right).$$

$$-\frac{\hbar_2\omega_2}{2} - m\omega_1^2 v^2 + m\omega_1^4 x^2 + \frac{\hbar_2\omega_2}{2} = \hbar_2 v\frac{\partial\varphi_2}{\partial x} + m\omega_1^4 x^2,$$

$$-m\omega_1^2 v^2 + m\omega_1^4 x^2 = -m\omega_1^2 v^2 + m\omega_1^4 x^2.$$

At $n = 1$ the potentials $Q_{1,0}(x)$ and $U_1(x)$ are as follows

$$U_1(x) = \frac{m\omega_1^2 x^2}{2}, \tag{6.53}$$

$$Q_{1,0}(x) = \frac{\alpha_1}{2\beta_1}\left(S_{1,0}'' + \frac{1}{2}|S_{1,0}'|^2\right) = \frac{\alpha_1}{2\beta_1}\left(-\frac{\omega_1^2}{\sigma_v^2} + \frac{\omega_1^4 x^2}{2\sigma_v^4}\right) = \frac{\hbar_1\omega_1}{2} - \frac{m\omega_1^2 x^2}{2},$$

where

$$S_{1,0}(x) = -\frac{\omega_1^2 x^2}{2\sigma_v^2} + const, \quad \frac{\partial S_{1,0}}{\partial x} = -\frac{\omega_1^2 x}{\sigma_v^2}, \quad \frac{\partial^2 S_{1,0}}{\partial x^2} = -\frac{\omega_1^2}{\sigma_v^2}.$$

Having the energy of the ground state denoted $\mathcal{E}_{1,0} = \frac{\hbar_1\omega_1}{2}$, for the quantum potential we obtain the expression

$$Q_{1,0}(x) = \mathcal{E}_{1,0} - \frac{m\omega_1^2 x^2}{2}. \tag{6.54}$$

Taking into account (6.49), the condition (5.10) is as follows

$$U_1 + Q_{1,0} + {}^{(1)}c_1 = \frac{\gamma_1}{2\alpha_1\beta_1}\int_{-\infty}^{+\infty}\left(-\frac{2\alpha_2}{\gamma_2}\left(\frac{m\omega_1^2}{\hbar_2}xv + \frac{\mathcal{E}_2}{\hbar_2}t\right) + {}^{(1)}c_2\right)dv, \tag{6.55}$$

where $^{(1)}c_2$ and $^{(1)}c_1$ are some functions (4.11) to be defined. The expression (6.55) becomes a valid identitcal equation if we suppose $^{(1)}c_2$ and $^{(1)}c_1$ as follows

$$^{(1)}c_2 = \frac{2\alpha_2}{\gamma_2}\frac{\mathcal{E}_2}{\hbar_2}t, \quad ^{(1)}c_1 = -\mathcal{E}_{1,0}.$$

According to (5.7) and (6.49), the field $\vec{E}_1$ is of the form

$$\vec{E}_1(\vec{r},t) = \int_{-\infty}^{+\infty} \vec{E}_2(\vec{r},\vec{v},t)dv = \vec{e}_x \frac{\omega_1^2}{\gamma_2} \int_{-\infty}^{+\infty} v dv = \vec{0}. \tag{6.56}$$

As a result, the motion equation (4.14) is of the form

$$\frac{d}{dt}\langle \vec{v} \rangle = -\gamma_1 \vec{E}_1 = \vec{0}. \tag{6.57}$$

According to (6.3), (6.27), the equation (6.57) is a valid identical equation as $\langle \vec{v} \rangle = \vec{0}$. $\langle \vec{v} \rangle = \vec{0}$. For the quantum pressure (4.21) and the quantum pressure force (4.23), we obtain (ground state)

$$^{(1)}P^{(q)} = -\alpha_1^2 f_{1,0}\frac{\partial^2 S_{1,0}}{\partial x^2} = \frac{\sigma_v^2}{\sqrt{2\pi}\sigma_r}e^{-\frac{x^2}{2\sigma_r^2}},$$

$$-\frac{1}{m}\frac{\partial Q_{1,0}}{\partial x} = -\frac{1}{f_{1,0}}\frac{\partial ^{(1)}P^{(q)}}{\partial x} = \omega_1^2 x.$$

The generalized Hamilton-Jacobi equation (3.19)/(3.20), the expressions for the generalized Lagrangian (3.11) and the generalized Hamiltonian (3.14)/(3.15) in the case $n=1$ coincide with the corresponding classical representations.

**Conclusion**

The approach considered in this paper is internally coherent and in the particular case at $n=1$ becomes in the classical apparatus of the quantum mechanics and statistical physics.

The results of the paper can be used in constructing models that take into account higher orders of kinematic characteristics.